\documentclass[final,notitlepage,12pt]{article}  
\usepackage{amsfonts}
\usepackage{amssymb}
\usepackage{graphicx}   
\usepackage{amsmath} 
\usepackage{latexsym}
\usepackage{rotating}
\usepackage{xcolor}
\usepackage{setspace}

\newtheorem{theorem}{Theorem}[section]

\newtheorem{definition}[theorem]{Definition}
\newtheorem{example}[theorem]{Example}

\newtheorem{lemma}[theorem]{Lemma}

\newtheorem{proposition}[theorem]{Proposition}

\DeclareMathOperator{\diag}{diag}
\DeclareMathOperator{\rank}{rank}

\numberwithin{equation}{section}
\input{tcilatex}
\begin{document}

\title{\textbf{CAN  PRINCIPAL  COMPONENT  ANALYSIS PRESERVE THE SPARSITY IN FACTOR LOADINGS?}}
\author{Jie Wei$^{a}$, Yonghui Zhang$^{b}$\thanks{%
We are deeply indebted to the Co-editor and three referees for their many constructive comments, which improve the paper substantially. We are grateful to  Simon Freyaldenhoven and Takashi Yamagata for
helpful discussions as well as seminar and conference participants at Peking
University, Renmin University, 2023 IAAE and 2023 IPDC. We  also thank  Simon Freyaldenhoven, Cheng Yong Tang, and
Yoshimasa Uematsu for sharing their codes.
Wei gratefully
acknowledges the financial support from the Program for HUST Academic
Frontier Youth Team (2017QYTD13) and the Fundamental Research Funds for the
Central Universities (HUST: 2022WKYXZX014). Zhang gratefully acknowledges
the financial support from National Natural Science Foundation of China
(Projects No.71973141 and No.71873033). Address correspondence to: Yonghui Zhang, School of
Economics, Renmin University of China, Beijing, China;  e-mail: yonghui.zhang@ruc.edu.cn. All errors are the authors' sole
responsibilities.  Matlab codes for our paper are available upon request.
} \\
$^{a}$School of Economics, Huazhong University of Science and
Technology, China\\
$^{b}$School of Economics, Renmin University of China, China}
\maketitle

\begin{abstract}
This paper studies the principal component analysis (PCA) estimation of weak
factor models with sparse loadings. We uncover an intrinsic \textit{%
near-sparsity preservation} property for the PCA estimators of loadings,
which comes from the approximately (block) \textit{upper triangular}
structure of the rotation matrix. It suggests an asymmetric relationship
among factors: the sparsity of the rotated loadings  for  a stronger factor can be
contaminated by the loadings from weaker ones, but the  sparsity of the rotated loadings of a weaker
factor is almost unaffected by  the loadings of  stronger ones. Then we
propose a simple alternative to the existing penalized approaches to
sparsify the loading estimators by screening out the small PCA loading
estimators directly, and construct consistent estimators for factor
strengths. The finite sample performance of the proposed estimators is
investigated via a set of Monte Carlo simulations. An application to the
FRED-QD dataset reveals the general sparsity pattern in factor loadings as
well as their dynamic features.


\noindent \textbf{JEL\ Classification:} C12, C53, C55

\end{abstract}

\section{INTRODUCTION}

Factor models have been widely used in economics and finance. In a factor
model, 
\begin{equation}
X_{it}=\lambda _{i}^{0\prime }F_{t}^{0}+e_{it},\text{ }i=1,\ldots
,N,t=1,\ldots ,T,  \label{m0}
\end{equation}%
where $X_{it}$ is the observed data for the $i$th individual at time $t$, $%
F_{t}^{0}$ is an $r\times 1$ vector of latent factors, $\lambda _{i}^{0}$ is
the corresponding $r\times 1$ vector of factor loadings, and $e_{it}$ is the
idiosyncratic error with possible weak dependence across $i$ or/and over $t$%
. Until very recently, a large body of literature on factor models has built
on the assumption that all factors are strong in the sense that both $%
T^{-1}\sum_{t=1}^{T}F_{t}^{0}F_{t}^{0\prime }\ $and $N^{-1}\sum_{i=1}^{N}%
\lambda _{i}^{0}\lambda _{i}^{0\prime }$ converge to some positive definite
(p.d.) matrices. Bai and Ng (2002) and Bai (2003) establish the asymptotics
of principal component (PC) estimators when both $N$ and $T$ are large. It
is well-known that the linear factor space can be estimated consistently by
the PC method with rate $\min \left( N^{1/2},T^{1/2}\right) $, and both
estimated factors and loadings are consistent up to some rotation matrix.
However, the assumption of strong factors may fail due to the sparsity in
loadings or many nonzero but small loadings. In that case, we generally get a weak factor model where $N^{-1}\sum_{i=1}^{N}  \lambda _{i}^{0}\lambda _{i}^{0\prime }$ tends to be singular  while $T^{-1}\sum_{t=1}^{T}F_{t}^{0}F_{t}^{0\prime }$ converges to a p.d. matrix.  In this paper, we focus on the weak factor models with sparse loadings.

Many empirical studies support the wide existence of sparse factor loadings.
For instance, both Stock and Watson (2002) and Ludvigson and Ng (2009) find
that the extracted PC factors from a large set of macroeconomic variables
can only fit several variables' time series observations well, suggesting a
sparsity structure in loadings; also see Kristensen (2017), Freyaldenhoven
(2022), and Uematsu and Yamagata (2023a, b, UY hereafter). Moreover, hierarchical or group
factor models, where factors have nonzero loadings only for some specific
cross-sectional units or group, also echo the sparsity in loadings. For
example, oil supply shock only affects industrial production sectors but not
others; size-sorted portfolios are influenced by the size factor, but
momentum-sorted portfolios may not be. Additionally, weak factors may emerge
as unsystematic risk due to market incompleteness (Dello-Preite et al.,
2024). For more examples of sparsity in loadings, see Ando and Bai (2017)
and Choi et al. (2021).

When loadings are sparse enough, what we encounter turns to be a weak factor
model. A \textit{sparse weak} factor only affects a small subset of
individuals and the model with sparse weak factors is called
\textquotedblleft \textit{sparse weak factor model}\textquotedblright .%
\footnote{%
We formally define the sparse weak factor model in Section 2.} The
sparsity structure in loadings has attracted a lot of research interests.
For example, Pelger and Xiong (2022) consider the sparse approximation to
factor models, and Freyaldenhoven (2023) and Despois and Doz (2023) study
the identification of factor models based on sparsity. In the presence of
sparse weak factors, Giglio et al. (2023b) show that the prediction based on
factor-augmented regression (FAR), a popular method for macroeconomic
forecasting, is inconsistent; also see Chao and Swanson (2022) and Chao et
al. (2022) for inconsistent prediction in factor-augmented vector
autoregression (FAVAR). Giglio et al. (2023a) show that in the 3-step
estimation of risk premium in Giglio and Xiu (2021), sparse loadings raise a
severe attenuation bias. For extremely weak factor models, Onatski (2012)
shows that the PC estimator is inconsistent.\footnote{%
In Onatski (2012), factors are extremely weak in the sense that eigenvalues
corresponding to the factors are of the same order of magnitude as those of
the idiosyncratic component.} For general weak factors, Bai and Ng (2023)
re-investigate the asymptotics for PCA based on the singular value
decomposition. In a linear regression framework with interactive fixed
effects such as Bai (2009) and Su and Chen (2013), when some factors are
weak, Armstrong et al. (2023) show that previously developed estimators and
confidence intervals (CIs) might be heavily biased, and then propose
improved estimators and bias-aware CIs that are uniformly valid regardless
of whether the factors are strong or not.

To avoid the problems caused by sparse weak factors, two main solutions have
been proposed. One is to recover the sparsity in loadings with penalties.
UY (2023a) propose a sparse orthogonal factor regression
(SOFAR) estimator using $\ell _{1}$-regularization, and UY (2023b) further provide an inferential procedure to determine whether each
component of the loadings is zero or not with false discovery rate (FDR)
control. The other is to pursue a strong factor model via screening out
irrelevant individuals. 
Giglio et al. (2023a, b) propose a supervised PCA method for two distinct purposes: selecting test assets to evaluate new factors and choosing predictors to forecast a target macroeconomic variable. By eliminating assets that lack exposure to the factor under evaluation or dropping predictors that are uncorrelated to the target variable, each factor within the  model becomes strong.
Similar ideas are also employed in Chao and Swanson (2022) and Chao et al. (2022) in FAVAR.

The regularization-based approaches are often relatively complicated and
carry heavy computational burdens due to the search of  tuning parameters or
involving complex iterative algorithms. In contrast, we directly investigate
the properties of the PCA method and propose to recover the sparsity in
loadings via screening  the PC estimator of loadings. There are several possible
contributions for our paper to the existing literature. \textit{First}, we
show the PC estimators for loadings are almost \textit{sparsity-preserving}.
To the best of our knowledge, our paper is the first one to reveal this
intrinsic sparsity preservation feature of the PCA estimator for factor
models. This result is in sharp contrast to the prevailing understanding
that the sparsity of loadings cannot be preserved due to a matrix rotation;
see Bailey et al. (2021, BKP hereafter) and Freyaldenhoven (2023). \textit{%
Second}, given the (almost) invariance of sparsity and the consistency of PC
estimators, we can easily\textit{\ recover the sparsity} of loadings with
proper screening to the PC estimator for loadings. The revelation of sparsity is
important since it can lead to a better interpretation and understanding of,
e.g., which factors are relatively important to which individuals. Another appealing
advantage of our method is that the computational burden is almost negligible
compared to the regularization-based procedures. We only need to run regular
PCA once and then apply our screening device to the estimated loadings
directly. \textit{Third}, based on the sparsified loading estimators, we can
straightforwardly estimate distinct factor strengths, and establish
consistency of the strength estimators. We allow for different factor
strengths for different factors and further improve BKP (2021) where only
the strength of the strongest latent factor is identified without using complicated sequential procedures. 
The estimated factor strengths as indicators can provide more valuable guidance for forecasting and policy making. 
\textit{Fourth}, for sparse weak factor models with mixed sparsity
degrees, we establish the large sample properties for PC estimators of
factors, loadings, and common components. As complements to Bai and Ng
(2023), we provide the uniform convergence results for PC estimates which
are useful in further analysis such as constructing diffusion indices and
factor instrumental regressions. \textit{Lastly}, given relatively limited works on determining the number of factors for weak factor
models such as Freyaldenhoven (2022) and Guo et al. (2022), we complement the existing literature by providing a simple method. For weak
factors, we choose the number of factors using singular value thresholding
(SVT), which is easier to apply and compute, and justify the validity of the
method.

Lastly, we need to mention the work by UY (2023a)  which is most closely related to our study.  Similar to our work, they also aim  to recover the 
sparsity of loadings matrix. In particular, their parameter of interest or \textit{pseudo-true} parameter is 
the representation (but not necessary the true DGP) of $\Lambda^{0}$ and $F^{0}$ in which they  satisfy $%
T^{-1}F^{0\prime }F^{0}=I_{r} $ and $\Lambda^{0\prime}\Lambda^{0}$ being diagonal. Utilizing such a representation is equivalent to selecting a particular rotation that aligns the model with the constraints adopted in the  usual PCA estimator. More recently, Jiang et al.(2023) show that such pseudo-true parameters always exist and further  provide insights on  what the plain PC method really estimates. 
Different from UY (2023a), our work focuses on the  parameters ($\Lambda^{0}$, $F^{0}$) which yield the  sparsest representation of  weak factor model.

The rest of this paper is organized as follows. We formally introduce the
sparse weak factor models, i.e., the weak factor model caused by sparse
loadings, and the PC estimation in Section 2. In Section 3, we study
the asymptotics of PC estimators for factors, factor loadings and the common
components under the weak factor scheme, including consistency, asymptotic
distributions, and uniform convergence rates. Based on  the fact that the rotation
matrix is approximately (block) \textit{upper triangular} proved in Section
3, we propose a screening method to sparsify the estimated factor loadings
and provide consistent estimators for factor strengths in Section 4. Section
5 discusses the determination of the number of factors for sparse weak
factor models. Monte Carlo simulations and an empirical application are
reported in Sections 6 and 7, respectively. Section 8 concludes. All
technical proofs and additional simulation results are relegated to the
Online Appendix.

NOTATIONS. For a set $S$, let $|S|$ be its cardinality.\ Let $a\vee b=\max
\left\{ a,b\right\} $ and $a\wedge b=\min \left\{ a,b\right\} $ for two real
numbers $a$ and $b$. For two random sequences $\left\{ a_{n}\right\} $ and $%
\left\{ b_{n}\right\} ,$ $a_{n}\lesssim _{p}b_{n}$ denotes $a_{n}/b_{n}\ $is
stochastically bounded and $a_{n}\asymp _{p}b_{n}$ if $a_{n}\lesssim
_{p}b_{n}$ and $b_{n}\lesssim _{p}a_{n}.$ When $\left\{ a_{n}\right\} $ and $%
\left\{ b_{n}\right\} $ are deterministic$,$ $a_{n}\lesssim b_{n}$ denotes $%
a_{n}/b_{n}\ $is bounded and $a_{n}\asymp b_{n}$ if $a_{n}\lesssim b_{n}$
and $b_{n}\lesssim a_{n}.$ For a square matrix $A$, $\mu _{\min }\left(
A\right) $ and $\mu _{\max }\left( A\right) $ are the smallest and largest
eigenvalue of $A$, respectively. Let $\left\Vert B\right\Vert =\left[ \text{%
Tr}\left( BB^{\prime }\right) \right] ^{1/2}$ be the Frobenius norm of
matrix $B$ and $\left\Vert B\right\Vert _{sp}=\mu _{\max }^{1/2}\left(
B^{\prime }B\right) $ be its spectral norm. For a $n\times 1$ vector $a$,
its $\ell _{0}$-norm is $\left\Vert a\right\Vert _{0}=\sum_{j=1}^{n}\mathbf{1%
}\left( a_{j}\neq 0\right) $, where $\mathbf{1}\left( \cdot \right) $ is the
usual indicator function.

\section{THE SPARSE WEAK FACTOR MODELS}


\subsection{The Definition of Sparse Weak Factor Models}

In this paper, we consider the following factor model with a sparse representation %
\begin{equation}
X_{it}=\lambda _{i}^{0\prime }F_{t}^{0}+e_{it}=\sum_{k=1}^{r}\lambda
_{ik}^{0}F_{tk}^{0}+e_{it},  \label{m1}
\end{equation}%
where $\lambda _{i}^{0}=\left( \lambda _{i1}^{0},\ldots ,\lambda
_{ir}^{0}\right) ^{\prime }$ and $F_{t}^{0}=\left( F_{t1}^{0},\ldots
,F_{tr}^{0}\right) ^{\prime }.$\footnote{Note that while the model in (\ref{m1}) is a sparse representation of the true model or a structural model, this does not imply that the true model or the structural factor model must necessarily take the sparsest form of loadings. In reality, the true model can exhibit either a sparse or even a non-sparse (given the existence of one strong factor) structure in loadings. Given the rotation equivalence property of factor structure, we focus on the sparsest representation of the true factor model and directly make assumptions on it. } In a matrix form, the model in (\ref{m1})
can be written as $X=\Lambda^{0} F^{0 \prime }+e$, where $X$ and $e$ are both $%
N\times T$ matrices with $X(i,t)=X_{it}$ and $e(i,t)=e_{it}$, respectively, $%
\Lambda ^{0}=\left( \lambda _{1}^{0},\ldots ,\lambda _{N}^{0}\right)
^{\prime }$, and $F^{0}=\left( F_{1}^{0},\ldots ,F_{T}^{0}\right) ^{\prime }$%
. We assume that the factors are well-behaved in the sense that $%
T^{-1}\sum_{t=1}^{T}F_{t}^{0}F_{t}^{0\prime }\rightarrow _{p}\Sigma _{F}$ as 
$T\rightarrow \infty $ for some p.d. and finite matrix $\Sigma _{F}$. For the $k$th
factor, $k=1,...,r$, the $\ell _{0}$-norm of the $N\times 1$\ factor
loadings $\Lambda _{\cdot ,k}^{0}:=\left( \lambda _{1k}^{0},...,\lambda
_{Nk}^{0}\right) ^{\prime }$ is of order $N^{\alpha _{k}}$ for some $\alpha
_{k}\in \lbrack 0,1]$, that is, 
\begin{equation}
\left\Vert \Lambda _{\cdot ,k}^{0}\right\Vert _{0}=\sum_{i=1}^{N}\mathbf{1}%
\left( \lambda _{ik}^{0}\neq 0\right) \asymp N^{\alpha _{k}}.  \label{degree}
\end{equation}%
According to (\ref{degree}), $\alpha _{k}=1$ defines a strong factor which
affects almost all cross-sectional units; $\alpha _{k}=0$ leads to an
extremely weak factor which at most affects a finite number of units; any $%
\alpha _{k}\in (0,1)$ gives a sparse structure on the $N$ loadings. A larger 
$\alpha _{k}$ implies a stronger factor which affects more cross-sectional
units. So the strength of the $k$th factor or the sparsity degree in its
loadings can be represented by $\alpha _{k}$. As BKP (2021), we define the
parameter $\alpha _{k}$ as the \textit{factor strength }of the $k$th factor.
Further, we arrange the $r$ factors in (\ref{m1}) by their
factor strengths such that 
\begin{equation*}
\alpha _{1}\geq \alpha _{2}\geq \cdots \geq \alpha _{r-1}\geq \alpha _{r}.
\end{equation*}%
A decreasing sequence $\left\{ \alpha _{k}\right\} _{k=1}^{r}$ can capture
the sparsity structure of the factor models.

For any bounded and invertible rotation matrix $R\in \mathbb{R}^{r\times r}$, we denote
 observationally equivalent factors and loadings as: $F=F^{0}R$
and $\Lambda =\Lambda ^{0}\left( R^{\prime }\right) ^{-1}$. Consequently, we
can define the factor strength vector for $F$, analogously to \eqref{degree}%
, as $\alpha \left( R\right) :=\left( \alpha _{1}\left( R\right) ,...,\alpha
_{r}\left( R\right) \right) ^{\prime }$ with $\alpha _{j}\left( R\right) \in
\lbrack 0,1]$ for $j=1,...,r$, and $\alpha _{j}\left( R\right) \geq \alpha
_{k}\left( R\right) $ for $j<k$.
Due to the intrinsic identification indeterminacy of model \eqref{m1}, to characterize
our sparse weak factor model with well-defined factor strengths, we introduce the following definition of the sparsest
representation. %
%

\begin{definition}
\label{def:sparse-represenation} (The sparsest representations of a factor
model) The rotation matrix $R^{\ast }\in \mathbb{R}^{r\times r}$ gives
a sparest representation of factor loadings (up to order), if  the factors after rotation  $F^{*}=F^0R^*$ satisfies that $T^{-1}F^{*\prime}F^{*}$ is  p.d. and finite in limit, and each
element of factor strength vector $\alpha\left( R^{\ast }\right) $ cannot be
further reduced for any other rotation matrix, that is 
\begin{equation*}
\alpha _{k}\left( R^{\ast }\right) \leq \alpha _{k}\left( R\right) \text{
for any invertible and bounded }R\in \mathbb{R}^{r\times r}\text{ and }k=1,...,r.
\end{equation*}
\end{definition}

The sparsest representation indeed exists and is  well defined. To see this, consider the case with two factors. Let  $\Lambda$ be the $N\times 2$ loading matrix, which is just a generic  matrix and does not have to be in its sparsest representation yet. Suppose the factor strengths for $\Lambda$ are such that $\alpha_{1}\geq\alpha_{2}$.  Given the structure of nonzero loadings between factors 1 and 2, it is immediate to see that:  (i)  for any  rotation matrix $R$ we must have $\alpha_{1}(R) = \alpha_{1}$; (ii) for any rotation matrix $R$ we must have $\alpha_{2}(R) \leq \alpha_{2}$, and it is possible that there exists a rotation matrix $R$ such that   $\alpha_{2}(R) <\alpha_{2}$. We can seek the rotation $R^{\ast}$ turning $\alpha_{2}(R)$ as small as possible and get the factor strengths vector $(\alpha_{1}(R^{\ast}), \alpha_{2}(R^{\ast}) )$ for the sparsest representation. For the case with more than two factors  ($r>2$), the same logic proceeds: we always start with and rotate the weakest factor (measured by initial $\alpha_{k}$'s) to be as sparse as possible, and move to the second weakest one and so on in a sequential way.\footnote{In cases of factors with equal strength, we can just place them in any order and proceed with other factors as before.} Finally, we collect and rank all factor strengths in a descending order and stack them in the $r\times 1$ vector $\alpha$. In this way, we can end up with the sparsest representation in which each $\alpha_{k}$ cannot be further reduced for $k=1,\ldots, r$. The factor strength vector for the sparsest representation can be uniquely determined, but the sparsest representations with the same strength vector are not unique. 
 
\noindent \textbf{Remark 1.} (i) Definition \ref{def:sparse-represenation} pins
down uniquely the factor strengths for the sparsest representations, regardless of common or
distinct factor strengths (or mixed of both). We will show that all factor
strengths can be consistently estimated in Section \ref{sec:factor-strength}%
. Here, we need to mention that the sparsest representation is not unique because many rotations give the same factor strengths vector  $\alpha(R^{*})$.  (ii) Perhaps a more ambitious aim in sparse weak factor model is to accurately
recover the set of nonzero loadings. But exact recovery is in general not possible nor of much interest, due to potential small perturbations for sparsity. For instance, consider two
strong factor loadings such that $\Lambda _{i,j}^{0}\sim \text{Uniform}[1,2]$
independently across $i=1,\ldots ,N$ and across $j=1,2$, so that the
loadings of either factor are all strictly positive. Then we can always
rotate $(\Lambda _{\cdot ,1}^{0},\Lambda _{\cdot ,2}^{0})$ in one way to
obtain $(\Lambda _{\cdot ,1}^{a},\Lambda _{\cdot ,2}^{a})$ such that $%
\Lambda _{11}^{a}=0$ and $\Lambda _{12}^{a}\neq 0$; or we can rotate it in
the other way to obtain $(\Lambda _{\cdot ,1}^{b},\Lambda _{\cdot ,2}^{b})$
such that $\Lambda _{11}^{b}\neq 0$ and $\Lambda _{12}^{b}=0$.\footnote{%
Specifically, $(\Lambda _{\cdot ,1}^{0},\Lambda _{\cdot ,2}^{0})$ can be
rotated such that $\Lambda _{\cdot ,1}^{a}=\Lambda _{\cdot ,1}^{0}-(\Lambda
_{11}^{0}/\Lambda _{12}^{0})\Lambda _{\cdot ,2}^{0}$ and $\Lambda _{\cdot
,2}^{a}=\Lambda _{\cdot ,2}^{0}$; or $\Lambda _{\cdot ,1}^{b}=\Lambda
_{\cdot ,1}^{0}$ and $\Lambda _{\cdot ,2}^{b}=\Lambda _{\cdot
,2}^{0}-(\Lambda _{12}^{0}/\Lambda _{11}^{0})\Lambda _{\cdot ,1}^{0}$.}
Apparently, the sparsity of $\Lambda ^{0}$ is not exactly the same with that
of $\Lambda ^{a}$ or $\Lambda ^{b}$, and yet makes non-essential difference in the term of factor strength.
Nevertheless we show that it is possible to approximately recover the set of nonzero loadings well relatively to its factor strength (i.e.,  the total number of nonzero loadings) in Proposition \ref%
{prop:symm_diff}.

In the rest part of the paper, we assume the underlying components ($F^{0}$
and $\Lambda ^{0})$ have the sparsest representation with $R^{\ast }=I_{r}$
and factor strengths $\alpha \left( I_{r}\right) $, where $I_{r}\in \mathbb{R%
}^{r\times r}$ is the identity matrix. In other words, there are no other
rotation matrices that can further reduce any element of the factor strength
vector $\alpha \left( I_{r}\right) $. Given that two factors may possess the
same degree of strength, we distribute $\alpha _{j}$'s into $G$ groups\ so
that: $\alpha ^{\lbrack 1]}\equiv\left\{ \alpha _{1},\ldots ,\alpha
_{m_{1}}:\alpha _{1}=\cdots =\alpha _{m_{1}}\right\} ,$...$,$ $\alpha
^{\lbrack G]}=\left\{ \alpha _{m_{G-1}+1},\ldots ,\alpha _{r}:\alpha
_{m_{G-1}+1}=\cdots =\alpha _{r}\right\} $ with $G\leq r$ and $\alpha
_{m_{1}}>\alpha _{m_{2}}>\ldots >\alpha _{m_{G}}$, and $G=r$ when every
factor's strength is unique. For ease of notation, let us also define the
cardinality of $\alpha ^{\lbrack g]} $ by $r_{g}\equiv \lvert \alpha
^{\lbrack g]}\rvert ,$ for $g=1,\ldots ,G.$ Clearly, $r=\sum_{g=1}^{G}r_{g}$%
. In addition, let $\alpha _{\lbrack j]} $ be the same strength shared in
group $g$, i.e., $\alpha _{\lbrack g]}=\alpha _{m_{g}}$ for $g=1,\ldots ,G.$

As BKP (2021) point out, for a factor with an extremely weak signal with $%
\alpha _{k}\in (0,1/2)$, it cannot be identified without\textit{\ prior}
restrictions and is not relevant in most financial and macroeconomic
applications. Moreover, Freyaldenhoven (2022) discusses the reason why only
factors affecting proportionally more than $\sqrt{N}$ of individuals are
relevant in arbitrage pricing theory and aggregate fluctuations in
macroeconomics. Hence we restrict the factor strength so that $\alpha
_{k}\in \lbrack \underline{\alpha },1]$ for $k=1,\ldots ,r$ with some $%
\underline{\alpha }>1/2$.\footnote{%
Factors with the strength parameter larger than $1/2$ are called
\textquotedblleft semi-strong\textquotedblright\ factors in BKP (2021).}
Admittedly, the restriction on the weakest factor's strength is stronger
than that in Bai and Ng (2023) and UY (2023a, b).

\subsection{Empirical Relevance of the Sparse Representations}

What is the relevance of the sparse representation in empirical
applications, compared with alternative (observationally equivalent)
representations? Here we provide two important motivational examples in
studying such a sparse representation with $\Lambda ^{0}$ and $F^{0}$.

\begin{example}[Empirical asset pricing]
The sparse representation would help identify the strength of observed or
constructed factors robustly in empirical asset pricing models. For observed
factors such as the three factors in Fama and French (1993), the
Fama-MacBeth two-pass regression has been developed to estimate the loadings
and risk premia. Based on estimated loadings, the factor strengths can be
estimated (BKP, 2021), and can provide important information for identifying
useful factors from the \textquotedblleft factor zoo\textquotedblright\
(Cochrane, 2011). However, such a model is highly likely misspecified due to
omitting important factors; see Giglio and Xiu (2021) and Kim et al. (2024).
Instead, one could estimate the sparse latent factor model such as \eqref{m1} and
obtain the estimators for factors $\{\widehat{F}_{t}\}_{t=1}^{T}$ and
sparsified loadings estimator $\widehat{\Lambda }$. Then for a given
observed factor $g_{t}$, using the linear projection of $g_{t}$ on $\widehat{%
F}_{t}$, one could estimate the factor strength of $g_{t}$  and further evaluate its risk premium. 
\end{example}

\begin{example}[Detect relevant factors for individuals]
The sparse representation can be used to identify redundant factors for each individual. To fix
ideas and simplify, let us consider model \eqref{m1} with $r=2$ and factor
loadings 
\begin{equation}
\Lambda ^{\diamond }=%
\begin{pmatrix}
\Lambda _{\cdot ,1} & a\Lambda _{\cdot ,1} \\ 
b\Lambda _{\cdot ,2} & \Lambda _{\cdot ,2}%
\end{pmatrix}%
 \label{eq:Lambda_diamond}
\end{equation}%
with $ab\neq 0$, where  each of the four blocks is a $\frac{N}{2}\times 1$
vector and $N$ is assumed to be even.\footnote{%
The example used here just illustrates the idea since the two factors
involved are, strictly speaking, not weak by our previous definition.} For a given $i$ it is sufficient to capture
the time series variation of $X_{it}$ by only one factor. The sufficiency is
clearly expressed by $\Lambda ^{\diamond }$'s sparse representation via
rotation: 
\begin{equation}
\Lambda ^{0}=%
\begin{pmatrix}
\Lambda _{\cdot ,1} & 0 \\ 
0 & \Lambda _{\cdot ,2}%
\end{pmatrix}%
.  \label{eq:Lambda_0}
\end{equation}%
Clearly, the sparsity structure in  \eqref{eq:Lambda_0}  implies  that there is only one relevant factor for each unit.   In addition, identification of (ir)relevant factors for certain individuals may
play a fundamental role in solving the weak factor problem. As proposed
by Giglio et al. (2023a), the knowledge of factor relevancy helps screening
irrelevant individuals (test assets) for a given $F_{\cdot ,k}$, and
evaluate $F_{\cdot ,k}$ only using the relevant ones to which $F_{\cdot ,k}$
is pervasive.
\end{example}

%
%

\subsection{PC Estimation}

\label{sec:PCestimation}

To begin with, we assume that the true number of factors $r$ is known, and
leave the determination of $r$ in Section \ref{sec:no_f}. The estimation of
factors and their loadings is via the method of PC in minimizing 
\begin{equation*}
\underset{\Lambda ,F}{\min }\frac{1}{NT}\sum_{i=1}^{N}\sum_{t=1}^{T}\left(
X_{it}-\lambda _{i}^{\prime }F_{t}\right) ^{2},
\end{equation*}%
subject to the usual identification restriction that $F^{\prime }F/T=I_{r}$
and $\Lambda ^{\prime }\Lambda $ being diagonal. Since it is known that the
PC estimator $\widetilde{F}$ is identified up to a full rank rotation
matrix, the identification restriction $F^{\prime }F/T=I_{r}$ is only
employed to pin down $\widetilde{F}$ for the purpose of estimation.
The estimated factors, denoted by $%
\widetilde{F}$, is $\sqrt{T}$ times the eigenvectors corresponding to the $r$
largest eigenvalues of the $T\times T$ matrix $\frac{X^{\prime }X}{NT}$ in
decreasing order. Then $\widetilde{\Lambda }=X\widetilde{F}/T$, and $%
\widetilde{e}=X-\widetilde{\Lambda }\widetilde{F}^{\prime }$. Also, let 
\begin{equation*}
\widetilde{V}=\diag\left( \widetilde{V}_{1},\ldots,\widetilde{V}_{r}\right) 
\end{equation*}
be the $r\times r$ diagonal matrix consisting of the $r$ largest
eigenvalues of $\frac{X^{\prime }X}{NT}$ in decreasing order. We also define
the common component estimator $\widetilde{C}=\widetilde{\Lambda }\widetilde{%
F}^{\prime }$ as the estimator for $C^{0}=\Lambda ^{0}F^{0\prime }.$

\section{LARGE SAMPLE PROPERTIES FOR PC ESTIMATORS}

\label{sec:large_sample}

\subsection{Main Assumptions}

We first define a scale matrix $A$ as follows, 
\begin{equation*}
A=\diag\left( N^{\alpha _{1}},\ldots,N^{\alpha _{r}}\right) ,
\end{equation*}%
which will be frequently used in this paper, and define the non-null set
with regard to $\lambda _{ik}^{0} $'s for $k=1,\ldots, r$, as 
\begin{equation}  \label{eq:non-null_set}
\mathcal{L}_{k}^{0}\equiv\{i:\lambda _{ik}^{0}\neq 0,i=1,\ldots ,N\}.
\end{equation}

\noindent \textbf{Assumption 1.} $E\Vert F_{t}^{0}\Vert ^{4}<\infty $ for $%
t=1,...,T$ and $T^{-1}F^{0\prime }F^{0} \rightarrow _{p}\Sigma _{F}$ as $%
T\rightarrow \infty $ for some p.d. matrix $\Sigma _{F}$.

\noindent \textbf{Assumption 2.} For the factor loadings,

(i) $0<\underline{\lambda }\leq \left\vert \lambda _{ik}^{0}\right\vert \leq 
\overline{\lambda }<\infty $, $\forall i\in \mathcal{L}_{k}^{0}$, for $%
k=1,\ldots ,r$, where $\underline{\lambda}$ and $\overline{\lambda }$ are two constants%
;

(ii) For the $k$th factor, the number of nonzero factor loading is $%
n_{k}\equiv \left\Vert \Lambda _{\cdot ,k}^{0}\right\Vert _{0}\asymp
N^{\alpha _{k}}$\ for $\alpha _{k}\in \lbrack \underline{\alpha },1]$ with $%
\underline{\alpha }>1/2$ for $k=1,\ldots ,r;$  $%
\alpha =\left( \alpha _{1},...,\alpha _{r}\right) ^{\prime }$ is the factor strength vector of the 
sparsest representation of the factor model;

(iii) $A^{-1/2}\Lambda ^{0\prime }\Lambda ^{0}A^{-1/2}\rightarrow \Sigma
_{\Lambda }^{\ast }$ as $N\rightarrow \infty $ for some p.d. matrix $\Sigma
_{\Lambda }^{\ast }$.

\noindent \textbf{Assumption 3.} There exists a positive constant $M\leq
\infty $, such that for all $N$ and $T$,

(i) $E(e_{it})=0$ and $E(e_{it}^{2})<M$ for all $i$ and $t;$

(ii) $E(e_{s}^{\prime }e_{t}/N)=\gamma _{N}(s,t)$, $\left\vert \gamma
_{N}(s,s)\right\vert \leq M$ for all $s$, and $T^{-1}\sum_{s=1}^{T}%
\sum_{t=1}^{T}\left\vert \gamma _{N}(s,t)\right\vert \leq M$;

(iii) $E(e_{it}e_{jt})=\tau _{ij,t}$ with $\left\vert \tau
_{ij,t}\right\vert \leq \tau _{ij}$ for all $t$ with some $\tau _{ij}>0$. In
addition, for $\forall \mathcal{S}\subset \{1,\ldots ,N\}$, $%
\sum_{j=1}^{N}\tau _{ij}\mathbf{1}\left( j\in \mathcal{S}\right) \leq M$ and 
$\left\vert \mathcal{S}\right\vert
^{-1/2}T^{-1/2}\sum_{j=1}^{N}\sum_{t=1}^{T}\left( e_{it}e_{jt}-\tau
_{ij,t}\right) \mathbf{1}\left( j\in \mathcal{S}\right) =O_{p}\left(
1\right) ;$

(iv) For every $(t,s)$, $E\left\vert N^{-1/2}\sum_{i=1}^{N}\left[
e_{is}e_{it}-E(e_{is}e_{it})\right] \right\vert ^{4}\leq M$;

(v) $\left( n_{k}T\right)
^{-1/2}\sum_{i=1}^{N}\sum_{t=1}^{T}F_{t}^{0}\lambda _{ik}^{0}e_{it}=O_{p}(1)$
for $k=1,\ldots ,r$;

(vi) $\max_{s}\sum_{t=1}^{T}\left\Vert \gamma _{N,F}(s,t)\right\Vert
^{2}\leq M$ and $\max_{s}E\left\Vert \varpi (s)\right\Vert ^{2}\leq M$,
where $\gamma _{N,F}(s,t)=N^{-1}E(F_{t}^{0}e_{t}^{\prime }e_{s})$ and $%
\varpi (s)=\left( NT\right) ^{-1/2}\sum_{i=1}^{N}\sum_{t=1}^{T}\left[
F_{t}^{0}e_{it}e_{is}-E(F_{t}^{0}e_{it}e_{is})\right] $;

(vii) $E\left( N^{-1}\sum_{i=1}^{N}\left\Vert
T^{-1/2}\sum_{t=1}^{T}F_{t}^{0}e_{it}\right\Vert ^{2}\right) \leq M$;

(viii) The eigenvalues of $\Sigma _{\Lambda }^{\ast }\Sigma _{F}$ are\
distinct.

\noindent \textbf{Assumption 4. }$\left\Vert e\right\Vert
_{sp}^{2}=O_{p}\left( \max \left\{ N,T\right\} \right) .$

Assumption 1 imposes a moment condition on the factors and requires the
existence of a p.d. probability limit of $T^{-1}F^{0\prime }F^{0}$.
Assumptions 2(i)-(ii) impose boundedness conditions on nonzero loadings and specify
the sparsest structure in loadings. The deterministic loadings in 2(ii) can be
relaxed to be stochastic with some additional moment conditions. Assumption
2(iii) requires the matrix $\Sigma _{\Lambda }^{\ast }$ to be p.d. but 
\textit{not} necessarily diagonal, and is thus not restrictive. Assumption
3(i)-(iii) impose moment conditions on errors and allow for weak
cross-sectional/serial dependence as Bai (2003). Note that 3(iii) is weaker
than Bai (2003) and thus generalizes the counterpart under a strong factor
model. 3(vi) is standard in panel factor models, and it is not redundant
since we do not assume $F_{t}^{0}$ and $e_{s}$ are independent. Assumption 4
is also adopted by Bai and Ng (2002, 2023). It surely holds for
independently identically distributed (iid) data with uniformly bounded $4$th moments, and may also hold for weakly dependent data across $i$ and $t$.

Our paper focuses on the sparse loading case with loadings being either zero or bounded away from zero as specified in Assumption 2(i). It is not as general as Bai and Ng (2023) considering the assumption on $\frac{\Lambda^{0 \prime} \Lambda^0}{N^{\alpha}}$ which accommodates two cases---sparse loadings and shrinking loadings: the former  have many zero loadings and the remaining loadings are non-degenerate,  while the latter are dense but all loadings are shrinking to zero. In addition, it can certainly also contain a mixture of the two cases in real applications.  Nevertheless, the sparse factor loadings are more  relevant in estimating important parameters, e.g., Giglio et al. (2023a), and it is also commonly used to interpret latent factors (Ludvigson and Ng, 2019).

\subsection{Consistency and Limiting Distributions}

We first present one of the key interesting results for PC estimation of
weak factor models. Recall that  $%
\widetilde{V}=\diag\left( \widetilde{V}_{1},\ldots ,\widetilde{V}_{r}\right) 
$ is the diagonal matrix consisting of the $r$ largest
eigenvalues of $\frac{X^{\prime }X}{NT}$ in decreasing order. 
We show that the eigenvalue matrix $\widetilde{V}$ preserves the
magnitude of factor strength, as stated in Proposition \ref{prop: V_rate}
below.

\begin{proposition}
\label{prop: V_rate}Under Assumptions 1-4, $\widetilde{V}_{k}\asymp
_{p}N^{\alpha _{k}-1}$ for $k=1,\ldots ,r.$
\end{proposition}

\noindent \textbf{Remark 2. }Unlike in a strong factor model, the diagonal
elements of matrix $\widetilde{V}$ vanish at various rates determined by
their corresponding factor strengths. It raises more challenges to our
asymptotic theory later for terms involving $\widetilde{V}^{-1}$ frequently, whereas
in a strong factor model all these terms are $O_{p}(1)$.

Next, we turn to the consistency and asymptotic
distributions for PC estimators. We introduce several rotation
matrices as below. Define
\begin{eqnarray*}
H &=&\frac{\Lambda ^{0\prime }\Lambda ^{0}}{N}\frac{F^{0\prime }\widetilde{F}
}{T}\widetilde{V}^{-1},\text{ }H_{1}=\Lambda ^{0\prime }\Lambda ^{0}( 
\widetilde{\Lambda }^{\prime }\Lambda ^{0})^{-1},\text{ }H_{2}=\left(
F^{0\prime }F^{0}\right) ^{-1}F^{0\prime }\widetilde{F},\text{ } \\
H_{3} &=&Q^{-1}\text{ with\ }Q=\frac{\widetilde{F}^{\prime }F^{0}}{T},\text{
and }H_{4}=\Lambda ^{0\prime }\widetilde{\Lambda }(\widetilde{\Lambda }
^{\prime }\widetilde{\Lambda })^{-1},
\end{eqnarray*}
where we have suppressed the dependence on sample sizes for these matrices to ease
burden of notations. The introduction of different rotation matrices are mainly for the convenience in the proofs of different results. We will show the asymptotic equivalence of the matrices 
$H$ and $H_{j}$ for $j=1,2,3,4,$ which also generalizes Lemma 3 of Bai and
Ng (2023). These rotation matrices and the equivalence results will play an
indispensable role in establishing the convergence rates and the asymptotic
distributions of our PC estimators. Before doing so, we would like to first
introduce an interesting finding related to the matrix $Q$ defined above.
One more additional condition is needed.

\noindent \textbf{Assumption 5. }$N^{1-\alpha _{r}}/T^{1/2}\rightarrow 0$.

It is worth mentioning that Bai and Ng (2023) impose a weaker condition ($%
N^{1-\alpha _{r}}/T\rightarrow 0)$ for consistency. They do not impose $%
N^{1-\alpha _{r}}/T^{1/2}\rightarrow 0$ until in proving distributional
theory. However, we impose Assumption 5 at an early stage for showing
Proposition \ref{prop:Q_upper_trian}, which plays a fundamental role in the
sparsity recovery in Section \ref{sec:sparsity_reveal}. Besides, with
Proposition \ref{prop:Q_upper_trian}, we can obtain a sharper result for
convergence rate; see Proposition \ref{prop: loading_consistent} later. Also
note that under $N/T\rightarrow c\in \left( 0,\infty \right) $ as in
Freyaldenhoven (2022), Assumption 5 is trivial.

\begin{proposition}
\label{prop:Q_upper_trian}Under Assumptions 1-5, the $r\times r$ matrix $%
Q=T^{-1}\widetilde{F}^{\prime }F^{0}$ is a full rank matrix with probability
approaching 1
and 
\begin{equation*}
Q_{lk}\equiv Q\left( l,k\right)
\begin{cases}
 \asymp _{p}N^{\alpha _{l}-\alpha _{k}}\text{
	for }1\leq k\leq l\leq r, \\
=O_{p}(1) \;\text{for }1\leq l<k\leq r.%
\end{cases}%
\end{equation*}
\end{proposition}

\noindent \textbf{Remark 3. }Proposition \ref{prop:Q_upper_trian} implies a
very important property of $Q$ serving as the rotation matrix: $Q$ is (block)%
\textit{\ upper triangular} (with different factor strengths) asymptotically.%
\footnote{%
In the special case of multiple factors with the \textit{same} strength, $Q$
is no longer a (block) upper triangular matrix. However, this would not
alter the main conclusions in this paper.} The property is helpful in
working with the sparse weak factor models. For more discussions, see the
next subsection.

The following proposition states the convergence rates of $\widetilde{F}$.

\begin{proposition}
\label{prop: factor_consistent} Under Assumptions 1-5, 
\begin{equation*}
\frac{1}{T}\left\Vert \widetilde{F}-F^{0}H\right\Vert ^{2}=O_{p}\left(
N^{\alpha _{1}-2\alpha _{r}}+N^{2\left( 1-\alpha _{r}\right) }T^{-2}\right)
=o_{p}\left( 1\right) .
\end{equation*}
\end{proposition}

\noindent \textbf{Remark 4. }The associated convergence rate for $\widetilde{F}$  in Proposition %
6.i of  Bai and Ng (2023) is $%
O_{p}\left( N^{-\alpha _{r}}+N^{2\left( 1-\alpha _{r}\right) }T^{-2}\right)
=o_{p}\left( 1\right) $, which is a bit better than ours. The reason is that
they use a slightly different rotation matrix from ours $H$, as indicated by
the comment below their Proposition 6. We employ the rotation matrix $H$
because it is more closely related to $Q$,  which will play a significant role in our subsequent analysis.

The convergence rate for factor loading estimate is provided in the next
proposition.

\begin{proposition}
\label{prop: loading_consistent}Under Assumptions 1-5, 
\begin{equation*}
\frac{1}{N}\left\Vert \widetilde{\Lambda }-\Lambda ^{0}Q^{\prime
}\right\Vert ^{2}=O_{p}\left( N^{-\alpha _{r}-1}+T^{-1}\right) =o_{p}\left(
1\right) .
\end{equation*}
\end{proposition}

\noindent \textbf{Remark 5. }Note that our rate in Proposition \ref{prop:
loading_consistent} is better than what is stated in Proposition 6.ii of  Bai and Ng (2023). The reason is that we have utilized the upper triangularity of
rotation $Q$ in our proof, which helps sharpen the bound for estimation
errors.

Before proceeding, we present the equivalence results of $H$ and $H_{j}$
for $j=1,2,3,4.$ Define $\gamma _{NT}\equiv \max \left( N^{ \frac{\alpha_{1} 
}{2} -\alpha _{r}}T^{-1/2},N^{1-\alpha _{r}}T^{-1},N^{\alpha_{1}-2\alpha
_{r}}\right) . $

\begin{lemma}
\label{lm: equivalent_H} Under Assumptions 1-5, $H_{j}=H+O_{p}\left( \gamma
_{NT}\right) $ for $j=1,2,3,4.$
\end{lemma}

Recall that $H_{3}=Q^{-1}$. Lemma \ref{lm: equivalent_H} thus implies
that $Q^{-1}=H+O_{p}\left( \gamma _{NT}\right) .$ This equivalence
will be used in showing the consistency of the estimator $\widetilde{C}$ for the common component. Lemma \ref{lm: equivalent_H} generalizes Lemma 1 of Bai and
Ng (2019) for strong factor models, and Lemma 3 of Bai and Ng (2023) for
weak factor models with a common factor strength.

\begin{proposition}
\label{prop:cc_consistent}Under Assumptions 1-5, 
\begin{equation*}
\frac{1}{NT}\left\Vert \widetilde{C}-C^{0}\right\Vert ^{2}=O_{p}\left(
N^{2\left( \alpha _{1}-\alpha _{r}\right) -1}+T^{-1}\right) =o_{p}\left(
1\right) .
\end{equation*}
\end{proposition}

\noindent \textbf{Remark 6.} It is noted in Proposition \ref%
{prop:cc_consistent} that the convergence rate of $\widetilde{C}$ in general
depends on weak factor strengths, or precisely, the discrepancy of strengths
between the strongest and weakest factors. It is only when all factor
strengths are the same that the convergence rate is maximal and coincides
with that under strong factor models. Also note that this result agrees with
Proposition 3 of Bai and Ng (2023) when all the factors have the same factor strength.

The following assumption is used to establish the limiting distribution of
PC\ estimators for weak factor models.

\noindent \textbf{Assumption 6. }The following hold for each $i$ and $t$ as $%
\left( N,T\right) \rightarrow \infty $

(i) $A^{-1/2}\Lambda ^{0\prime }e_{t}\overset{d}{\rightarrow }N\left(
0,\Gamma _{t}\right) ,$ where $\Gamma _{t}=\lim_{N\rightarrow \infty
}A^{-1/2}\sum_{i=1}^{N}\sum_{j=1}^{N}\lambda _{i}^{0}\lambda _{j}^{0\prime
}E\left( e_{it}e_{jt}\right) A^{-1/2}.$

(ii) $T^{-1/2}\sum_{t=1}^{T}F_{t}^{0}e_{it}\overset{d}{\rightarrow }N\left(
0,\Phi _{i}\right) ,$ where $\Phi _{i}=\lim_{T\rightarrow \infty
}T^{-1}\sum_{s=1}^{T}\sum_{t=1}^{T}E\left( F_{t}^{0}F_{s}^{0\prime
}e_{it}e_{is}\right) .$

One more regularity condition is imposed on sample size ($N,T$) and the
weakest factor strength $\alpha _{r}$ to guarantee the distributional
theory, which is also employed by Assumption C'(iv) in Bai and Ng (2023).

\noindent \textbf{Assumption 7. } $N^{\frac{3}{2}-\alpha
_{r}}T^{-1}\rightarrow 0$.

\begin{theorem}
\label{thm: Ft_dist}Under Assumptions 1-7, 
\begin{equation*}
A^{1/2}\left( \widetilde{F}_{t}-H_{4}^{\prime }F_{t}^{0}\right) \overset{d}{
\rightarrow }N\left( 0,\Psi ^{0}\Gamma _{t}\Psi ^{0\prime }\right) ,
\end{equation*}
where $\Psi ^{0}=\limfunc{plim}_{N,T\rightarrow \infty }N^{-1}A^{1/2} 
\widetilde{V}^{-1}QA^{1/2}.$
\end{theorem}

\noindent \textbf{Remark 7.} (i)\textbf{\ }As shown in the proof of Theorem %
\ref{thm: Ft_dist}, the matrix$\ \Psi ^{0}$ is (block) diagonal. In
particular, if no two factors have the same strength, i.e., $\alpha _{k}\neq
\alpha _{l}$ for $k\neq l,$ then $\Psi ^{0}$ is an exactly diagonal matrix
so that $\Psi ^{0\prime }=\Psi ^{0}.$ (ii) Theorem \ref{thm: Ft_dist} also
reveals that the $k$th factor $\widetilde{F}_{tk}$ is asymptotically
normally distributed with convergence rate of $N^{\alpha _{k}/2}.$

\noindent \textbf{Assumption 8. }$\sqrt{T}N^{\alpha_{1}-2\alpha
_{r}}\rightarrow 0$.

Assumption 8 parallels Assumption C(ii) of Bai and Ng (2023), and a smaller
gap between $\alpha _{1}$ and $\alpha _{r}$ makes the assumption more likely
to hold.

\begin{theorem}
\label{thm: lamdi_dist}Under Assumptions 1-6 and 8, 
\begin{equation*}
\sqrt{T}\left( \widetilde{\lambda }_{i}-Q\lambda _{i}^{0}\right) \overset{d}{
\rightarrow }N\left( 0,Q^{\prime -1}\Phi _{i}Q^{-1}\right) .
\end{equation*}
\end{theorem}

\noindent \textbf{Remark 8. }Theorem \ref{thm: lamdi_dist} reveals that the
asymptotic distribution of $\widetilde{\lambda }_{i}$ is the same under both
strong and weak factors and invariant to factor strengths.

With Theorems \ref{thm: Ft_dist} and \ref{thm: lamdi_dist}, we come to the
limiting distribution of $\widetilde{C}_{it}.$ To this end, recall $r_{G}$
from Section \ref{sec:PCestimation}, and define the $r\times r$ matrix $%
S^{\dagger }=\diag(\underbrace{0,\ldots ,0}_{r-r_{G}},\underbrace{1\ldots ,1}%
_{r_{G}})$.

\begin{theorem}
\label{thm: Cit_dist}Under Assumptions 1-8, 
\begin{equation*}
\frac{\widetilde{C}_{it}-C_{it}^{0}}{\sqrt{N^{-\alpha
_{r}}V_{it}+T^{-1}U_{it}}}\overset{d}{\rightarrow }N\left( 0,1\right) ,
\end{equation*}
where $V_{it}=\lambda _{i}^{0\prime }S^{\dagger }\Sigma _{\Lambda }^{\ast
-1}\Gamma _{t}\Sigma _{\Lambda }^{\ast -1}S^{\dagger }\lambda _{i}^{0}$ and $%
U_{it}=F_{t}^{0\prime }\Sigma _{F}^{-1}\Phi _{i}\Sigma _{F}^{-1}F_{t}^{0}.$
\end{theorem}

\noindent \textbf{Remark 9. }Theorem \ref{thm: Cit_dist} implies that $%
\widetilde{C}_{it}-C_{it}^{0}=O_{p}\left( \max \left\{ N^{-\alpha
_{r}/2},T^{-1/2}\right\} \right) .$ As for the asymptotic covariance matrices
for $\widetilde{F}_{t}$ and $\widetilde{\lambda }_{i},$ Bai and Ng (2023) have
proposed consistent estimators assuming cross-sectional or serial independence
for $\left\{ e_{it}\right\} .$ Under weakly serial dependence, Bai (2003)
proposes a consistent Newey-West HAC estimator for the asymptotic covariance
of $\widetilde{\lambda }_{i}.$ For estimating the asymptotic covariance of $%
\widetilde{F}_{t}$ under weakly cross-sectional (CS) independence$,$ Bai and
Ng (2006) propose a consistent CS-HAC estimator under covariance
stationarity with $E\left( e_{it}e_{jt}\right) =\sigma _{ij}$ for all $t$'s$%
. $ One could follow the aforementioned approaches to formulate consistent
estimators for the asymptotic covariance matrices of $\widetilde{F}_{t}$ and 
$\widetilde{\lambda }_{i},$ which would lead to a consistent variance
estimator for $\widetilde{C}_{it}$. For hypothesis testing, with $\widetilde{
F}_{t}$ for instance, there is no need to know factor strengths$,$ as the
feasible estimator for the variance of $\widetilde{F}_{t}$ automatically
accommodates factor strengths,\ and\ is thus adequate for such a purpose.%
\footnote{%
See the discussion under Proposition 4 of Bai and Ng (2023) for the case
with a homogeneous factor strength, and one can easily show that the
argument also works under heterogeneous factor strengths.}

\subsection{Uniform Convergence Rates}

In this subsection, we establish uniform convergence rate results for $%
\widetilde{F}_{t}$, $\widetilde{\lambda}_{i},$ and $\widetilde{C}_{it}$ over 
$i$ or (and) $t.$ These results can be exploited in recovering model
sparsity in Section \ref{sec:sparsity_reveal}, in the factor-augmented
forecast regression, and are perhaps also of independent interest.

Given that dependence is allowed across both $i$ and $t$, we first define a
strong mixing condition, generalized over $i$ and $t,$ similar to Ma et al.
(2021). Suppose that there is some labelling of the cross-sectional units $%
i_{l_{1}},\ldots ,i_{l_{N}},$ whose generic index we denote by $i^{\ast },$
such that the CS dependence decays with distance $\left\vert i^{\ast
}-j^{\ast }\right\vert .$ Then we define a mixing rate applied for the
random field $\left\{ G_{i^{\ast }t}:1\leq i^{\ast }\leq N,1\leq t\leq
T\right\} ,$ where $G_{i^{\ast }t}\equiv \left( F_{t}^{0\prime },e_{i^{\ast
}t}\right) ^{\prime }.$ For $S_{1},S_{2}\subset \left[ 1,\ldots ,N\right]
\times \left[ 1,\ldots ,T\right] ,$ let%
\begin{equation*}
\alpha \left( S_{1},S_{2}\right) \equiv \sup \left\{ \left\vert P\left(
A\right) P\left( B\right) -P\left( A\cap B\right) \right\vert :A\in \sigma
\left( G_{i^{\ast }t},\left( i^{\ast },t\right) \in S_{1}\right) ,B\in
\sigma \left( G_{i^{\ast }t},\left( i^{\ast },t\right) \in S_{2}\right)
\right\} ,
\end{equation*}%
where $\sigma \left( \cdot \right) $ denotes a sigma-field. Then the $\alpha 
$-mixing coefficient of $\left\{ G_{i^{\ast }t}\right\} $ is defined as 
\begin{equation*}
\alpha \left( k\right) \equiv \sup \left\{ \alpha \left( S_{1},S_{2}\right)
:d\left( S_{1},S_{2}\right) \geq k\right\} ,
\end{equation*}%
where $d\left( S_{1},S_{2}\right) \equiv \min \left\{ \sqrt{\left(
t-s\right) ^{2}+\left( i^{\ast }-j^{\ast }\right) ^{2}}:\left( i^{\ast
},t\right) \in S_{1},\left( j^{\ast },s\right) \in S_{2}\right\} .$

The definition of $\alpha \left( k\right) $ generalizes the usual one in the
time series context. In particular, when $\alpha \left( k\right) $ is
applied to a (single or vector of) time series, it coincides with the
usual one defined by, e.g., Fan et al. (2011).  For the purpose of
estimation, we do not need to know the true labelling $\left\{ i^{\ast
}\right\} .$ Ma et al. (2021) show that their inference is valid as long as
the number of mis-assigned indices is $o\left( N^{1/2}\right) .$ In
conducting inference, our approach is effective with the true labelling $%
\left\{ i^{\ast }\right\} $ being completely unknown, and thus further relaxes the
assumption by Ma et al. (2021).

We now specify the additional assumption for establishing the uniform convergence rates.

\noindent \textbf{Assumption 9.\ }(i) $\left\{ e_{t}\right\} _{t\geq 1}$ and 
$\left\{ F_{t}^{0}\right\} _{t\geq 1}$ are both stationary and ergodic;

(ii) There exist $s_{2}>0$ and $K>0$ such that $\forall t\in \mathbb{Z}%
^{+}, $ $\alpha \left( t\right) \leq \exp \left( -Kt^{s_{2}}\right) ;$

(iii) There exist $s_{1}>0,$ $b_{1}>0,$ $s_{3}>0,$ $b_{2}>0,$ satisfying $%
3s_{1}^{-1}+s_{2}^{-1}>1$ and $3s_{3}^{-1}+s_{2}^{-1}>1$, such that $\forall
w>0,$ $P\left( \left\vert e_{it}\right\vert >w\right) \leq \exp \left(
-\left( w/b_{1}\right) ^{s_{1}}\right) $ and $P\left( \left\vert
F_{tk}^{0}\right\vert >w\right) \leq \exp \left( -\left( w/b_{2}\right)
^{s_{3}}\right) $ for $k=1,...,r;$

Assumption 9 imposes restrictions on the mixing rate decay and tail bounds
on factors and errors, as well as a mild rate restriction on sample size ($%
N,T$), which is also employed in Fan et al. (2011).

From the previous section, we see that the PC estimators $\widetilde{F}_{t}$
and $\widetilde{\lambda}_{i}$ are both consistent up to a certain rotation
matrix. So to better state the uniform convergence result, we define the
rotated factor and factor loading by $F_{t}^{\ast }=H_{4}^{\prime }F_{t}^{0}$ and $\lambda _{i}^{\ast }\equiv Q\lambda
_{i}^{0}$, respectively$.$

\begin{theorem}
\label{thm: Uniform} Under Assumptions 1-9,

(i) $\sup_{i}\left\vert \widetilde{\lambda }_{ik}-\lambda _{ik}^{\ast
}\right\vert =O_{p}\left( \sqrt{(\ln N)/T}\right) $ for $k=1,...,r;$

(ii) $\sup_{t}\left\vert \widetilde{F}_{tk}-F_{tk}^{\ast }\right\vert
=O_{p}\left( N^{-\alpha _{k}/2}\sqrt{\ln T}\right) $ for $k=1,...,r;$

(iii) $\sup_{i,t}\left\vert \widetilde{C}_{it}-C_{it}^{0}\right\vert
=O_{p}\left( \left( \ln T\right) ^{1/s_{3}}\sqrt{(\ln N)/T}+N^{-\alpha
_{r}/2}\sqrt{\ln T}\right) .$
\end{theorem}

\noindent \textbf{Remark 10. }In Theorem \ref{thm: Uniform}, result (i)
implies that the estimation errors for factor loadings are
dominated by $\sqrt{(\ln N)/T}$ across all $i$. This result is very useful to explore
loading sparsity and factor strength in the next section. For each factor,
result (ii) provides the uniform convergence rate which depends on the
factor strength. Result (iii) establishes the uniform convergence rate for
the common component estimators and the rate is determined by the smallest
factor strength $\alpha _{r}$ and a parameter $s_{3}$ which controls the
probability tail bound of factors.

\section{REVELATION OF SPARSITY WITH PC ESTIMATORS}

\label{sec:sparsity_reveal}

It is well known that latent factor models are subject to a rotational
indeterminacy. This identification issue is unwanted to reveal the loading
sparsity structure. Specifically, the large sample properties of PC
estimators in Section \ref{sec:large_sample} indicates that loadings are
identified only up to a rotation matrix, and it is believed that rotation in
general plays a deterrent role in revealing sparsity. As a result,
interpretation of factors becomes intimidating, which is usually done by
associating a factor with cross-sectional units of nonzeros estimated
loadings. For example, Ludvigson and Ng (2009) write,

\textquotedblleft \textit{Moreover, we caution that any labeling of the
factors is imperfect, because each is influenced to some degree by all the
variables in our large dataset and the orthogonalization means that no one
of them will correspond exactly to a precise economic concept like output or
unemployment, which are naturally correlated.}\textquotedblright\ 

\noindent However, we show that even in the presence of the rotation brought
in by PC, the sparsity in factor loadings can still be preserved. This
justifies the ad hoc manner of interpretation of a factor with only a small
set of observed variables correlated with the PC factor in many empirical
applications; see Pelger and Xiong (2022).

\subsection{Preservation of Sparsity Degree by the PC Rotation}

\label{sec:PC_rot}

The success of sparsity preservation by PC largely hinges on the property of
rotation matrix $Q$. Proposition \ref{prop:Q_upper_trian} implies that $Q$
is a non-strictly (block) \textit{upper triangular} matrix with full rank in
probability.\footnote{%
A (lower, upper) triangular matrix is strictly (lower, upper) triangular if
its diagonal elements are zero, see Abadir and Magnus (2005, page 17).} In
particular, 
\begin{equation*}
Q_{lk}%
\begin{cases}
\asymp _{p}N^{\alpha _{l}-\alpha _{k}},\;\text{for }1\leq k<l\leq r \\ 
\asymp _{p}1,\;\text{for }1\leq k=l\leq r \\ 
=O_{p}(1),\;\text{for }1\leq l<k\leq r%
\end{cases}%
.
\end{equation*}

The particular form of matrix $Q$ gives rise to sparsity recovery as $N$
grows. To illustrate, consider a simple two-factor model with factor
strengths $\alpha _{1}=0.9$ and $\alpha _{2}=0.7$. Now 
\begin{equation}
Q=\left( 
\begin{array}{cc}
Q_{11} & Q_{12} \\ 
Q_{21} & Q_{22}%
\end{array}%
\right) \asymp _{p}\left( 
\begin{array}{cc}
1 & O_{p}(1) \\ 
N^{-0.2} & 1%
\end{array}%
\right) .  \label{Q-matrix}
\end{equation}%
According to Proposition \ref{prop: loading_consistent}, the PC estimator of
loading matrix converges to the rotated $\Lambda ^{0}$ as $\Lambda ^{\ast
}=\Lambda ^{0}Q^{\prime }$, which can be written as 
\begin{equation*}
\Lambda ^{\ast }=\left( \Lambda _{\cdot 1}^{\ast },\Lambda _{\cdot 2}^{\ast
}\right) =\left( \Lambda _{\cdot 1}^{0},\Lambda _{\cdot 2}^{0}\right) \left( 
\begin{array}{cc}
Q_{11} & Q_{21} \\ 
Q_{12} & Q_{22}%
\end{array}%
\right) =\left( Q_{11}\Lambda _{\cdot 1}^{0}+Q_{12}\Lambda _{\cdot
2}^{0},Q_{21}\Lambda _{\cdot 1}^{0}+Q_{22}\Lambda _{\cdot 2}^{0}\right) .
\end{equation*}%
Given the approximately upper triangular structure of $Q$ in (\ref{Q-matrix}%
), we have 
\begin{equation}
\left\Vert \Lambda _{\cdot 1}^{\ast }\right\Vert _{0}=\left\Vert
Q_{11}\Lambda _{\cdot 1}^{0}+Q_{12}\Lambda _{\cdot 2}^{0}\right\Vert
_{0}\asymp _{p}N^{0.9}+N^{0.7}\asymp _{p}N^{0.9}.  \label{Lambda1}
\end{equation}%
So the rotated loadings for factor 1 preserve its sparsity degree. Given
that $\Lambda _{\cdot 2}^{\ast }=Q_{21}\Lambda _{\cdot 1}^{0}+Q_{22}\Lambda
_{\cdot 2}^{0}$, the sparsity degree of factor 2 is contaminated by $%
Q_{21}\Lambda _{\cdot 1}^{0}$. Nevertheless, the contamination is vanishing
due to diminishing $Q_{21}$. Note that each nonzero element in $%
Q_{12}\Lambda _{\cdot 2}^{0}$ is of order $O_{p}(1)$ and the total number of
nonzero elements is of order $N^{0.7};$ the total number of nonzero
elements in $Q_{21}\Lambda _{\cdot 1}^{0}$ is of order $N^{0.9}$, but each
nonzero element is of order $O_{p}\left( N^{-0.2}\right) $ and converges to
zero. So the first component $Q_{21}\Lambda _{\cdot 1}^{0}$ goes to a zero
vector asymptotically. Alternatively, we define an \textit{adaptive measure
for sparsity degree} for a generic $n\times 1$ vector $a$ as $\left\Vert
a\right\Vert _{\kappa _{N},0}=\sum_{j=1}^{n}\mathbf{1}\left( |a_{j}|\geq
\kappa _{N}\right) $, where $\kappa _{N}=N^{-\varrho }$ for some $\varrho
\in \left( 0,0.2\right) .$ Then 
\begin{equation}
\left\Vert \Lambda _{\cdot 2}^{\ast }\right\Vert _{\kappa _{N},0}=\left\Vert
Q_{21}\Lambda _{\cdot 1}^{0}+Q_{22}\Lambda _{\cdot 2}^{0}\right\Vert
_{\kappa _{N},0}\asymp _{p}N^{0.7}.  \label{Lambda2}
\end{equation}%
The statement in \eqref{Lambda1} also certainly holds if the $\ell _{0}$%
-norm is replaced with $\left\Vert \cdot \right\Vert _{\kappa _{N},0}$. This
result suggests that the PC rotation can preserve sparsity degree, up to
negligible terms, so that the factor strengths remain unchanged.

To have a better understanding of how the PC rotation acts upon the sparse
loading $\Lambda ^{0}$, we illustrate with the two-factor model as above via
a simulation. The data generating process is similar to that in Section \ref%
{sec:MC}.\footnote{%
The only exception is that $\mathcal{L}_{k}^{0}$ is randomly determined
across $i$ in Section \ref{sec:MC}, while it is fixed as prescribed here for
illustration.} We set $(N,T)=(200,200)$ and simulate $r=2 $ positively
correlated factors.

\begin{figure}[tbh]
\begin{center}
\includegraphics[width=1.1\textwidth, trim = 00 50 -50 -00]
		{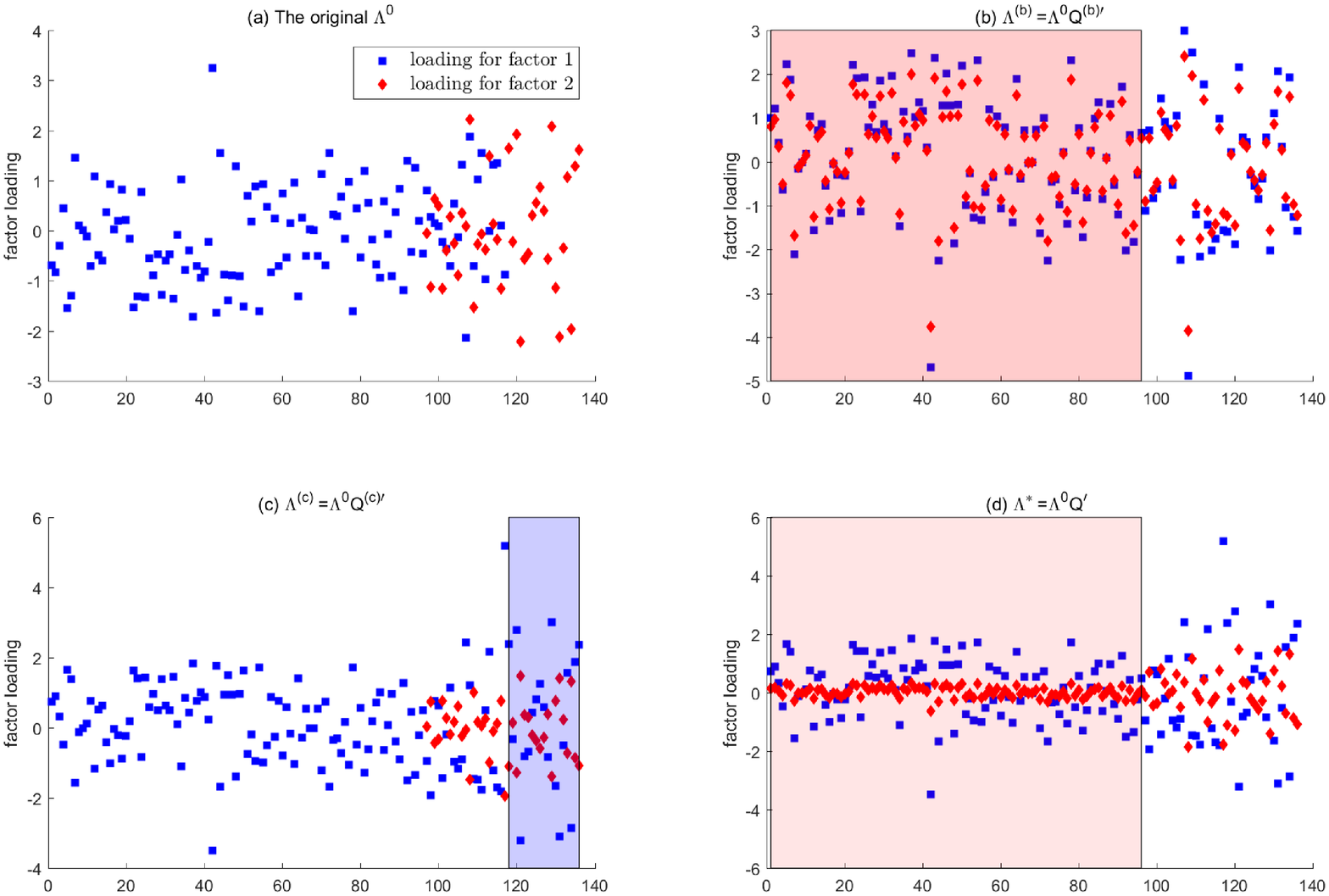}
\end{center}
\caption{Illustration of sparsity degree preservation by PC rotation}
\label{fg:PC_rotation}
\end{figure}

Panel (a) of Figure \ref{fg:PC_rotation} shows underlying $\Lambda ^{0}$ in
which $\mathcal{L}_{1}^{0}=\{i:i=1,\ldots ,117\}$, and $\mathcal{L}%
_{2}^{0}=\{i:i=97,\ldots ,136\}$. The rotation matrix is calculated as $%
Q=\left( 
\begin{array}{rr}
-1.07 & 1.46 \\ 
-0.19 & -0.67%
\end{array}%
\right) .$ Panel (b) illustrates a common concern that $\Lambda ^{0}$
rotated by an arbitrary nonsingular matrix, say $Q^{(b)}$, would become less
sparse. To see it clearly, we randomly generate each element of $Q^{(b)}$ as 
$\mathcal{N}(0,1)$ so that $Q^{(b)}=\left( 
\begin{array}{rr}
-1.44 & -0.98 \\ 
-1.16 & -0.75%
\end{array}%
\right) $. It is obvious in Panel (b) that $\mathcal{L}_{2}^{0}$ is
\textquotedblleft inflated\textquotedblright\ considerably (by the shaded
area) such that the two factors turn out to have the same strength. Panel
(c) displays the rotation effect by a stylized exactly upper triangular
matrix $Q^{(c)}$. $Q^{(c)}$ here is set the same as $Q$ except $%
Q_{21}^{(c)}=0$, i.e., $Q^{(c)}=\left( 
\begin{array}{cc}
-1.07 & 1.46 \\ 
0 & -0.67%
\end{array}%
\right) $. Clearly, the upper triangular $Q^{(c)}$ resolves inflation to $%
\mathcal{L}_{2}^{0}$. On the other hand, we see in panel (c) that $\mathcal{L%
}_{1}^{0}$ is somewhat inflated due to the loading on $\{i:i=118,\ldots
,136\}$ (by the shaded area) being mixed with factor 2's nonzero loadings. Nevertheless, the total number of zero loadings in factor 1 distorted by factor 2  is  negligible relative to factor 1's pervasiveness, since factor 2 is relatively weaker. Finally, panel (d) displays $\Lambda ^{\ast }$
rotated by the real $Q$ in the simulation. The absolute value of $Q_{21}$ is
indeed very small as expected in a finite sample ($Q_{21}\asymp
_{p}200^{-0.2}$), though it is not exactly 0. By virtue of this feature, the
inflated part of $\Lambda _{\cdot ,2}^{\ast }$ (by the shaded area) is
uniformly small and vanishing at a certain rate, which can be screened off
together with estimation errors to reveal sparsity. We formalize the idea in
the next subsection.

\subsection{Sparsity Recovery with PC Estimators}

\label{sec: Sparsity_recovery}

Our previous analysis in Section \ref{sec:PC_rot} provides a positive
identification result for $\Lambda ^{\ast }$ rotated by $Q$. In practice, $%
\Lambda ^{\ast }$ is not observable, and the available is the PC estimator $%
\widetilde{\Lambda }$. In this subsection, we will present how the recovery
of sparsity, not just its degree, is achieved by working with $\widetilde{%
\Lambda }$. Undoubtedly we need to take into account of the \textit{%
estimation error} consisting of $\widetilde{\lambda }_{i}-\lambda _{i}^{\ast
}=T^{-1}H^{\prime }F^{0\prime }e_{i}+T^{-1}(\widetilde{F}-F^{0}H)^{\prime
}e_{i}$. Theorem \ref{thm: Uniform} (i) lends us a hand stating that the
estimation errors are also uniformly vanishing, and thus justifies the
sparsity recovery via screening the PC estimator $\widetilde{\lambda }_{i}$.

We shall show that the set $\mathcal{L}_{k}^{0}$ characterizing sparsity for
factor $k$ can be recovered well approximately with regulated PC estimators.
When the absolute value of $\widetilde{\lambda }_{ik}$ is small, we can set
the factor loading as 0. So we choose the nonzero factor loadings with a
threshold value $c_{\lambda ,NT}$: 
\begin{equation}
\widehat{\lambda }_{ik}=\widetilde{\lambda }_{ik}\mathbf{1}\left\{
\left\vert \widetilde{\lambda }_{ik}\right\vert >c_{\lambda ,NT}\right\} \ 
\text{where }c_{\lambda ,NT}=\frac{1}{\sqrt{\ln \left( NT\right) }}.\text{ }
\label{eq:sparsified_lmd}
\end{equation}%
Let us define the estimated support for the $k$th factor loading by $%
\widehat{\mathcal{L}}_{k}=\{i:\widehat{\lambda }_{ik}\neq 0\}.$ To evaluate
the sparsity recovery accuracy of $\widehat{\mathcal{L}}_{k}$, we denote the 
\textit{symmetric difference} between the true nonzero set $\mathcal{L}%
_{k}^{0}$ and its estimator $\widehat{\mathcal{L}}_{k}$ by 
\begin{equation*}
\mathcal{L}_{k}^{0}\triangle \widehat{\mathcal{L}}_{k}=\left( \mathcal{L}%
_{k}^{0}\backslash \widehat{\mathcal{L}}_{k}\right) \cup \left( \widehat{%
\mathcal{L}}_{k}\backslash \mathcal{L}_{k}^{0}\right) ,
\end{equation*}%
where $\mathcal{L}_{k}^{0}\backslash \widehat{\mathcal{L}}_{k}$ is the false
negative set and $\widehat{\mathcal{L}}_{k}\backslash \mathcal{L}_{k}^{0}$
is the false positive set. Clearly, $\mathcal{L}_{k}^{0}\triangle \widehat{%
\mathcal{L}}_{k}$ summarizes two types of errors.

To accommodate factors with identical strength, we define a set 
\begin{equation}
\omega \left( k\right) =\left\{ l:\alpha _{l}=\alpha _{k},1\leq l\leq
r\right\} .  \label{eq:omega_k}
\end{equation}%
For any $l\in \omega \left( k\right) $, $\alpha _{l}=\alpha _{k}$.

\begin{proposition}
\label{prop:symm_diff} Under Assumptions 1-5 and 8-9, for $k=1,\ldots ,r$,
we have:

(i) if $\alpha _{k}$ is unique, then $\frac{\lvert \mathcal{L}%
_{k}^{0}\triangle \widehat{\mathcal{L}}_{k}\rvert }{N^{\alpha _{k}}}%
=o_{p}(1);$

(ii) if $\alpha _{k}$ is not unique such that $|\omega \left( k\right) |\geq
2$, then $\frac{\lvert \mathcal{L}_{k}^{0}\triangle \widehat{\mathcal{L}}%
_{k}\rvert }{N^{\alpha _{k}}}=o_{p}(1)$ if $\frac{\left\vert \cup
_{k^{\prime }\in \omega \left( k\right) }\mathcal{L}_{k^{\prime
}}^{0}\backslash \mathcal{L}_{k}^{0}\right\vert }{N^{\alpha _{k}}}=o(1);$
otherwise, $\frac{\lvert \mathcal{L}_{k}^{0}\triangle \widehat{\mathcal{L}}%
_{k}\rvert }{N^{\alpha _{k}}}=O_{p}(1)$.
\end{proposition}

\noindent \textbf{Remark 11.} In general, we cannot recover the set $%
\mathcal{L}_{k}^{0}$ sharply in the sense that $\lvert \mathcal{L}%
_{k}^{0}\triangle \widehat{\mathcal{L}}_{k}\rvert =o_{p}(1).$ We have $%
\lvert \mathcal{L}_{k}^{0}\triangle \widehat{\mathcal{L}}_{k}\rvert
=O_{p}(N^{\alpha _{k}})$ or $o_{p}(N^{\alpha _{k}})$ depending on whether $%
\alpha _{k}$ is unique or the overlap degree of loading supports from the
same strength group. In general, the recovered sparsity based on PC
estimates is perhaps less sparse than the true underlying sparsity for
factor $k$. These results seem weak but they are general enough to identify
the factor strengths. For\ the special case of nested sparsity, i.e., $%
\mathcal{L}_{k}^{0}\supseteq \left( \cup _{l\geq k}\mathcal{L}%
_{l}^{0}\right) \cup \left( \cup _{k^{\prime }\in \omega \left( k\right) }%
\mathcal{L}_{k^{\prime }}^{0}\right) ,$ we can show that the recovery is
almost sharp in the sense that $\lvert \mathcal{L}_{k}^{0}\triangle \widehat{%
\mathcal{L}}_{k}\rvert =O_{p}(1).$\

\noindent \textbf{Remark 12.}\label{rmk:r5} (i) In the
sparsity recovery, $c_{\lambda ,NT}$ plays the key role of screening off
noises due to rotation and estimation errors. We show that the noises are $%
O_{p}\left( N^{-\zeta }+\left[ (\ln N)/T\right] ^{1/2}\right) $ uniformly,
where $\zeta $ is the minimum discrepancy between distinct factor strengths,
i.e., $\zeta =\min_{1\leq k\leq G-1}\left( \alpha _{\lbrack k]}-\alpha
_{\lbrack k+1]}\right) $. So we set $c_{\lambda ,NT}=\left[ \ln \left(
NT\right) \right] ^{-1/2}$ to dominate the noise. (ii) Alternatively, one
can use threshold $\widetilde{c}_{\lambda ,NT}=c\left[ \ln \left( NT\right) %
\right] ^{-1/2}$ with a tuning parameter $c$. We have tried $c$ over the
range $[0.5, 1.5]$, and found the results robust to the turning parameter;
moreover, $c$ can be determined by some adaptive methods such as cross
validation. Given that the data are all standardized as in Bai and Ng
(2002), we focus on the simple threshold $c_{\lambda ,NT}$. In both
simulations and empirical application, our tuning parameter-free threshold
value $c_{\lambda ,NT}$ works reasonably well. Note that a simple threshold
value is also adopted in Fan et al. (2015).

\noindent \textbf{Remark 13.}  Note that the uniform order of noises depends on the minimum discrepancy between different factor strengths; see the previous remark. So the recovery of loading sparsity works better when the gaps between factor strengths are bigger. When the factor strengths of, say $\alpha_{j}$ and $\alpha_{k}$, become very close, our method is likely to render the same sparsity for both factor loadings, and may not perform well in distinguishing these two. Nevertheless, perhaps it does not hurt much since the true loading sparsity for factor $j$ and $k$ here is possibly very similar in the first place. Also notice that in this case it will not cause a problem for working with factor $l$ with $\alpha_{l}$ far away from $\alpha_{j}$ and $\alpha_{k}$.

\noindent \textbf{Remark 14.} (i) Given sparsified loading estimator $%
\widehat{\lambda }_{ik}$, it is natural to consider further updating the
estimator of factors. However, it is easy to show that the updated
estimators of factors have the same convergence rates and asymptotical
distributions as simple PC estimators. It may be interesting to compare
their finite sample performance and further investigate their higher order
asymptotics. We leave it for further research. (ii) As the loadings would be
of interest in many empirical applications, one may consider further
refining the PC estimate to reduce its contamination, to achieve more
precise estimation and sparsity recovery. Such consideration is reasonable
especially when strengths of some factors are close, as loading vectors of
these two groups are more intermingled. This may call for a more complicated
procedure, e.g., in further combining SOFAR under its rotation-specific
assumptions, and raise considerable theoretical challenges. We also leave it
for further research.

Before closing Section \ref{sec: Sparsity_recovery}, we would like to
mention that UY's (2023a) SOFAR estimator recovers
sparsity alternatively by $\ell _{1}$ regularization. Interestingly, if we
contrast the estimate of factor loadings by PC (their Figures 2 and 7) with
that by SOFAR (their Figures 9 and 11) in their empirical results, we
immediately realize that PC and SOFAR estimates are almost identical, except
for many small noises introduced by PC. This observation suggests the
validity of combining PC estimate with proper screening to recover sparsity,
which in fact underlines the method used in our paper.

\subsection{Estimating Factor Strengths}

\label{sec:factor-strength}

BKP (2021) use the estimated factor strength to measure  the pervasiveness of the unobserved macroeconomic shocks. They focus on the identification and estimation of the largest factor strength. The reason is  that the PC method can identify the latent factors only up to a non-singular rotation matrix,  and is thus supposed to recover the strength only for the strongest factor. To further identify and  estimate the  strength for weaker factors, they propose  some sequential procedures using weighted cross-sectional averages (CSAs); for more discussions, see Section 4 of BKP (2021). 
Yet with the property of sparsity preservation for the PC method, we 
can consistently estimate factor strengths for \textit{all} factors with various
strengths and then unveil the pervasiveness of all factors completely. We
hope that our proposed estimator of factor strength is of independent
interest.

We can understand how influential\ each factor is by studying its strength.
Recall that $\alpha _{k}$ is such that $\sum_{i=1}^{N}\lambda _{ik}^{0\prime}\lambda _{ik}^{0}\asymp N^{\alpha _{k}}.$ Now for $k=1,...,r,$ define
\begin{equation*}
\widehat{D}_{k}=\sum_{i=1}^{N}\mathbf{1}\left\{ \left\vert \widetilde{%
\lambda }_{ik}\right\vert >c_{\lambda ,NT}\right\} ,
\end{equation*}%
where $c_{\lambda ,NT}=\left[ \ln \left( NT\right) \right] ^{-1/2}$. Our estimator for $\alpha _{k}$ is simply given by
\begin{equation} \label{eq:strength_est}
\widehat{\alpha }_{k}=\frac{\ln \widehat{D}_{k}}{\ln N}.
\end{equation}%
Note that the above estimator $\widehat{\alpha }_{k}$ in \eqref{eq:strength_est} is also employed by  UY (2023a) for latent factors and by BKP (2021) primarily  for observed factors.

To gain some insights on the effectiveness of $\widehat{\alpha }_{k}$, it is
easy to see that Proposition \ref{prop:symm_diff} implies that the recovered
set $\widehat{\mathcal{L}}_{k}$ also approximates the union of $\mathcal{L}%
^{0}_{k}$ and $\mathcal{L}^{0}_{l}$, for factor $F^{0}_{.,l}$ being weaker
than $F^{0}_{.,k}$. So, for instance, it is legitimate to have $\lvert 
\widehat{\mathcal{L}} _{k}\triangle \left(\cup_{l=k, \ldots, r} \mathcal{L}%
_l^0\right)\rvert / N^{\alpha _{k}} =o_{p}(1)$ in Proposition \ref%
{prop:symm_diff} when $\alpha_{k}$ is unique. This implication together with
Proposition \ref{prop:symm_diff} (i) delivers a bound when $\alpha_{k}$ is
unique as follows, 
\begin{equation}
\sum_{i=1}^{N}\mathbf{1}\left\{ \lambda _{ik}^{0}\neq 0\right\} +
o_{p}(N^{\alpha_{k}}) \leq \widehat{D}_{k}\leq \sum_{l=k}^{r}\sum_{i=1}^{N}%
\mathbf{1}\left\{ \lambda _{il}^{0}\neq 0\right\} + o_{p}(N^{\alpha_{k}}) ,
\label{eq:strength_inq}
\end{equation}%
which guarantees the consistency of $\widehat{\alpha}_{k}$.
When $\alpha_{k}$ is not unique, one would replace $o_{p}(N^{\alpha_{k}})$
with $O_{p}(N^{\alpha_{k}})$ in \eqref{eq:strength_inq}, which suffices for
the consistency of $\widehat{\alpha}_{k}$. The following theorem gives the consistency of factor strength estimators.

\begin{theorem}
\label{thm: factor_strength}Under Assumptions 1-5 and 8-9, for $k=1,...,r,$ $%
\widehat{\alpha }_{k}\overset{p}{\rightarrow }\alpha _{k}.$
\end{theorem}

\noindent \textbf{Remark 15.} (i) Theorem \ref{thm: factor_strength} shows that
our proposed estimator can consistently estimate strengths for all factors. The intuition behind this
consistency result is similar to that behind Proposition \ref{prop:symm_diff}%
. Again, the key is to realize that the rotated loading matrix $\Lambda
_{\cdot ,k}^{\ast }$ is able to preserve strength $\alpha _{k}$ up to the
given threshold $c_{\lambda ,NT}$, and also that the errors induced by
estimation are also uniformly dominated by $c_{\lambda ,NT}$. (ii) The above theorem provides only consistency results for the factor strength estimators. Following BKP(2021), we can further establish the convergence rate and limiting distribution for the strength estimators with more complicated conditions. 
(iii) In practice, one can rely on the estimated factor strength to determine a latent factor is strong or weak.  As argued by BKP (2021),  the precise estimation of factor strength relies on a large cross-section  sample  size $N$. So we follow BKP (2021) and suggest a conservative way to treat a factor with (estimated) strength above 0.95 as strong enough, while a factor with strength below 0.90 as weak enough. 

\section{THE DETERMINATION OF THE NUMBER OF FACTORS}

\label{sec:no_f}

The determination of the number of factors has been of long-standing
interest in the literature of factor models. Various selection criteria have
been proposed for strong factor models, e.g., Bai and Ng (2002), Onatski
(2010), Ahn and Horenstein (2013), Lu and Su (2017), Wei and Chen (2020) and
Fan et al. (2022). As for consistent selection of the number of factors in
weak factor models, UY (2023a) show that the edge
distribution (ED) estimator by Onatski (2010) is consistent, and Onatski
(2015) proposes selecting the number of factors based on the approximations
to the squared error of the least squares estimator of the common component
under both strong and weak factor asymptotics. Freyaldenhoven (2022) devises
a statistic in combining both eigenvalues and eigenvectors of the covariance
matrix, to enhance its discriminatory power in distinguishing factors
stronger than a certain threshold, assuming that $N$ and $T$ grow
proportionally. Guo et al. (2022) exploit a data-driven adaptive penalty of
factor strength for information criteria, to select weak factors and
meanwhile avoid overfitting.

We propose determining the number of factors based on SVT, as discussed by
Bai and Ng (2019) and Freyaldenhoven (2022). The procedure is very simple
and works in the same spirit as in Bai and Ng (2002). To determine the
number of factors in $F^{0}$, we use the following estimator of $r$, 
\begin{equation}
\widehat{r}=\max \left\{ k:\widetilde{V}_{k}^{r_{\max }}\geq \widehat{\sigma 
}^{2}N^{-1/2}\left( \ln \ln N\right) ^{1/2}\right\} ,  \label{eq:no_factor}
\end{equation}%
where $\widehat{\sigma }^{2}$ is a consistent estimator of $%
(NT)^{-1}\sum_{i=1}^{N}\sum_{t=1}^{T}E\left( e_{it}^{2}\right) $, $r_{\max }$
is a large bounded positive integer such that $r\leq r_{\max }$, and $%
\widetilde{V}_{k}^{r_{\max }}$ is the $k$th diagonal element of $\widetilde{V%
}^{r_{\max }}\equiv \diag\left( \widetilde{V}_{1},\ldots ,\widetilde{V}%
_{r_{\max }}\right) $ being an $r\times r$ diagonal matrix consisting of the 
$r_{\max }$ largest eigenvalues of $X^{\prime }X/(NT)$ in decreasing order.
As for $\widehat{\sigma }^{2}$ in practice, we compute it similarly to Bai
and Ng (2002) as $\widehat{\sigma }^{2}=\frac{1}{NT}\sum_{i=1}^{N}%
\sum_{t=1}^{T}\left( X_{it}-\widetilde{\lambda }_{i}^{r_{\max }\prime }%
\widetilde{F}_{t}^{r_{\max }}\right) ^{2}$, where the superscript $r_{\max }$
signifies the allowance of $r_{\max }$ factors in the estimation. We set $r_{\max}=8$ later in our numerical studies. To check the robustness of $r_{\max }$, we also include results under different values of $r_{\max }$ in the Online Appendix.

\noindent \textbf{Remark 16. }The estimator proposed in \eqref{eq:no_factor}
is very similar to the \textquotedblleft $BN_{\sqrt{N}}$\textquotedblright\
estimator proposed in equation (16) of Freyaldenhoven (2022). Specifically,
their defined $g(N)$ composing the SVT equals to $\left( \ln \ln N\right)
^{1/2}$ in our setting.

\begin{theorem}
\label{thm: number of factors}Under Assumptions 1-5 and 8-9, $\widehat{r}%
\overset{p}{ \rightarrow }r.$
\end{theorem}

\noindent \textbf{Remark 17. }Proving Theorem \ref{thm: number of factors}
is to check two conditions to hold in probability: (i) $\widehat{r}\geq r$
and (ii) $\widehat{r}\leq r.$ Condition (i) is actually implied by
Proposition \ref{prop: V_rate}. We prove condition (ii) by contradiction,
and this relies on a sharper bound obtained under $\left\Vert e\right\Vert
_{sp}^{2}=O_{p}\left( \max \left\{ N,T\right\} \right) $ imposed by
Assumption 4. This assumption is also needed for Bai and Ng (2002) in
determining the number of factors via information criteria even in strong
factor models.

\noindent \textbf{Remark 18. }The ED estimator proposed by Onatski (2010) is
based on the fact that all the \textquotedblleft
systematic\textquotedblright\ eigenvalues diverge to infinity, whereas any
finite number of the largest \textquotedblleft
idiosyncratic\textquotedblright\ eigenvalues cluster around a single point.
Onatski (2010) determines the number of factors by separating the diverging
eigenvalues with a wedge parameter $\delta $, and proves the consistency of
the ED estimator only requiring that $\min_{1\leq k\leq r}\alpha _{k}>0$.
However, the ED estimator hinges on the idiosyncratic terms being Gaussian,
or being independent cross-sectionally or over time in case of non-Gaussian,
and is thus restrictive in applications with macroeconomics and finance. As
for choosing $\delta $, Onatski (2010) approximates the upper bound of
eigenvalue differences by an OLS estimate which is then doubled to formulate 
$\delta $. His final estimator of $r$ is obtained via iterations given $%
\delta $. In contrast, our estimator of $r$ is more straightforward to use.

\noindent \textbf{Remark 19. }So far, the sparsity in this paper refers to
the \textit{exact} sparsity. We can also consider \textit{approximate}
sparsity under which loadings may contain considerable nonzero yet small
entries. Such settings are considered and allowed for in Lettau and Pelger
(2020) and Bai and Ng (2023). All of our results will continue to hold if
the small entries decay sufficiently fast. For example, it is not hard to
show that the extension works if we replace the exact sparse loadings $%
\lambda _{ik}^{0}$ with the approximate sparse ones $\lambda
_{ik}^{\triangle }:=\lambda _{ik}^{0}\mathbf{1}\left\{ \lambda _{ik}^{0}\neq
0\right\} +N^{-\beta }\mathbf{1}\left\{ \lambda _{ik}^{0}=0\right\} $, for $%
\beta >1/2$.

\section{MONTE CARLO SIMULATIONS}

\label{sec:MC}

In this section, we study the finite sample performance of our proposed
estimators for sparsity-induced weak factor models. It includes comparison
of simple PC estimators with various regularized estimators, in the estimation
of factors, loadings, common components, the number of factors, as well as
the factor strengths.

\subsection{The Data Generating Process}

We consider the following data generating process (DGP):%
\begin{eqnarray*}
X_{it} &=&\lambda _{i}^{0\prime }F_{t}^{0}+e_{it},\text{ }i=1,...,N,\text{ }%
t=1,...,T, \\
F_{1t}^{0} &=&0.5F_{1,t-1}^{0}+u_{1t},\text{ and }F_{kt}^{0}=\left(
-0.8\right) ^{k}F_{1t}^{0}+u_{kt},\text{ }k=2,...,r.
\end{eqnarray*}%
The simulated factors are correlated with each other with various degrees.
We let $u_{kt}$ be mutually uncorrelated $N\left( 0,1\right) $ for $%
k=1,...,r $. To specify the factor loadings $\lambda _{ik}^{0},$ for each $%
k=1,...,r,$ we first \textit{randomly} select $\lfloor N^{\alpha
_{k}}\rfloor $ of $\left\{ \lambda _{ik}^{0}\right\} _{i=1}^{N}$ and specify
them as iid $N\left( 0,1\right) $, and then set the rest of $\left\{ \lambda
_{ik}^{0}\right\} _{i=1}^{N}$ as zero.

For the $N\times 1$ vector $e_{t},$ we specify the (marginal) distribution
of $e_{it}$ as the student-$t\left( 5\right) $ to allow for heavy tails. The
cross-sectional dependence across $e_{it}$ is admitted through the $N\times
N $ covariance matrix $\Sigma _{e}$ as follows. $\Sigma _{e}=\diag\left\{
\Sigma _{1},...,\Sigma _{N/4}\right\} $ as a block-diagonal matrix with $%
4\times 4$ blocks located along the main diagonal. Each $\Sigma _{i}$ is
assumed to be $I_{4}$ initially. We then randomly choose $\lfloor
N^{0.3}\rfloor $ blocks among them and make them non-diagonal by setting $%
\Sigma _{i}\left( m,n\right) =0.5^{\left\vert m-n\right\vert }$.\ The design
of cross-sectional dependence follows Fan et al. (2015) except that the
dependence is stronger here.

We have tried simulations with the number of factors $r=3$ and $5.$ For $%
r=3, $ we set $\alpha =\left( 0.9,0.75,0.6\right) ;$ while for $r=5,$ we set 
$\alpha =\left( 1,0.9,0.8,0.7,0.6\right) $. To preserve space, we only include results under $r=3$ in the main text and delegate the results under $r=5$ to the Online Appendix. The replication number of
simulations is set as 2000.

\subsection{Simulation Results}

We first compare different methods to determine the number of factors with
ours (WZ). The alternative selection rules range from the $IC_{p1}$ by Bai
and Ng (2002, BN), Guo et al. (2022, GCT), Freyaldenhoven (2022, FR), ED by
Onatski (2010), and Ahn and Horenstein (2013, AH). $r_{\max}$ is set to be 8
if needed. The root mean square error (RMSE) and bias of the estimated
number of factors by each method are reported in Tables \ref%
{tab:no_factor_r3}. It is obvious that GCT, FR
and AH are all subject to underestimation of $r$ in the presence of weak
factors. ED and BN are not very bad but they are outperformed by our
proposed method, since they tend to over- and under-estimate $r$,
respectively. The ED estimator is not as effective as found previously in
weak factor models, e.g., by UY (2023a), implying that the
ED performance may be sensitive to the choice of the wedge parameter $\delta$%
, as remarked in Section 5. Our proposed estimator of factor numbers is
outstanding against all alternatives at almost all sample sizes.

\begin{table}[tbph]
\caption{Estimating the number of factors when $r=3$}
\label{tab:no_factor_r3}\centering
\resizebox{\linewidth}{!}{
		\begin{tabular}{ccccccccccccccc}
			\hline
			&       & \multicolumn{6}{c}{RMSE}                      &       & \multicolumn{6}{c}{Bias} \\
			\cline{3-8}\cline{10-15}    $N$     & $T$     & WZ    & BN    & GCT   & FR    & ED    & AH    &       & WZ    & BN    & GCT   & FR    & ED    & AH \\
			\hline
			100   & 100   & 0.291  & 0.676  & 0.939  & 1.962  & 0.415  & 1.998  &       & 0.078  & -0.446  & -0.882  & -1.944  & 0.034  & -1.998  \\
			& 200   & 0.213  & 0.425  & 0.867  & 1.967  & 0.412  & 2.000  &       & -0.028  & -0.174  & -0.752  & -1.950  & 0.083  & -2.000  \\
			& 400   & 0.269  & 0.311  & 0.782  & 1.972  & 0.952  & 1.999  &       & -0.070  & -0.094  & -0.608  & -1.959  & 0.322  & -1.998  \\
			200   & 100   & 0.183  & 0.594  & 0.912  & 1.968  & 0.282  & 2.000  &       & 0.028  & -0.347  & -0.832  & -1.944  & 0.058  & -2.000  \\
			& 200   & 0.143  & 0.224  & 0.722  & 1.949  & 0.289  & 2.000  &       & -0.015  & -0.046  & -0.519  & -1.904  & 0.063  & -2.000  \\
			& 400   & 0.180  & 0.092  & 0.454  & 1.961  & 0.397  & 2.000  &       & -0.033  & -0.008  & -0.199  & -1.927  & 0.075  & -2.000  \\
			400   & 100   & 0.140  & 0.676  & 0.928  & 1.950  & 0.225  & 2.000  &       & 0.018  & -0.455  & -0.862  & -1.902  & 0.048  & -2.000  \\
			& 200   & 0.077  & 0.217  & 0.632  & 1.795  & 0.288  & 2.000  &       & -0.004  & -0.045  & -0.397  & -1.609  & 0.064  & -2.000  \\
			& 400   & 0.102  & 0.032  & 0.225  & 1.583  & 0.261  & 1.999  &       & -0.010  & 0.000  & -0.012  & -1.253  & 0.055  & -1.998  \\
			\hline
	\end{tabular}	}
\end{table}


To study the finite sample performance of PC estimators under weak factor
models, we compare PC regression with the sparse orthogonal factor
regression (SOFAR) proposed by UY (2023a, b), assuming the
true number of factors to be known. Based on Lasso penalization on factor
loadings for sparsity, UY (2023a) develop the Adaptive (Ada) SOFAR
estimator. They later make inference based on the SOFAR estimator by
introducing the Debiased (Deb) SOFAR estimator to recover its asymptotic
normality. Furthermore, they address the multiple testing problem of loading
sparsity and construct the Resparsified (Res) SOFAR estimator to fulfill FDR
control. Note that the various SOFAR estimators in UY (2023b) come from penalized regression which targets sparsity explicitly.
Thus it seems natural to suggest that one should be better off using their
approaches to deal with weak factor models than using PC. However, as we
have mentioned before, the simple PC estimators enjoy a nice property of
automatic sparsity recognition, so that PC could even outperform the
complicated SOFAR.

To measure the performance of factor and loading estimators, whose
asymptotic properties are well studied in previous Section \ref%
{sec:large_sample}, we follow Doz et al. (2012) to use the trace statistics:%
\footnote{%
Uematsu and Yamagata (2023a) evaluate performance of estimators by the $\ell
_{2}-$norm losses: $\left\Vert \sum_{k=1}^{r}N_{k}^{-1/2}\left[ \func{abs}
\left( \widehat{\Lambda }_{\cdot ,k}\right) -\func{abs}\left( \Lambda
_{\cdot ,k}^{0}\right) \right] \right\Vert _{2}$ and $\left\Vert
\sum_{k=1}^{r}T^{-1/2}\left[ \func{abs}\left( \widehat{F}_{\cdot ,k}\right)
- \func{abs}\left( F_{\cdot ,k}^{0}\right) \right] \right\Vert _{2}$. Such
norm losses are more relevant when the factors and loadings are identified
up to column-wise sign indeterminacy, rather than just rotation
indeterminacy, and additional restrictions are required, as we explain right
below. We instead employ the trace statistics whose validity does not rely
on such restrictions, and they demonstrate how effectively estimators of
factors (loadings) span the same space as latent factors (loadings).} 
\begin{equation*}
TR^{F}=\frac{\func{Tr}\left( F^{0\prime }\widetilde{F}\left( \widetilde{F}
^{\prime }\widetilde{F}\right) ^{-1}\widetilde{F}^{\prime }F^{0}\right) }{ 
\func{Tr}\left( F^{0\prime }F^{0}\right) },\quad TR^{\Lambda }=\frac{\func{%
Tr }\left( \Lambda ^{0\prime }\widetilde{\Lambda }\left( \widetilde{\Lambda }
^{\prime }\widetilde{\Lambda }\right) ^{-1}\widetilde{\Lambda }^{\prime
}\Lambda ^{0}\right) }{\func{Tr}\left( \Lambda ^{0\prime }\Lambda
^{0}\right) }.
\end{equation*}%
We also report the root mean squared errors in estimating the common
component $C_{it}$ (RMSE$^{C}$). All results are included in Tables \ref%
{tab:FM_estimate_r3}. The Res estimator is
under the FDR rate $q=0.1$.\footnote{%
We have calculated the Res estimator when $q=0.05$, and the results are very
close to reported here.} It is interesting to see that the PC estimators of
factors and common components are almost always better than any of SOFAR
type estimators. Although the Resparsified SOFAR estimator for $\Lambda $
outperforms PC under bigger sample sizes, the margin is small. The SOFAR
estimators are constructed under pseudo-true loadings that are sparse and
meanwhile the pseudo-true factors that are \textit{orthogonal}. As commented by Bai and Ng (2023, page 1906),  this approach hinges on a relatively strong
restriction. The restriction, however, does not
agree with our DGP with correlated factors and sparse loadings, which might
account for the worse performance of SOFAR. In an additional experiment
whose results are not reported here, we modify our DGP with factors being
indeed orthogonal while keeping the rest unchanged, and find that the SOFAR estimators are much closer to or
even outperforms the PC estimator in estimating factors evaluated by $TR^{F}$%
, though their $RMSE^{C}$ are still bigger in general. Hence regression by
PC (with screening) seems a better choice given its robustness and easy
implementation. On the other hand, it is worth mentioning that the
restriction of $\alpha _{r}>1/2$ required for PC estimators in this paper is
stronger than what is required in UY (2023a, b) for their
SOFAR estimators.

While the results reported in Tables \ref{tab:FM_estimate_r3} are under known numbers of factors, we also experiment
with estimated numbers of factors by each proposed approach, which is more
realistic and reflects more precisely how the estimation of factor numbers
may affect consequent estimators, and report the results in Appendix \ref%
{sec:additional_MC}. The results suggest that the conclusion above basically
still holds, except for the comparison of RMSE$^{C}$ when both $N$ and $T$
are relatively small.

\begin{table}[tbph]
\caption{Estimation of factor models when $r=3$}
\label{tab:FM_estimate_r3}\centering
\resizebox{\linewidth}{!}{
		\begin{tabular}{cccccccccccccccc}
			\hline
			&       & \multicolumn{4}{c}{$TR^{F}$}     &       & \multicolumn{4}{c}{$TR^{\Lambda}$}     &       & \multicolumn{4}{c}{$RMSE^C$} \\
			\cline{3-6}\cline{8-11}\cline{13-16}    $N$     & $T$     & PC    & Ada   & Deb   & Res   &       & PC    & Ada   & Deb   & Res   &       & PC    & Ada   & Deb   & Res \\
			\hline
			100   & 100   & \textbf{0.924 } & 0.898  & 0.909  & 0.909  &       & \textbf{0.718 } & 0.622  & 0.709  & 0.713  &       & \textbf{0.973 } & 1.020  & 1.005  & 1.005  \\
			& 200   & \textbf{0.936 } & 0.918  & 0.924  & 0.924  &       & 0.786  & 0.720  & 0.783  & \textbf{0.788 } &       & \textbf{0.953 } & 0.988  & 0.967  & 0.967  \\
			& 400   & \textbf{0.943 } & 0.935  & 0.938  & 0.938  &       & 0.830  & 0.806  & 0.833  & \textbf{0.837 } &       & \textbf{0.948 } & 0.970  & 0.958  & 0.957  \\
			200   & 100   & \textbf{0.955 } & 0.934  & 0.942  & 0.942  &       & 0.745  & 0.647  & 0.765  & \textbf{0.786 } &       & \textbf{0.886 } & 0.941  & 0.894  & 0.893  \\
			& 200   & \textbf{0.964 } & 0.953  & 0.957  & 0.957  &       & 0.811  & 0.752  & 0.826  & \textbf{0.841 } &       & 0.881  & 0.910  & 0.879  & \textbf{0.879 } \\
			& 400   & \textbf{0.969 } & 0.964  & 0.966  & 0.966  &       & 0.852  & 0.838  & 0.861  & \textbf{0.871 } &       & \textbf{0.872 } & 0.892  & 0.877  & 0.877  \\
			400   & 100   & \textbf{0.969 } & 0.951  & 0.956  & 0.956  &       & 0.750  & 0.660  & 0.767  & \textbf{0.803 } &       & \textbf{0.813 } & 0.866  & 0.820  & 0.818  \\
			& 200   & \textbf{0.976 } & 0.966  & 0.970  & 0.970  &       & 0.816  & 0.764  & 0.828  & \textbf{0.853 } &       & \textbf{0.806 } & 0.844  & 0.812  & 0.812  \\
			& 400   & \textbf{0.980 } & 0.975  & 0.977  & 0.977  &       & 0.858  & 0.846  & 0.864  & \textbf{0.879 } &       & \textbf{0.802 } & 0.822  & 0.805  & 0.804  \\
			\hline
	\end{tabular}}
\end{table}

Given our emphasis on sparsity recovery in this paper, we are also curious
about how well sparsity can be recovered based on PCA. To address this
question, we look into the false discover rate (FDR) and power performance
for recover sparsity of the loading matrix, following UY (2023b). To define the two terms, let $\mathcal{S}$ denote an index set of
nonzero elements (e.g., $\mathcal{L}^{0}_{k}$), and $\widehat{\mathcal{S}}$
be a set discovered by some procedure (e.g., $\widehat{\mathcal{L}}_{k}$).
Then, 
\begin{equation*}
\mathrm{FDR}=\mathbb{E}[\mathrm{FDP}] \quad \text { with } \quad \mathrm{FDP}
=\frac{\left|\mathcal{S}^c \cap \widehat{\mathcal{S}}\right|}{|\widehat{ 
\mathcal{S}}| \vee 1},
\end{equation*}
and 
\begin{equation*}
\text { Power }=\mathbb{E}\left[\frac{|\mathcal{S} \cap \widehat{\mathcal{S}}
|}{|\mathcal{S}| \vee 1}\right].
\end{equation*}
Specifically, $\text{FDR}_{k}$ and $\text{Power}_{k}$ denote the results
for sparsity for factor $k$'s loadings, while $\overline{\text{FDR}}$ and $%
\overline{\text{Power}}$ are for the overall factors' loadings.\footnote{%
For our specific purpose, $\text{FDR}_{k}$ and $\text{Power}_{k}$ are
perhaps more relevant than the ``maximum cosine similarity'' in
Freyaldenhoven (2023), which measures the correlation between the estimated
loading and the true one for each factor.} Tables \ref{tab:FDR_r3} shows that the results of FDR and power are better for stronger
factors using either the PC or SOFAR based method. Admittedly, the PC based
method is inferior to SOFAR based ones at relatively small or medium sample
sizes. This is expected as the SOFAR based methods explicitly target
sparsity by additionally employing regularization or FDR control, while the
simple PC estimators are inevitably subject to contamination distortion
discussed in Section 4.1. Nevertheless, it is a bit surprising to see that
FDR is abnormally high when either $N$ or $T$ is large for SOFAR, while it
is decreasing with sample sizes for PC.\footnote{%
We have also compared with the debiased SOFAR and the method proposed by
Freyaldenhoven (2023). Each of the two methods results in extremely high FDR.%
}

\begin{table}[htbp]
\caption{FDR and Power under $r=3$ }
\label{tab:FDR_r3}\centering
\begin{tabular}{ccccccccccc}
\hline
$N$ & $T$ & \multicolumn{1}{c}{$\text{FDR}_{1}$} & \multicolumn{1}{c}{$\text{
FDR}_{2}$} & \multicolumn{1}{c}{$\text{FDR}_{3}$} & \multicolumn{1}{c}{$%
\overline{\text{FDR}}$} &  & \multicolumn{1}{c}{$\text{Power}_{1}$} & 
\multicolumn{1}{c}{$\text{Power}_{2}$} & \multicolumn{1}{c}{$\text{Power}
_{3} $} & \multicolumn{1}{c}{$\overline{\text{Power}}$} \\ \hline
&  & \multicolumn{9}{c}{Panel A: PC+Screening} \\ \cline{3-11}
100 & 100 & 0.208 & 0.412 & 0.448 & 0.295 &  & 0.806 & 0.434 & 0.492 & 0.657
\\ 
& 200 & 0.212 & 0.403 & 0.431 & 0.292 &  & 0.845 & 0.445 & 0.532 & 0.688 \\ 
& 400 & 0.201 & 0.396 & 0.416 & 0.281 &  & 0.887 & 0.462 & 0.556 & 0.721 \\ 
200 & 100 & 0.215 & 0.430 & 0.438 & 0.297 &  & 0.838 & 0.450 & 0.535 & 0.694
\\ 
& 200 & 0.213 & 0.399 & 0.369 & 0.280 &  & 0.872 & 0.474 & 0.579 & 0.727 \\ 
& 400 & 0.206 & 0.390 & 0.358 & 0.271 &  & 0.909 & 0.497 & 0.612 & 0.760 \\ 
400 & 100 & 0.209 & 0.419 & 0.355 & 0.277 &  & 0.852 & 0.492 & 0.623 & 0.734
\\ 
& 200 & 0.205 & 0.404 & 0.255 & 0.257 &  & 0.893 & 0.507 & 0.677 & 0.770 \\ 
& 400 & 0.201 & 0.405 & 0.245 & 0.254 &  & 0.923 & 0.517 & 0.687 & 0.793 \\ 
&  & \multicolumn{9}{c}{Panel B: SOFAR\_Adaptive} \\ \cline{3-11}
100 & 100 & 0.018 & 0.085 & 0.252 & 0.080 &  & 0.737 & 0.729 & 0.723 & 0.733
\\ 
& 200 & 0.011 & 0.084 & 0.267 & 0.079 &  & 0.801 & 0.790 & 0.776 & 0.794 \\ 
& 400 & 0.203 & 0.483 & 0.687 & 0.403 &  & 0.908 & 0.736 & 0.675 & 0.827 \\ 
200 & 100 & 0.017 & 0.064 & 0.188 & 0.056 &  & 0.730 & 0.709 & 0.717 & 0.723
\\ 
& 200 & 0.010 & 0.054 & 0.170 & 0.046 &  & 0.803 & 0.786 & 0.781 & 0.795 \\ 
& 400 & 0.208 & 0.511 & 0.635 & 0.396 &  & 0.905 & 0.755 & 0.771 & 0.848 \\ 
400 & 100 & 0.202 & 0.361 & 0.348 & 0.249 &  & 0.785 & 0.408 & 0.543 & 0.662
\\ 
& 200 & 0.202 & 0.456 & 0.470 & 0.307 &  & 0.856 & 0.623 & 0.660 & 0.775 \\ 
& 400 & 0.204 & 0.543 & 0.578 & 0.374 &  & 0.897 & 0.757 & 0.786 & 0.849 \\ 
&  & \multicolumn{9}{c}{Panel C: SOFAR\_Resparsified} \\ \cline{3-11}
100 & 100 & 0.046 & 0.177 & 0.338 & 0.142 &  & 0.785 & 0.774 & 0.781 & 0.781
\\ 
& 200 & 0.036 & 0.129 & 0.317 & 0.115 &  & 0.851 & 0.843 & 0.826 & 0.845 \\ 
& 400 & 0.212 & 0.532 & 0.712 & 0.435 &  & 0.930 & 0.787 & 0.717 & 0.860 \\ 
200 & 100 & 0.063 & 0.211 & 0.346 & 0.157 &  & 0.764 & 0.727 & 0.756 & 0.753
\\ 
& 200 & 0.035 & 0.126 & 0.273 & 0.099 &  & 0.849 & 0.843 & 0.843 & 0.847 \\ 
& 400 & 0.218 & 0.558 & 0.665 & 0.435 &  & 0.925 & 0.792 & 0.826 & 0.876 \\ 
400 & 100 & 0.221 & 0.588 & 0.690 & 0.416 &  & 0.838 & 0.493 & 0.613 & 0.725
\\ 
& 200 & 0.216 & 0.605 & 0.721 & 0.440 &  & 0.886 & 0.509 & 0.648 & 0.764 \\ 
& 400 & 0.215 & 0.610 & 0.739 & 0.463 &  & 0.920 & 0.613 & 0.760 & 0.824 \\ 
\hline
\end{tabular}%
\end{table}

 Moreover, in additional simulation we try to work with very small gaps of factor strengths and present the FDR and power results for sparsity recovery in Table 13 in the Online Appendix, which compares with Table 3 in the paper with larger gaps of $\alpha=(0.9,0.75,0.6)$. The results show that under the smaller gaps of strength, for factor 1 FDR is increasing from around 0.2 to around 0.3, although its power is quite robust; for factor 2 there is only mild deterioration for its FDR and power; the weakest factor 3 suffers most from the shrinking gaps. So perhaps we need to be cautious to work with a very weak factor whose strength is close to others'.

We further investigate the estimated factor strength by various methods in
Tables \ref{tab:fct_strength_r3}. All
estimates become less accurate as the true strength degree decreases, as
expected. It is admitted that our proposed strength estimate (PC+Screening)
suffers more of overestimation with fairly weak factors, e.g., those with $%
\alpha_{k}=0.6$. This is likely attributed to the contamination discussed in
Section 4.1, and also echos the findings in Tables \ref{tab:FDR_r3}. Otherwise, our estimate is comparable to, and sometimes even
better than, those from SOFAR which is designed deliberately for sparsity
recovery, especially for strength $\alpha_{k} \in [0.7,0.9]$. The message
here again delivers the usefulness of our factor strength estimator, which
works reasonably well while avoids involving heavy computation as in SOFAR.
In addition, given the discussion in Remark 12, we also replace the
screening value $c_{\lambda,NT}$ with $\widetilde{c}_{\lambda,NT}=c[\ln (N
T)]^{-1 / 2}$, and report the estimated factor strengths with $c=0.8$ and $%
1.2$ in Appendix E.2 which do not change much.

\begin{table}[htbp]
\caption{Estimation of factor strength when $r=3$ with $\protect\alpha %
=\left( 0.9,0.75,0.6\right)$}
\label{tab:fct_strength_r3}\centering
\begin{tabular}{ccccccccc}
\hline
&  & \multicolumn{3}{c}{RMSE} &  & \multicolumn{3}{c}{Bias} \\ 
\cline{3-5}\cline{7-9}
$N$ & $T$ & $\widehat{\alpha}_{1}$ & $\widehat{\alpha}_{2}$ & $\widehat{
\alpha} _{3}$ &  & $\widehat{\alpha}_{1}$ & $\widehat{\alpha}_{2}$ & $%
\widehat{\alpha }_{3}$ \\ \hline\cline{3-9}\hline
&  & \multicolumn{7}{c}{Panel A: PC+Screening} \\ \cline{3-9}
100 & 100 & \textbf{0.014 } & \textbf{0.047 } & 0.138 &  & \textbf{0.000 } & 
\textbf{0.006 } & 0.100 \\ 
& 200 & \textbf{0.014 } & \textbf{0.048 } & 0.169 &  & \textbf{0.002 } & 
\textbf{0.003 } & 0.073 \\ 
& 400 & \textbf{0.014 } & \textbf{0.049 } & 0.208 &  & \textbf{0.003 } & 
\textbf{0.007 } & \textbf{0.048 } \\ 
200 & 100 & \textbf{0.010 } & \textbf{0.048 } & 0.126 &  & 0.002 & \textbf{\
0.028 } & 0.111 \\ 
& 200 & \textbf{0.009 } & \textbf{0.045 } & 0.138 &  & \textbf{0.002 } & 
\textbf{0.023 } & 0.091 \\ 
& 400 & \textbf{0.009 } & \textbf{0.045 } & 0.166 &  & \textbf{0.004 } & 
\textbf{0.026 } & \textbf{0.075 } \\ 
400 & 100 & \textbf{0.007 } & \textbf{0.053 } & 0.114 &  & \textbf{0.002 } & 
\textbf{0.044 } & 0.101 \\ 
& 200 & \textbf{0.006 } & 0.048 & 0.103 &  & \textbf{0.002 } & 0.040 & 0.079
\\ 
& 400 & \textbf{0.006 } & \textbf{0.052 } & 0.115 &  & \textbf{0.003 } & 
\textbf{0.042 } & \textbf{0.065 } \\ 
&  & \multicolumn{7}{c}{Panel B: SOFAR\_Debiased} \\ \cline{3-9}
100 & 100 & 0.017 & 0.094 & \textbf{0.085 } &  & 0.004 & -0.060 & \textbf{\
-0.019 } \\ 
& 200 & 0.023 & 0.065 & \textbf{0.106 } &  & 0.019 & 0.028 & \textbf{0.067 }
\\ 
& 400 & 0.030 & 0.103 & \textbf{0.162 } &  & 0.028 & 0.096 & 0.146 \\ 
200 & 100 & 0.012 & 0.095 & \textbf{0.059 } &  & \textbf{0.000 } & -0.071 & 
\textbf{-0.021 } \\ 
& 200 & 0.017 & 0.052 & \textbf{0.084 } &  & 0.014 & 0.025 & \textbf{0.062 }
\\ 
& 400 & 0.025 & 0.095 & \textbf{0.148 } &  & 0.024 & 0.091 & 0.138 \\ 
400 & 100 & 0.010 & 0.102 & \textbf{0.050 } &  & -0.004 & -0.085 & \textbf{\
-0.033 } \\ 
& 200 & 0.013 & \textbf{0.043 } & \textbf{0.054 } &  & 0.011 & \textbf{0.023 
} & \textbf{0.036 } \\ 
& 400 & 0.020 & 0.088 & \textbf{0.114 } &  & 0.019 & 0.086 & 0.105 \\ 
&  & \multicolumn{7}{c}{Panel C: SOFAR\_Resparsified} \\ \cline{3-9}
100 & 100 & 0.024 & 0.067 & 0.095 &  & 0.020 & 0.044 & 0.064 \\ 
& 200 & 0.030 & 0.103 & 0.130 &  & 0.028 & 0.095 & 0.106 \\ 
& 400 & 0.036 & 0.137 & 0.179 &  & 0.035 & 0.135 & 0.164 \\ 
200 & 100 & 0.017 & 0.062 & 0.096 &  & 0.015 & 0.057 & 0.085 \\ 
& 200 & 0.025 & 0.099 & 0.129 &  & 0.024 & 0.098 & 0.121 \\ 
& 400 & 0.031 & 0.130 & 0.162 &  & 0.031 & 0.129 & 0.157 \\ 
400 & 100 & 0.013 & 0.057 & 0.090 &  & 0.012 & 0.054 & 0.081 \\ 
& 200 & 0.021 & 0.093 & 0.104 &  & 0.020 & 0.092 & 0.098 \\ 
& 400 & 0.027 & 0.123 & 0.140 &  & 0.026 & 0.123 & 0.136 \\ \hline
\end{tabular}%
\end{table}

\section{EMPIRICAL APPLICATION}

We apply our approach to explore potential weak factors on macroeconomic
indicators. We use data from the FRED-QD as a quarterly database for
macroeconomic research (McCracken and Ng, 2021). The original data set
consists of 248 quarterly frequency series dating back to 1959:Q2. By
disregarding series with missing observations, we end up with $N=181$
series. The series are classified into 13 groups: NIPA; Industrial
Production; Employment and Unemployment; Housing, Inventories, Orders, and
Sales; Prices; Earnings and Productivity; Interest Rates; Money and Credit;
Household Balance Sheets; Exchange Rates; Stock Markets; and Non-Household
Balance Sheets.\footnote{%
The original FRED-QD data have one additional group named \textquotedblleft
Other\textquotedblright , containing only two series. However, both series
are dropped due to missing observations, leaving us 13 groups in use.}

For each series used in our sample, after having made the decision that the
series should be managed in levels or log-levels, the transformation codes
are first and second differences based on whether the series is $I(0)$, $I(1$%
), or $I(2)$ as suggested by McCracken and Ng (2021). Due to the
transformation, two initial observations are dropped so each series starts
from 1959:$Q4$. We also normalize each series to have zero mean and unit
variance.

\begin{figure}[htbp]
\includegraphics[width=\textwidth, trim = 0 0 0 0]
	{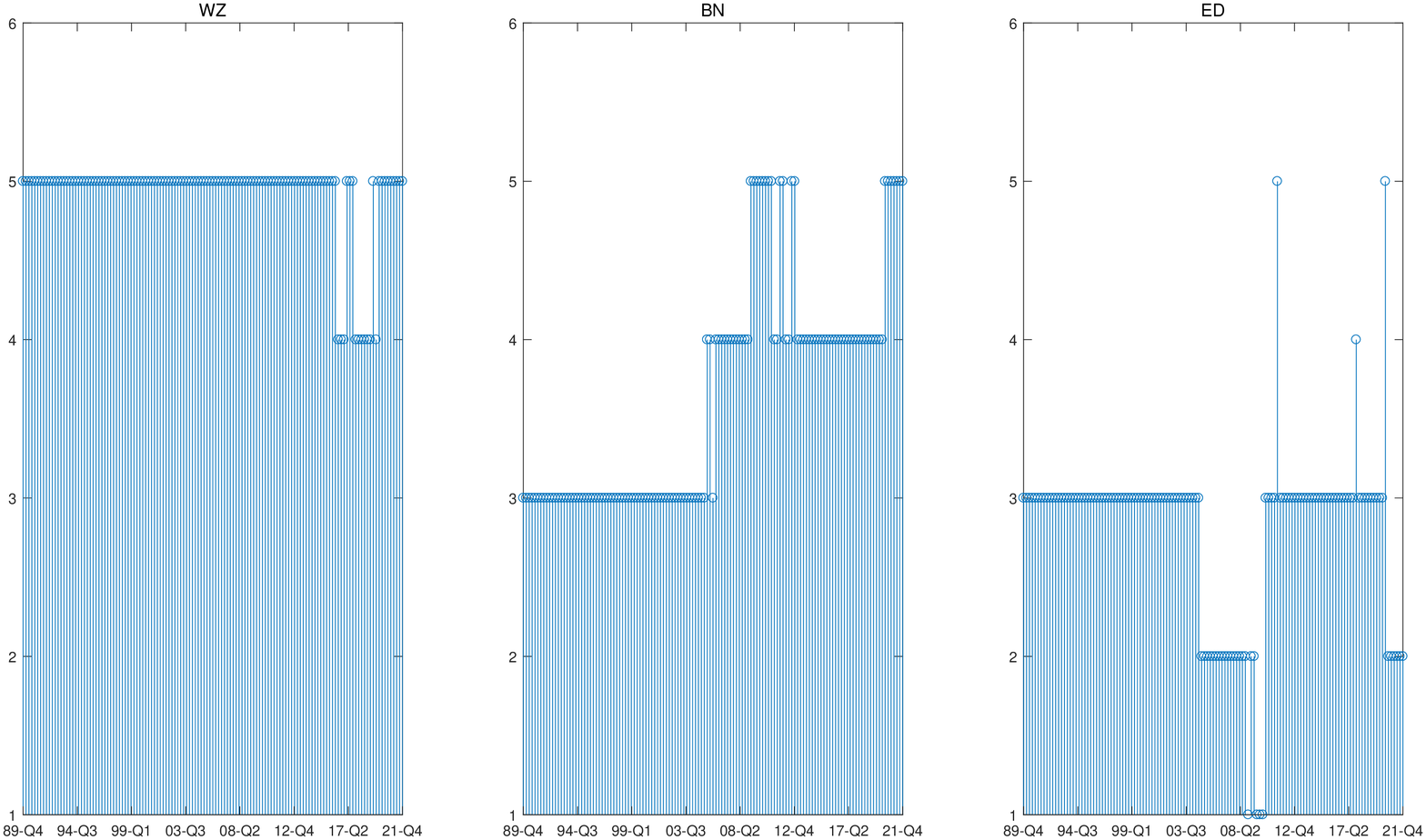}
\caption{Estimated factor numbers by rolling windows}
\label{fg:fn}
\end{figure}

We depict the numbers of weak factors estimated by our method (WZ for
short), Bai and Ng's (2002) $IC_{p1}$ (BN for short) and edge distribution
(ED) estimator by Onatski (2010) in Figure \ref{fg:fn}. The estimation is
implemented under a rolling window scheme with a fixed length 120. The time
on the x-axis denotes the right end point of the window interval. For WZ,
the large panel has five factors in most of time periods. There are a few
windows taking four factors, in which BN also reports the same number. On
the other hand, BN only estimates three factors for the first half of
rolling windows. For the second half, BN finds five factors in only a few
window intervals. Onatski (2010) suggests that the ED method is expected to
well detect present weak factors. However, the result does not seem to agree
with this: ED outputs three factors most of time, and the number drops to
two or even one and oscillates substantially at certain periods. There are
only three times when ED reports estimated factor numbers bigger than three.

\begin{figure}[tbph]
\includegraphics[width=\textwidth]
	{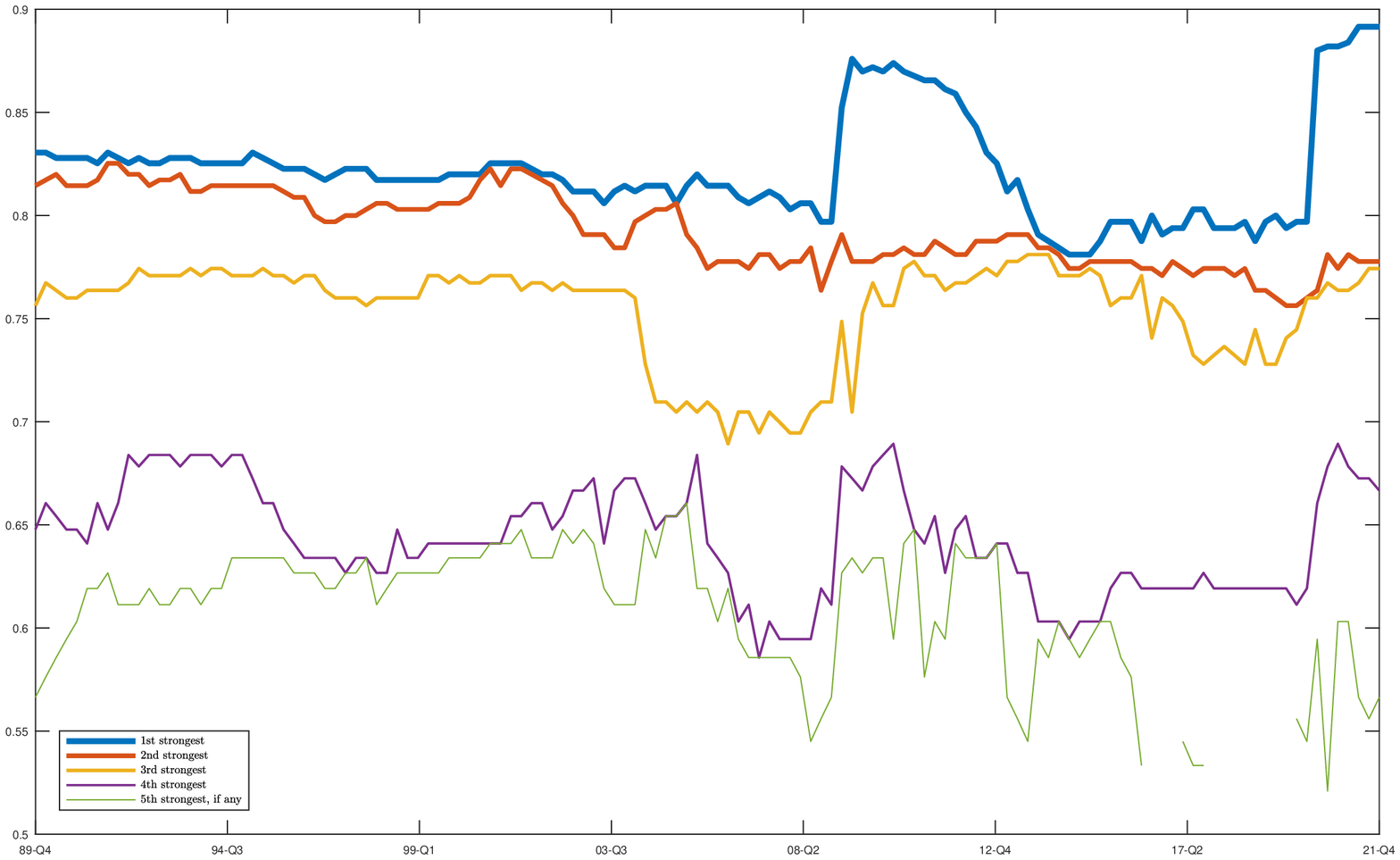}
\caption{Estimated factor strength by rolling windows}
\label{fg:strength}
\end{figure}

Next, to have an idea of estimated factor strength, for each of identified
weak factors we further draw its estimated factor strength over rolling
windows in Figure \ref{fg:strength}. In this way we can study the dynamic of
factor strengths, as it is recently found that factor models may exhibit
time variation in loading parameters; see Ma and Su (2018), Ma et al. (2020)
and Fu et al. (2023). For most of window intervals ending from 1989:Q4 to
2021:Q4, we have five weak factors. Figure \ref{fg:strength} demonstrates a
clear sparsity structure of latent factors. The first two strongest factors
seem to have very close strength around 0.8 most of the time, although $%
\widehat{\alpha }_{1}$ may spike to being close to 0.9 a few times. $%
\widehat{\alpha }_{3}$ ranges from 0.7 to 0.8. $\widehat{\alpha }_{4} $
fluctuates between 0.6 to 0.7, and interestingly it seems to spike
simultaneously with $\widehat{\alpha }_{1}$, while to reach bottom
simultaneously with $\widehat{\alpha }_{3}$. $\widehat{\alpha }_{5}$ is
around 0.6 and moves close to $\widehat{\alpha }_{4}$ up to 2014. However, $%
\widehat{\alpha }_{5}$ drops even to 0 (a reduced factor) during the last 6
years. Factor strengths play a crucial role in the identifying and
estimating risk premia (Pesaran and Smith, 2019) and in factor augmented
regression (Chao et al., 2022). BKP (2021) also points out:
\textquotedblleft The 1st strength of macroeconomic shocks is also of
special interest, as its value has important bearing on forecasting and
policy analysis.\textquotedblright

\begin{figure}[htbp]
\includegraphics[width=\textwidth, height=0.9\textheight, trim = 0 0 0 50]
	{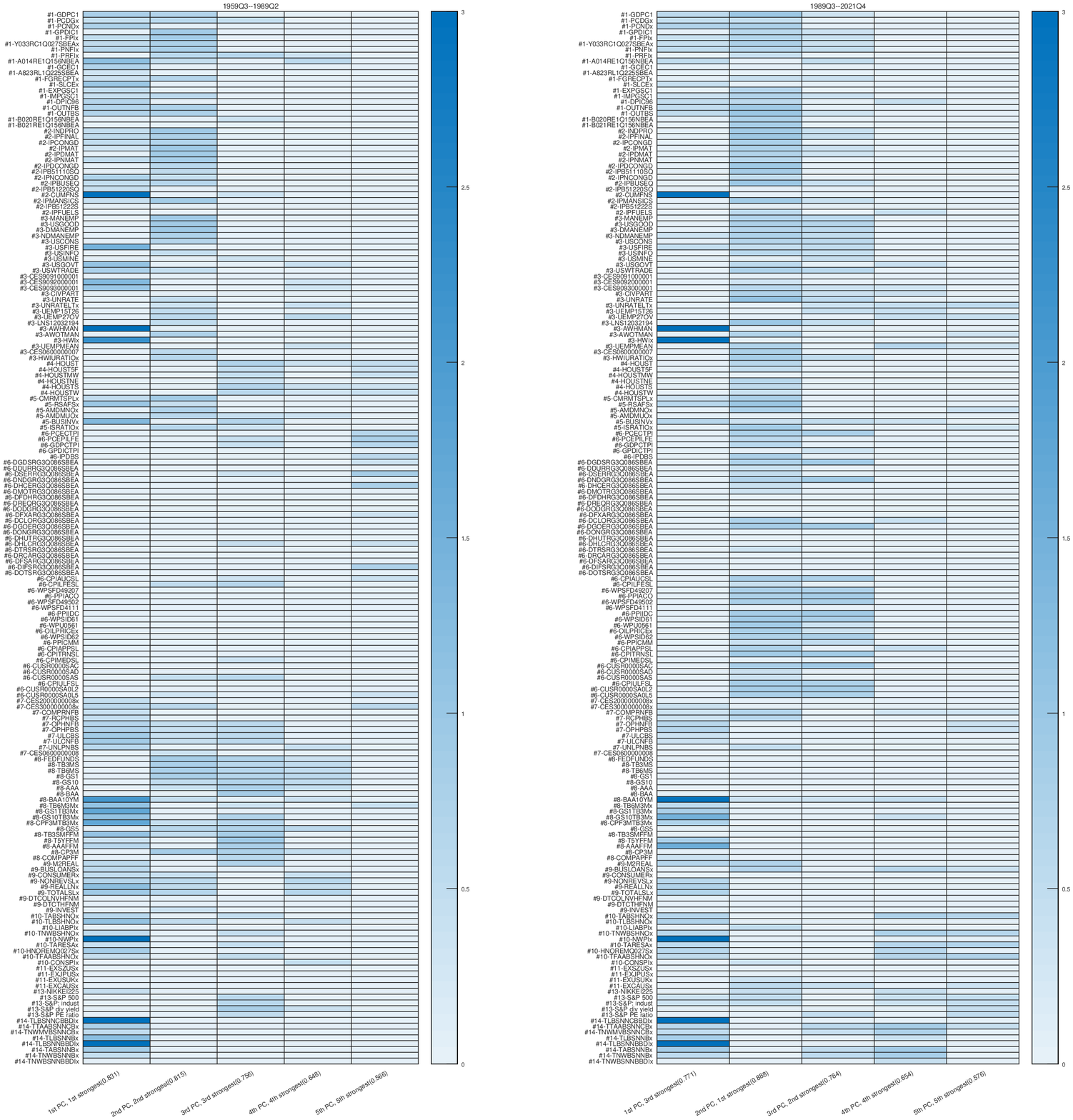}
\caption{Heat map for recovered factor loading sparsity}
\label{fg:heatmap}
\end{figure}

We are also interested in factor pervasiveness in the cross-sectional
dimension, that is, what are those specific series exposed to a given weak
factor, and the relative influence each factor exerts over space. We can
infer the degree of influence roughly by looking at the sparsity of $%
\widehat{\lambda }_{ik}$ over $i$, given the implication by Proposition \ref%
{prop:symm_diff}. We represent distribution of $\lvert \widehat{\lambda }%
_{ik}\rvert $ via a heat map of Figure \ref{fg:heatmap}. We consider two
subperiods of roughly the same length: 1959:Q4-1989:Q2 and 1989:Q3-2021:Q4.
For each subperiod, we detect 5 weak factors. Each row ($i$) represents a
series and we add its associated group number ($\#1-\#13$) in front of its
name. Each column ($k$) represents a principle component extracted as a
latent factor, and we also signify its rank by factor strength. In
particular, the estimated factor strengths are 0.831, 0.815, 0.756, 0.648,
and 0.566 for 1959:Q3-1989:Q2, and 0.888, 0.784, 0.771, 0.654, 0.576 for
1989:Q3-2021:Q4, as put in the parenthesis. The darkness of each cell
indicates the absolute value of $\widehat{\lambda }_{ik}$. Given that $%
\lvert \widehat{\lambda }_{ik}\rvert \leq 3$ mostly across $i$ and $k$, we
right censor $\widehat{\lambda }_{ik}$ at 3 to obtain a sharper
visualization of the heat map.

%

Some interesting results are the following. (i) The 1st PC factor is not
necessarily the strongest one, as the latter only depends on the support of
loadings while the former depends on both the support and magnitude of
(non-null) loadings. (ii) While the top two factor strengths are close
during the first subperiod, they move farther away from each other during
the second subperiod. The rest of three factor strengths are fairly stable
over the two subperiods. (iii) The 2nd PC factor is loaded by many new
series in the second half of periods, most of which belong to group 6
(Prices); the 3rd PC factor also gains additional influence, mostly from
groups 1, 2 and 3 (NIPA; Industrial Production; Employment and
Unemployment), giving rise to its incremental strength, although it loses
influence on a few series belonging to groups 8 (Interest Rates) and 10
(Household Balance Sheets). On the other hand, the 1st PC factor is
decreasing in its pervasiveness in the second half. These movements suggest
that the interpretation of underlying latent factors may change over time in
the long run. (iv) It may be also worth noticing that while some series are
not influenced by any weak factors in subperiod 1, e.g., many of those from
group 6, they start to load on some factors in subperiod 2.

The other point we can take away from Figure \ref{fg:heatmap} is to relate
it to the solution for the weak factor problem proposed by Giglio et al.
(2023a). The recovered sparsity, although contaminated to some extent, may
provide valuable information to their screening-based approach.
Specifically, we can drop unit $i$ with $\widehat{\lambda}_{ik}=0$ to deal
with weak factor $k$ in the spirit of Giglio et al. (2023a), as our theory
suggests it is more likely that $\{i:\widehat{\lambda}_{ik} = 0\} \subseteq
\{i:\lambda _{ik}^{0} = 0\} $.

\section{CONCLUSIONS}

This paper analyzes the well known approach of PCA, and derives several
novel properties of it to deal with weak latent factor models with sparse
factor loadings. It unveils an interesting and fundamental fact that the PC
estimators can preserve the sparsity in estimated factor loadings for
sparsity-induced weak factor models. This fact facilitates the derivation of
asymptotic properties of PC estimators, enables us to recover the sparsity
of loadings, and estimate the strengths of each factor. In addition, the
determination of the number of factors in weak factor models is also
investigated. The numerical studies confirm that our proposed approach works
reasonably well in finite sample, and an empirical application to FRED-QD
data set shows that our method is useful to detect factor strengths, loading
sparsity and their dynamics.

Our PCA based estimators of weak factor models belong to unsupervised PCA.
Conceptually, one can apply supervised PCA (e.g., Huang et al., 2022) to
weak factor models and obtain more efficient estimation and inference. We
agree that supervision with PCA would further improve the performance of our
proposed estimators by better exploiting information available, and yet also
raise up additional complexity. So we leave it to future research. \newline
\bigskip

\textbf{References}\setlength{\baselineskip}{11pt}

\begin{description}
\item Abadir, K. \& J. Magnus (2005)\textit{\ Matrix Algebra}. Cambridge
University Press.

\item Ahn, S.C. \&   A. R. Horenstein (2013) Eigenvalue ratio test for the
number of factors. \textit{Econometrica} 81(3), 1203--1227.

\item Armstrong, T.B., M. Weidner, \&  A. Zeleneev (2023) Robust estimation
and inference in panels with interactive fixed effects. \textit{Working Paper%
}, University of  Southern California.

\item Ando, T. \& J. Bai (2017) Clustering huge number of financial time
series: A panel data approach with high-dimensional predictors and factor
structures. \textit{Journal of the American Statistical Association}
112(519), 1182--1198.

\item Bai, J. (2003) Inferential theory for factor models of large
dimensions. \textit{Econometrica} 71(1), 135--171.

\item Bai, J. (2009) Panel data models with interactive fixed effects. 
\textit{Econometrica} 77(4), 1229--1279.

\item Bai, J. \&  S. Ng (2002) Determining the number of factors in
approximate factor models. \textit{Econometrica} 70(1), 191--221.

\item Bai, J. \&  S. Ng (2006). Confidence intervals for diffusion index
forecasts and inference for factor-augmented regressions.\textit{\
Econometrica} 74(4), 1133-1150.

\item Bai, J. \&  S. Ng  (2019) Rank regularized estimation of approximate
factor models. \textit{Journal of Econometrics} 212(1), 78--96.

\item Bai, J. \&   S. Ng (2023) Approximate factor models with weaker
loadings. \textit{Journal of Econometrics} 235(2), 1893--1916.

\item Bailey, N., G. Kapetanios, \&  M.H.  Pesaran (2021) Measurement of
factor strength: Theory and practice. \textit{Journal of Applied Econometrics%
} 36(5), 587--613.

\item Chao, J.C., Y. Liu, \&   N.R. Swanson (2022) Consistent factor
estimation and forecasting in factor-augmented VAR models. \textit{Working
Paper}, Maryland University.

\item Chao, J.C.\&  N.R.  Swanson (2022) Selecting the relevant variables
for factor estimation in FAVAR models. \textit{Working Paper}, Maryland
University.

\item Choi, I., R. Lin, \&   Y. Shin (2023) Canonical correlation-based
model selection for the multilevel factors. \textit{Journal of Econometrics} 233(1), 22-44.

\item Cochrane, J. (2011) Presidential address: Discount rates. \textit{%
Journal of Finance} 66(4), 1047-1108.

\item Dello-Preite, M., R. Uppal, P. Zaffaroni, \&  I. Zviadadze (2024)
Cross-sectional asset pricing with unsystematic risk. \textit{Working Paper}, Imperial College London. Available at
SSRN 4135146

\item Despois, T.  \&  C. Doz (2023). Identifying and interpreting the
factors in factor models via sparsity: Different approaches. \textit{Journal of Applied Econometrics} 38(4), 451-667.

\item Doz, C.,  D. Giannone, \&  L. Reichlin (2012) A quasi-maximum
likelihood approach for large, approximate dynamic factor models. \textit{%
The Review of Economics and Statistics} 94(4), 1014--1024.

\item Fama, E.\&  K. French (1993). Common risk factors in the returns on
stocks and bonds. \textit{Journal of Financial Economics} 33(1), 3-56.

\item Fan, J., J. Guo, \&  S. Zheng (2022) Estimating number of factors by
adjusted eigenvalues thresholding. \textit{Journal of the American
Statistical Association} 117(538), 852--861.

\item Fan, J., Y. Liao, \&  M.  Mincheva (2011) High-dimensional covariance
matrix estimation in approximate factor models. \textit{The} \textit{Annals
of Statistics} 39(6), 3320--3356.

\item Fan, J., Y. Liao, \&  J. Yao (2015) Power enhancement in
high-dimensional cross sectional tests. \textit{Econometrica} 83(4),
1497--1541.

\item Freyaldenhoven, S. (2022) Factor models with local factors determining
the number of relevant factors. \textit{Journal of Econometrics}  229(1),
80--102.

\item Freyaldenhoven, S. (2023) Identification through sparsity in factor
models: The $\ell _{1}$-rotation criterion. \textit{Working Paper}, Federal Reserve Bank of Philadelphia.

\item Fu, Z., L. Su, \& X. Wang (2023) Estimation and Inference on
time-varying FAVAR models. \textit{Journal of Business \& Economic
Statistics} 42(2), 533--547.

\item Giglio, S. \&  D. Xiu (2021) Asset pricing with omitted factors. 
\textit{Journal of Political Economy} 129(7), 1947--1990.

\item Giglio, S., D. Xiu,  \&  D. Zhang (2023a) Test assets and weak
factors. \textit{Forthcoming in The Journal of Finance}.

\item Giglio, S., D. Xiu,  \&  D. Zhang  (2023b) Prediction when factors are
weak. \textit{Working Paper}, NBER.

\item Guo, X., Y. Chen,  \&  C.Y. Tang (2023) Information criteria for
latent factor models: A study on factor pervasiveness and adaptivity. 
\textit{Journal of Econometrics} 233(1), 237-250.

\item Huang, D., F. Jiang,  K. Li,  G. Tong, \&  G. Zhou (2022) Scaled
PCA: A new approach to dimension reduction. \textit{Management Science}
68(3), 1678--1695.


\item Jiang, P., Y. Uematsu,  \&  T. Yamagata(2023) Revisiting asymptotic
theory for principal component estimators of approximate factor models.
\textit{Working Paper}, Hitotsubashi University. Available at arXiv:2311.00625v1.

\item Kim, S., V. Raponi,  \& P. Zaffaroni (2024) Testing for weak factors
in asset pricing. \textit{Working Paper}, Imperial College Business School. Available at SSRN 4819759.

\item Kristensen, J.T. (2017) Diffusion indexes with sparse loadings.\textit{%
\ Journal of Business \& Economic Statistics} 35(3), 434--451.


\item Lettau, M. \& M. Pelger (2020) Factors that fit the time series and
cross-section of stock returns. \textit{The Review of Financial Studies}
33(5), 2274--2325.

\item Lu, X. \& L.Su  (2017) Determining the number of groups in latent
panel structures with an application to income and democracy. \textit{%
Quantitative Economics} 8, 729--760.

\item Ludvigson, S.C. \& S. Ng (2009) Macro factors in bond risk premia. 
\textit{The Review of Financial Studies} 22(12), 5027-5067.

\item Ma, S., W. Lan, L. Su, \& C. Tsai (2020) Testing alphas in
conditional time-varying factor models with high-dimensional assets. \textit{%
Journal of Business \& Economic Statistics} 38(1), 214--227.

\item Ma, S., O. Linton, \& J. Gao (2021) Estimation and inference in
semiparametric quantile factor models. \textit{Journal of Econometrics}
222(1), 295--323.

\item Ma, S.\& L. Su (2018) Estimation of large dimensional factor
models with an unknown number of breaks. \textit{Journal of Econometrics}
207(1), 1--29.

\item McCracken, M. \& S. Ng (2021). FRED-QD:\textit{\ }A quarterly
database for macroeconomic research.\textit{\ Federal Reserve Bank of St.
Louis Review} 103(1), 1--44.

\item Onatski, A. (2010) Determining the number of factors from empirical
distribution of eigenvalues. \textit{The Review of Economics and Statistics} 92(4), 1004--1016.

\item Onatski, A. (2012) Asymptotics of the principal components estimator
of large factor models with weakly influential factors.\textit{\ Journal of
Econometrics} 168(2), 244--258.

\item Onatski, A. (2015) Asymptotic analysis of the squared estimation error
in misspecified factor models. \textit{Journal of Econometrics} 186(2),
388--406.

\item Pelger, M. \& R. Xiong (2022) Interpretable sparse proximate factors
for large dimensions. \textit{Journal of Business \& Economic Statistic}
40(4), 1642--1664.

\item Pesaran, M.H. \& R. Smith (2019) The role of factor strength and
pricing errors for estimation and inference in asset pricing models. \textit{%
CESifo Working Paper} No.7919, University of Southern California.

\item Rapach, D. \& G. Zhou. (2021) Sparse macro factors. \textit{Working
Paper},  Washington University in St. Louis.


\item Stock, J.H. \& M.W. Watson (2002) Macroeconomic forecasting using
diffusion indexes. \textit{Journal of Business} \& \textit{Economic
Statistics} 20(2), 147--162.

\item Su, L. \& Q. Chen (2013) Testing homogeneity in panel data models
with interactive fixed effects. \textit{Econometric Theory} 29, 1079--1135.

\item Uematsu, Y. \& T. Yamagata  (2023a) Estimation of sparsity-induced
weak factor models. \textit{Journal of Business \& Economic
Statistics} 41(1), 126--139.

\item Uematsu, Y. \& T. Yamagata (2023b) Inference in sparsity-induced
weak factor models. \textit{Journal of Business \&  Economic
Statistics}  41(1), 213--227.

\item Wei, J. \&  H. Chen (2020) Determining the number of factors in
approximate factor models by twice \textit{K}-fold cross validation. \textit{%
Economics Letters} 191, 109149.
\end{description}

\bigskip \newpage \appendix
\setcounter{page}{1}

\begin{center}

\textbf{Online Appendix to “CAN PRINCIPAL COMPONENT ANALYSIS PRESERVE THE SPARSITY IN FACTOR LOADINGS”}

Jie Wei$^{a}$, Yonghui Zhang$^{b}$


$^{a}$School of Economics, Huazhong University of Science and Technology, China

$^{b}$School of Economics, Renmin University of China, China

%
%
\end{center}

%




\section{Proofs of main results in Section \protect\ref{sec:large_sample}}

To start with, we present several useful lemmas which will be used
frequently in the proofs of main results. Their proofs can be found in
Section \ref{sec:auxiliary}. We will use the fact that $\Vert B\Vert
_{sp}\leq \Vert B\Vert _{F}\leq \Vert B\Vert _{sp}\func{rank}(B)^{1/2}$ over
several places in our proofs.

\begin{lemma}
\label{lm:LeF} Under Assumptions 1-5, $\Lambda^{0 \prime} e \widetilde{F} =
O_{p} \left( N^{\alpha_{1}/2} T^{1/2} + N^{\alpha_{1}-\alpha_{r}} T + N^{1+%
\frac{\alpha_{1}}{2}-\alpha_{r}} \right). $
\end{lemma}

\begin{lemma}
\label{lm:Xi} Under Assumptions 1-5, $\Xi =\widetilde{V}^{-1}QAN^{-1}$ is a
full rank (block) lower triangular matrix, and $\Xi =O_{p}(1).$
\end{lemma}

\begin{lemma}
\label{lm:elffv} Under Assumptions 1-5, $\frac{1}{NT}\left\Vert e^{\prime
}\Lambda ^{0}F^{0\prime }\widetilde{F}\widetilde{V}^{-1}\right\Vert
=O_{p}\left( N^{-\frac{\alpha _{r}}{2}}T^{\frac{1}{2}}\right) .$
\end{lemma}

\begin{lemma}
\label{lm:eelffv} Under Assumptions 1-5, $\frac{1}{NT}ee^{\prime }\Lambda
^{0}F^{0\prime }\widetilde{F}\widetilde{V}^{-1}=A^{-\frac{1}{2}}O_{p}(N+T).$
\end{lemma}

\begin{lemma}
\label{lm:G_full_rank} Under Assumptions 1-5, the $r\times r$ matrix $Q
\equiv \widetilde{F}^{\prime} F^{0 }/T$ is of full rank $r$ with probability approaching 1.
\end{lemma}

Next we provide the proofs for the main results on PC estimator.

\noindent \textbf{Proof of Proposition \ref{prop: V_rate}. }Given $X=\Lambda
^{0}F^{0\prime }+e,$ we get 
\begin{equation*}
\frac{1}{NT}X^{\prime }X=\frac{1}{NT}F^{0}\Lambda ^{0\prime }\Lambda
^{0}F^{0\prime }+\frac{1}{NT}F^{0}\Lambda ^{0\prime }e+\frac{1}{NT}e^{\prime
}\Lambda ^{0}F^{0\prime }+\frac{1}{NT}e^{\prime }e.
\end{equation*}%
Multiplying both sides by $N/N^{\alpha _{r}}$,%
\begin{equation*}
\frac{1}{N^{\alpha _{r}}T}X^{\prime }X=\frac{1}{N^{\alpha _{r}}T}F^{0\prime
}\Lambda ^{0\prime }\Lambda ^{0}F^{0\prime }+\frac{1}{N^{\alpha _{r}}T}%
F^{0}\Lambda ^{0\prime }e+\frac{1}{N^{\alpha _{r}}T}e^{\prime }\Lambda
^{0}F^{0\prime }+\frac{1}{N^{\alpha _{r}}T}e^{\prime }e.
\end{equation*}%
Note that the Frobenius norm of the second and third terms on the RHS is $%
o_{p}\left( 1\right) $ as%
\begin{eqnarray*}
\frac{1}{N^{\alpha _{r}}T}\left\Vert F^{0}\Lambda ^{0\prime }e\right\Vert
&\leqslant &\frac{1}{N^{\alpha _{r}}T}\left\Vert F^{0}\right\Vert \left\Vert
\Lambda ^{0\prime }e\right\Vert =\frac{1}{N^{\alpha _{r}}T}O_{p}\left( T^{%
\frac{1}{2}}\right) O_{p}\left( N^{\frac{\alpha _{1}}{2}}T^{\frac{1}{2}%
}\right) \\
&=&O_{p}\left( N^{\frac{\alpha _{1}}{2}-\alpha _{r}}\right) =o_{p}(1).
\end{eqnarray*}%
This implies that 
\begin{equation*}
\frac{1}{N^{\alpha _{r}}T}\left\Vert F^{0}\Lambda ^{0\prime }e\right\Vert
_{sp}=\frac{1}{N^{\alpha _{r}}T}\left\Vert e^{\prime }\Lambda ^{0}F^{0\prime
}\right\Vert _{sp}=o_{p}(1).
\end{equation*}%
For the fourth term,%
\begin{equation*}
\frac{1}{N^{\alpha _{r}}T}\left\Vert e^{\prime }e\right\Vert _{sp}=\frac{1}{%
N^{\alpha _{r}}T}O_{p}(N+T)=O_{p}\left( N^{1-\alpha _{r}}T^{-1}+N^{-\alpha
_{r}}\right) =o_{p}(1).
\end{equation*}%
Hence, the $r$ eigenvalues of $\frac{1}{N^{\alpha _{r}}T}X^{\prime }X$ are
asympotically equal to the $r$ eigenvalues of $\frac{1}{N^{\alpha _{r}}T}%
F^{0}\Lambda ^{0\prime }\Lambda ^{0}F^{0\prime }$ in probability. Given that
the eigenvalues of matrix $AB$ and those of $BA$ are identical, the $r$
eigenvalue of $\frac{1}{N^{\alpha _{r}}T}X^{\prime }X$ are determined in
probability by 
\begin{equation*}
\left( \frac{1}{N^{\alpha _{r}}}\Lambda ^{0\prime }\Lambda ^{0}\right)
\left( \frac{1}{T}F^{0\prime }F^{0}\right) =\frac{1}{N^{\alpha _{r}}}%
A^{1/2}\left( A^{-1/2}\Lambda ^{0}\Lambda ^{0}A^{-1/2}\right) A^{1/2}\left( 
\frac{1}{T}F^{0\prime }F^{0}\right) .
\end{equation*}%
By Assumptions 1-2, both $A^{-1/2}\Lambda ^{0\prime }\Lambda ^{0}A^{-1/2}$
and $T^{-1}F^{0\prime }F^{0}$ converge to some p.d. matrices. It follows
that the $k$th eigenvalue of $\frac{1}{N^{\alpha _{r}}T}X^{\prime }X$
satisfies 
\begin{equation*}
\mu _{k}\left( \frac{1}{N^{\alpha _{r}}T}X^{\prime }X\right) \asymp
_{p}N^{-\alpha _{r}}A_{k}=N^{-\alpha _{r}}N^{\alpha _{k}},\text{ for }%
k=1,\cdots ,r.
\end{equation*}%
Thus, 
\begin{equation*}
\widetilde{V}_{k}=\mu _{k}\left( \frac{1}{NT}X^{\prime }X\right) \asymp
_{p}N^{\alpha _{r}-1}N^{-\alpha _{r}}A_{k}=N^{\alpha _{k}-1},\text{ for }%
k=1,\cdots ,r.\qed
\end{equation*}%
\bigskip

\noindent \textbf{Proof of Proposition \ref{prop:Q_upper_trian}. } We first
define 
\begin{equation*}
\Sigma_{N,\Lambda} = \frac{\Lambda^{0 \prime} \Lambda^{0}}{N},
\end{equation*}
and it follows that $\Sigma _{N,\Lambda }(k,k)\asymp N^{\alpha _{k}-1}$. In
addition, it is easy to see that $\Sigma _{N,\Lambda }(l,k)=O\left(
N^{\alpha _{l}\wedge\alpha _{k} -1}\right) $. Recall that we have already
shown in proving Lemma \ref{lm:G_full_rank} that (a) $H$ is of full rank $r$
in probability$,$ and (b) $H=O_{p}(1)$. Also recall that $\widetilde{V}%
_{k}\asymp _{p}N^{\alpha _{k}-1}$. Hence for $H=\Sigma _{N,\Lambda }
Q^{\prime} \widetilde{V}^{-1}$ to hold with the asymptotic properties (a)
and (b) just mentioned above, the $r\times r$ matrix $Q$ must be such that%
\begin{equation}  \label{eq:Q_structure}
Q(l,k)=O_{p}\left( N^{\alpha _{l}-\alpha _{k}}\right) ,\text{ for }1\leq
k<l\leq r.
\end{equation}

To see why \eqref{eq:Q_structure} must hold, consider $1\leq k<l\leq r$, and
then we have 
\begin{align*}
H(k,l) & = \sum_{j=1}^{r} \Sigma_{N,\Lambda}(k,j) Q(l,j) (\widetilde{V}%
(l,l))^{-1} \\
& \asymp_{p} \Sigma_{N,\Lambda}(k,k) Q(l,k) N^{1-\alpha_{l}} \\
& \asymp_{p} N^{\alpha_{k}-1} N^{1-\alpha_{l}} Q(l,k).
\end{align*}
In order to have $H(k,l) = O_{p}(1)$, we must have \eqref{eq:Q_structure} to
hold.

This clearly demonstrates that matrix $Q$ is a (block) upper triangular
matrix in probability. Specifically, for $1\leq k<l\leq r$ such that $\alpha
_{l}<\alpha _{k},$ it must be that $Q(l,k)=O_{p}\left( N^{\alpha _{l}-\alpha
_{k}}\right)=o_{p}(1).$ Moreover, since $Q$ is also of full rank of $r$
asymptotically as proved in Lemma \ref{lm:G_full_rank}, we also have that $%
Q(k,k)\asymp _{p}1$.  Meanwhile, the whole matrix $Q \equiv \widetilde{F}^{\prime} F^0 / T = O_{p}(1)$. Therefore, $Q$ is a non-strictly (block) upper
triangular matrix. \ \qed\bigskip

\noindent \textbf{Proof of Proposition \ref{prop: factor_consistent}. }Note
that 
\begin{equation*}
\widetilde{F}-F^{0}H=\left( \frac{F^{0}\Lambda ^{0\prime }e\widetilde{F}}{NT}%
+\frac{e^{\prime }\Lambda ^{0}F^{0\prime }\widetilde{F}}{NT}+\frac{e^{\prime
}e\widetilde{F}}{NT}\right) \widetilde{V}^{-1}.
\end{equation*}%
This implies 
\begin{align*}
\frac{1}{\sqrt{T}}\left\Vert \widetilde{F}-F^{0}H\right\Vert & \leq \left\{
\left( \frac{\left\Vert F^{0}\right\Vert \Vert \widetilde{F}\Vert }{T}%
\right) \left( \frac{\left\Vert \Lambda ^{0\prime }e\right\Vert }{\sqrt{T}N}%
\right) \left\Vert \widetilde{V}^{-1}\right\Vert +\frac{\left\Vert e^{\prime
}\Lambda ^{0}F^{0\prime }\widetilde{F}\widetilde{V}^{-1}\right\Vert }{NT%
\sqrt{T}}+\frac{\left\Vert e^{\prime }e\widetilde{F}\right\Vert }{NT^{3/2}}%
\left\Vert \widetilde{V}^{-1}\right\Vert \right\} \\
& =O_{p}\left( \frac{\left\Vert \Lambda ^{0\prime }e\right\Vert }{\sqrt{T}N}%
\left\Vert \widetilde{V}^{-1}\right\Vert \right) +O_{p}\left( \frac{%
\left\Vert e^{\prime }\Lambda ^{0}F^{0\prime }\widetilde{F}\widetilde{V}%
^{-1}\right\Vert }{NT\sqrt{T}}\right) +O_{p}\left( \frac{\left\Vert
e^{\prime }e\widetilde{F}\right\Vert }{NT^{3/2}}\left\Vert \widetilde{V}%
^{-1}\right\Vert \right) .
\end{align*}%
Note that $\frac{1}{\sqrt{T}N}\left\Vert \Lambda ^{0\prime }e\right\Vert
=O_{p}\left( N^{\alpha _{1}/2-1}\right) .$ Also, $\frac{\left\Vert
ee^{\prime }\widetilde{F}\right\Vert }{NT^{3/2}}\leq \frac{\left\Vert
ee^{\prime }\right\Vert _{sp}\Vert \widetilde{F}\Vert }{NT^{3/2}}%
=O_{p}\left( \frac{1}{N}\right) +O_{p}(\frac{1}{T}).$ Together with Lemma %
\ref{lm:elffv} and Proposition \ref{prop: V_rate}, it leads to that%
\begin{equation*}
\frac{1}{T}\left\Vert \widetilde{F}-F^{0}H\right\Vert ^{2}=O_{p}\left(
N^{2\left( 1-\alpha _{r}\right) }\right) O_{p}\left( N^{\alpha
_{1}-2}+T^{-2}\right) =O_{p}\left( N^{\alpha _{1}-2\alpha _{r}}+N^{2\left(
1-\alpha _{r}\right) }T^{-2}\right) .\qed
\end{equation*}%
\bigskip

\noindent \textbf{Proof of Proposition \ref{prop: loading_consistent}. }%
First, note that 
\begin{equation*}
\widetilde{\Lambda }-\Lambda ^{0}Q^{\prime }=\frac{1}{T}e\left( \widetilde{F}%
-F^{0}H\right) +\frac{1}{T}eF^{0}H.
\end{equation*}%
Then let%
\begin{eqnarray*}
L_{1} &=&\frac{1}{T}e\left( \widetilde{F}-F^{0}H\right) \\
&=&\left( \frac{1}{NT}ee^{\prime }\Lambda ^{0}\right) \left( \frac{1}{T}%
F^{0^{\prime }}\widetilde{F}\right) \widetilde{V}^{-1}+\frac{1}{NT^{2}}%
eF^{0}\Lambda ^{0\prime }e\widetilde{F}\widetilde{V}^{-1}+\frac{1}{NT^{2}}%
ee^{\prime }e\widetilde{F}\widetilde{V}^{-1} \\
&\equiv &L_{11}+L_{12}+L_{13}.
\end{eqnarray*}%
Note 
\begin{equation*}
L_{11}=\frac{1}{NT}\left( ee^{\prime }\right) \Lambda ^{0}Q^{\prime }%
\widetilde{V}^{-1}=\frac{1}{NT}O_{p}(N+T)\left( \Lambda ^{0}A^{-\frac{1}{2}%
}\right) \left( A^{-\frac{1}{2}}N\right) \left( AN^{-1}Q^{\prime }\widetilde{%
V}^{-1}\right) .
\end{equation*}%
By Assumption 2, $\left\Vert \Lambda ^{0}A^{-\frac{1}{2}}\right\Vert
=O\left( 1\right) .$ By Lemma \ref{lm:Xi}, $AN^{-1}Q^{\prime }\widetilde{V}%
^{-1}=O_{p}\left( 1\right) .$ So, 
\begin{equation*}
\left\Vert L_{11}\right\Vert \leq \frac{1}{NT}O_{p}(N+T)O_{p}\left( A^{-%
\frac{1}{2}}N\right) =O_{p}\left( N^{-\alpha _{r}}+N^{1-\frac{\alpha _{r}}{2}%
}T^{-1}\right) .
\end{equation*}%
Next for $L_{12},$ $L_{12}=\frac{1}{NT^{2}}\left( eF^{0}\right) \left(
\Lambda ^{0^{\prime }}eF^{0}H\right) \widetilde{V}^{-1}+\frac{1}{NT^{2}}%
eF^{0}\Lambda ^{0^{\prime }}e\left( \widetilde{F}-F^{0}H\right) \widetilde{V}%
^{-1},$ so%
\begin{equation*}
\left\Vert L_{12}\right\Vert =\frac{1}{NT^{2}}O_{p}(\sqrt{NT})O_{p}\left( 
\sqrt{N^{\alpha _{1}}T}\right) O_{p}\left( N^{1-\alpha _{r}}\right) \left[
1+O_{p}\Vert \widetilde{F}-F^{0}H\Vert \right]
\end{equation*}%
Given that $\left\Vert \widetilde{F}-F^{0}H\right\Vert =O_{p}\left( \sqrt{%
TN^{\alpha _{1}-2\alpha _{r}}}+\sqrt{N^{2(1-2\alpha _{r})}T^{-1}}\right) $
by Proposition \ref{prop: factor_consistent}, we have%
\begin{equation*}
\left\Vert L_{12}\right\Vert =O_{p}\left( N^{\frac{\alpha _{1}}{2}-\alpha
_{r}+\frac{1}{2}}T^{-1}+N^{\alpha _{1}-2\alpha _{r}+\frac{1}{2}}T^{-\frac{1}{%
2}}\right) .
\end{equation*}%
Lastly for $L_{13},$ 
\begin{equation*}
\left\Vert L_{13}\right\Vert =\frac{1}{NT^{2}}\left\Vert e\right\Vert
_{sp}^{3}O_{p}\left( T^{\frac{1}{2}}\right) O_{p}\left( N^{1-\alpha
_{r}}\right) =O_{p}\left( T^{-\frac{3}{2}}N^{\frac{3}{2}-\alpha
_{r}}+N^{-\alpha _{r}}\right) .
\end{equation*}

Given Assumption 5$,\quad N^{\frac{3}{2}-\alpha _{r}}T^{-\frac{3}{2}}\ll
N^{1-\frac{\alpha _{r}}{2}}T^{-1}.$ Also, it's easy to see $N^{1-\frac{%
\alpha _{r}}{2}}\gg N^{\alpha _{1}-2\alpha _{r}+\frac{1}{2}}.$ So it follows
that 
\begin{equation}
L_{1}=O_{p}\left( N^{-\frac{\alpha _{r}}{2}}+N^{1-\frac{\alpha _{r}}{2}%
}T^{-1}+N^{\alpha _{1}-2\alpha _{r}+\frac{1}{2}}T^{-\frac{1}{2}}\right) .
\label{eq:L1}
\end{equation}%
Meanwhile, as $T^{-1}eF^{0}H=O_{P}\left( N^{\frac{1}{2}}T^{-\frac{1}{2}%
}\right) $ implied by Assumption 3 (vii), and $H=O_{p}\left( 1\right) $\ as
proved in Lemma \ref{lm:G_full_rank}\textbf{,} 
\begin{equation*}
\left\Vert \widetilde{\Lambda}-\Lambda ^{0}Q^{\prime }\right\Vert
=O_{p}\left( N^{-\frac{\alpha _{r}}{2}}+N^{1-\frac{\alpha _{r}}{2}}T^{-1}+N^{%
\frac{1}{2}}T^{-\frac{1}{2}}\right) =O_{p}\left( N^{-\frac{\alpha _{r}}{2}%
}+N^{\frac{1}{2}}T^{-\frac{1}{2}}\right) .
\end{equation*}%
So we have $N^{-1}\left\Vert \widetilde{\Lambda}-\Lambda ^{0}Q^{\prime
}\right\Vert ^{2}=O_{p}\left( N^{-\alpha _{r}-1}+T^{-1}\right) .$\qed\bigskip

\noindent \textbf{Proof of Lemma \ref{lm: equivalent_H}. }First, note that
from $\frac{1}{NT}X^{\prime }X\widetilde{F}=\widetilde{F}\widetilde{V},$ we
have $\frac{1}{NT}\left( \Lambda ^{0}F^{0\prime }+e\right) ^{\prime
}(\Lambda ^{0}F^{0\prime }+e)\widetilde{F}=\widetilde{F}\widetilde{V}.$ That
is, 
\begin{equation*}
\widetilde{F}\widetilde{V}=\frac{1}{NT}F^{0}\Lambda ^{0\prime }\Lambda ^{0}%
\widetilde{F}+\frac{1}{NT}F^{0}\Lambda ^{0\prime }e\widetilde{F}+\frac{1}{NT}%
e^{\prime }\Lambda ^{0}F^{0\prime }\widetilde{F}+\frac{1}{NT}e^{\prime }e%
\widetilde{F}.
\end{equation*}%
Thus, multiplying $\widetilde{F}^{\prime }/T$ on both sides of the above
equation leads to 
\begin{equation*}
\widetilde{V}=\frac{1}{T}\widetilde{F}^{\prime }F^{0}\frac{1}{N}\Lambda
^{0\prime }\Lambda ^{0}\frac{1}{T}F^{0\prime }\widetilde{F}+\frac{1}{NT^{2}}%
\widetilde{F}^{^{\prime }}F^{0\prime }\Lambda ^{0\prime }e\widetilde{F}+%
\frac{1}{NT^{2}}\widetilde{F}^{\prime }e^{\prime }\Lambda ^{0}F^{0\prime }%
\widetilde{F}+\frac{1}{NT^{2}}\widetilde{F}^{\prime }e^{\prime }e\widetilde{F%
},
\end{equation*}%
which gives rise to 
\begin{align*}
I_{r}& =QH+\frac{1}{NT^{2}}\widetilde{F}^{\prime }F^{0\prime }\Lambda
^{0\prime }e\widetilde{F}\widetilde{V}^{-1}+\frac{1}{NT^{2}}\widetilde{F}%
^{\prime }e^{\prime }\Lambda ^{0}F^{0\prime }\widetilde{F}\widetilde{V}^{-1}+%
\frac{1}{NT^{2}}\widetilde{F}^{\prime }e^{\prime }e\widetilde{F}\widetilde{V}%
^{-1} \\
& \equiv QH+b_{1}+b_{2}+b_{3},\text{ say.}
\end{align*}%
It is easy to see, by Lemme \ref{lm:LeF}, that 
\begin{align*}
b_{1} & = (NT)^{-1} N^{1-\alpha_{r}} O_{p}\left( \Lambda^{0 \prime} e 
\widetilde{F} \right) \\
& = N^{-\alpha_{r}} T^{-1} O_{p} \left( N^{\alpha_{1}/2} T^{1/2} +
N^{\alpha_{1}-\alpha_{r}} T + N^{1+\frac{\alpha_{1}}{2}-\alpha_{r}} \right)
\\
& = O_{p} \left( N^{\frac{\alpha_{1}}{2}-\alpha_{r}}T^{-1/2} +
N^{\alpha_{1}-2\alpha_{r}} + N^{1+\frac{\alpha_{1}}{2}-2\alpha_{r}}T^{-1}
\right) \\
& = O_{p} \left( N^{\frac{\alpha_{1}}{2}-\alpha_{r}}T^{-1/2} +
N^{\alpha_{1}-2\alpha_{r}} \right),
\end{align*}
where the last equality is due to Assumption 5. The same result also holds
for $b_{2}.$ Meanwhile, 
\begin{equation*}
b_{3}=O_{p}\left( 1\right) \frac{1}{NT}\left\Vert e^{\prime }e\right\Vert
_{sp}\widetilde{V}^{-1}=O_{p}\left( N^{-1}+T^{-1}\right) O_{p}\left(
N^{1-a_{r}}\right) =O_{p}\left( N^{-a_{r}}+N^{1-a_{r}}T^{-1}\right) .
\end{equation*}%
So it follows that 
\begin{equation}
I_{r}=QH+O_{p}\left( N^{\frac{\alpha _{1}}{2}%
-a_{r}}T^{-1/2}+N^{-a_{r}}+N^{1-a_{r}}T^{-1}\right) =QH+O_{p}\left( \gamma
_{NT}\right) ,  \label{eq:IQH}
\end{equation}%
implying that $Q=H^{-1}+O_{p}\left( \gamma _{NT}\right) .$

Second, notice that \eqref{eq:IQH} also implies that $H=Q^{-1}+O_{p}\left(
\gamma _{NT}\right) =H_{3}+O_{p}\left( \gamma _{NT}\right) .$

Third, recall by definition, $\widetilde{\Lambda }=\frac{1}{T}\Lambda
^{0}F^{0^{\prime }}\widetilde{F}+\frac{1}{T}e\widetilde{F}.$ Therefore, we
have 
\begin{equation*}
\frac{\Lambda ^{0\prime }\widetilde{\Lambda }}{N}=\frac{\Lambda ^{0\prime
}\Lambda ^{0}}{N}\frac{F^{0\prime }\widetilde{F}}{T}+\frac{\Lambda ^{0\prime
}e\widetilde{F}}{NT}.
\end{equation*}%
So it is followed by%
\begin{equation*}
\left( \Lambda ^{0\prime }\Lambda ^{0}\right) ^{-1}\Lambda ^{0\prime }%
\widetilde{\Lambda }=F^{0\prime }\widetilde{F}/T+\left( \Lambda ^{0\prime
}\Lambda ^{0}\right) ^{-1}\Lambda ^{0\prime }e\widetilde{F}/T=F^{0\prime }%
\widetilde{F}/T+\mathcal{R}_{1},
\end{equation*}%
where $\mathcal{R}_{1}\equiv \left( \Lambda ^{0^{\prime }}\Lambda
^{0}\right) ^{-1}\Lambda ^{0\prime }e\widetilde{F}/T=\left( \Lambda
^{0\prime }\Lambda ^{0}A^{-1}\right) ^{-1}A^{-1}\Lambda ^{0\prime }e%
\widetilde{F}/T.$ We further note that $A^{-1}\Lambda ^{0\prime }e\widetilde{%
F}/T= O_{p} \left( N^{-\alpha_{r}} T^{-1} \Lambda ^{0\prime }e\widetilde{F}
\right) =O_{p}\left( \gamma _{NT} \right)$ by Lemma \ref{lm:LeF}. The
results above imply that $T^{-1}\widetilde{F}^{\prime }F^{0}=\widetilde{%
\Lambda }^{\prime }\Lambda ^{0}\left( \Lambda ^{0\prime }\Lambda ^{0}\right)
^{-1}+O_{p}\left( \gamma _{NT}\right) .$ Again, recall the definitions that $%
Q=T^{-1}\widetilde{F}^{\prime }F^{0},$ and $H_{1}=\left( \Lambda ^{0\prime
}\Lambda ^{0}\right) ^{-1}\Lambda ^{0\prime }\widetilde{\Lambda }$, together
with $Q=H^{-1}+O_{p}\left( \gamma _{NT}\right) $ proved previously, we come
to that $H_{1}=H+O_{p}\left( \gamma _{NT}\right) .$

Fourth, notice immediately that $\Lambda ^{0\prime }\widetilde{\Lambda}%
/N=\left( \Lambda ^{0\prime }\Lambda /N\right) \left( F^{0\prime }\widetilde{%
F}/T\right) +\frac{1}{NT}\Lambda ^{0\prime }e\widetilde{F}$, and we
post-multiply $(\widetilde{\Lambda}^{\prime }\widetilde{\Lambda}/N)^{-1}$ on
both sides of the equation, making use of $\widetilde{\Lambda}^{\prime }%
\widetilde{\Lambda}/N=\widetilde{V}$, to get that%
\begin{eqnarray*}
H_{4} &\equiv &\left( \Lambda ^{0\prime }\widetilde{\Lambda}\right) \left( 
\widetilde{\Lambda}^{\prime }\widetilde{\Lambda}\right) ^{-1} \\
&=&H+\frac{1}{NT}\Lambda ^{0\prime }e\widetilde{F}\left( \frac{1}{N}%
\widetilde{\Lambda}^{\prime }\widetilde{\Lambda}\right) ^{-1} \\
&=& H + O_{p} \left( A^{-1} \Lambda ^{0\prime }e\widetilde{F} / T\right) \\
&=& H+O_{p}\left( \gamma _{NT}\right),
\end{eqnarray*}
where the last equality is by Lemma \ref{lm:LeF}.

Lastly, to deal with $H_{2}$, we again start by definition 
\begin{equation*}
\widetilde{F}=\frac{1}{N}X^{\prime }\widetilde{\Lambda }\widetilde{V}^{-1}=%
\frac{1}{N}F^{0}\Lambda ^{0\prime }\widetilde{\Lambda }\widetilde{V}^{-1}+%
\frac{1}{N}e^{\prime }\widetilde{\Lambda }\widetilde{V}^{-1}.
\end{equation*}%
Now, if we pre-multiply $\frac{1}{T}F^{0\prime }$ on both sides, we get%
\begin{eqnarray*}
\frac{1}{T}F^{0\prime }\widetilde{F} &=&\frac{1}{T}F^{0\prime }F^{0}\left( 
\frac{1}{N}\Lambda ^{0\prime }\widetilde{\Lambda }\right) \widetilde{V}^{-1}+%
\frac{1}{NT}F^{0\prime }e^{\prime }\widetilde{\Lambda }\widetilde{V}^{-1} \\
&=&\left( \frac{1}{T}F^{0\prime }F^{0}\right) \left( \frac{1}{N}\Lambda
^{0\prime }\widetilde{\Lambda }\right) \widetilde{V}^{-1}+\mathcal{R}_{2},
\end{eqnarray*}%
where 
\begin{eqnarray*}
\mathcal{R}_{2} &\equiv &\frac{1}{NT}F^{0\prime }e^{\prime }\left( 
\widetilde{\Lambda }-\Lambda ^{0}Q^{\prime }\right) \widetilde{V}^{-1}+\frac{%
1}{NT}F^{0\prime }e^{\prime }\Lambda ^{0}Q^{\prime }\widetilde{V}^{-1} \\
&=&\frac{1}{NT}F^{0\prime }e^{\prime }\left( \widetilde{\Lambda }-\Lambda
^{0}Q^{\prime }\right) \widetilde{V}^{-1}+\frac{1}{T}F^{0\prime }e^{\prime
}\Lambda ^{0}A^{-1}\left( AN^{-1}Q^{\prime }\widetilde{V}^{-1}\right) \\
&=&O_{p}\left( N^{-\frac{1}{2}}T^{-\frac{1}{2}}\right) O_{p}\left( N^{-\frac{%
\alpha _{r}}{2}}+N^{\frac{1}{2}}T^{-\frac{1}{2}}\right) O_{p}\left(
N^{1-\alpha _{r}}\right) +O_{p}\left( N^{-\frac{\alpha _{r}}{2}}T^{-\frac{1}{%
2}}\right) \\
&=&O_{p}\left( N^{\frac{1}{2}-\frac{3}{2}\alpha _{r}}T^{-\frac{1}{2}%
}+N^{1-\alpha _{r}}T^{-1}\right) +O_{p}\left( N^{-\frac{\alpha _{r}}{2}}T^{-%
\frac{1}{2}}\right) =O_{p}\left( \gamma _{NT}\right) ,
\end{eqnarray*}%
and the third equality is by Theorem \ref{prop: loading_consistent}\textbf{, 
}Lemma \ref{lm:Xi} and Assumption 3 (v). It follows that 
\begin{equation*}
\underbrace{\left( F^{0\prime }F^{0}\right) ^{-1}F^{0\prime }\widetilde{F}}%
_{:=H_{2}}=\underbrace{\left( \Lambda ^{0\prime }\widetilde{\Lambda }\right)
\left( \widetilde{\Lambda }^{\prime }\widetilde{\Lambda }\right) ^{-1}}%
_{:=H_{4}}+O_{p}\left( \gamma _{NT}\right) .
\end{equation*}%
Putting things together, we have shown (i) $H=H_{3}+O_{P}\left( \gamma
_{NT}\right) ,$ (ii) $H_{1}=H+O_{p}\left( \gamma _{NT}\right) ,$ (iii) $%
H_{4}=H+O_{p}\left( \gamma _{NT}\right) ,$ (iv) $H_{2}=H_{4}+O_{p}\left(
\gamma _{NT}\right) .$ In all, we conclude that $H_{l}=H+O_{p}\left( \gamma
_{NT}\right) ,$ $l=1,\cdots ,4.$\qed\bigskip

\noindent \textbf{Proof of Proposition \ref{prop:cc_consistent}. }Now we have%
\begin{eqnarray*}
\frac{1}{NT}\left\Vert \widetilde{C}-C^{0}\right\Vert ^{2} &=&\frac{1}{NT}%
\left\Vert \widetilde{\Lambda }\widetilde{F}^{\prime }-\Lambda ^{0}H^{\prime
-1}H^{\prime }F^{0\prime }\right\Vert ^{2} \\
&=&\frac{1}{NT}\left\Vert \left( \widetilde{\Lambda }-\Lambda ^{0}H^{\prime
-1}+\Lambda ^{0}H^{\prime -1}\right) \widetilde{F}^{\prime }-\Lambda
^{0}H^{\prime -1}H^{\prime }F^{0\prime }\right\Vert ^{2} \\
&\leq &\frac{2}{NT}\left\Vert \Lambda ^{0}H^{\prime -1}\left( \widetilde{F}%
-F^{0}H\right) ^{\prime }\right\Vert ^{2}+\frac{2}{NT}\left\Vert \widetilde{%
\Lambda }-\Lambda ^{0}H^{\prime -1}\right\Vert ^{2}\Vert \widetilde{F}\Vert
^{2} \\
&=&O_{p}\left( \frac{1}{N}\left\Vert \Lambda ^{0}\right\Vert ^{2}\right)
O_{p}\left( \frac{1}{T}\left\Vert \widetilde{F}-F^{0}H\right\Vert ^{2}\right)
\\
&&+O_{p}\left( \frac{1}{N}\left\Vert \widetilde{\Lambda }-\Lambda
^{0}Q^{^{\prime }}\right\Vert ^{2}\right) +O_{p}\left( \left\Vert
Q-H^{-1}\right\Vert ^{2}\right) \\
&=&O_{p}\left( N^{\alpha _{1}-1}\right) O_{p}\left( N^{\alpha _{1}-2\alpha
_{r}}+N^{2\left( 1-\alpha _{r}\right) }T^{-2}\right) \\
&&+O_{p}\left( N^{-\alpha _{r}-1}+T^{-1}\right) +O_{p}\left( N^{\alpha
_{1}-2\alpha _{r}}T^{-1}+N^{2\left( 1-\alpha _{r}\right) }T^{-2}+N^{-2\alpha
_{r}}\right) \\
&=&O_{p}\left( N^{2\left( \alpha _{1}-\alpha _{r}\right) -1}+T^{-1}\right) .
\end{eqnarray*}%
The second last equality is due to Propositions \ref{prop: factor_consistent}%
\textbf{, }\ref{prop: loading_consistent}\textbf{, }and Lemma \ref{lm:
equivalent_H}\textbf{. }The last equality is due to Assumption 5.\textbf{\ }%
\qed\bigskip

\noindent \textbf{Proof of Theorem \ref{thm: Ft_dist}. }Let us start from
the definition, $\widetilde{F}=\frac{1}{N}X^{\prime }\widetilde{\Lambda}%
\widetilde{V}^{-1}.$ Then by plugging $X=\Lambda ^{0}F^{0\prime }+e,$ we get 
$\widetilde{F}=\frac{1}{N}F^{0}\Lambda ^{0\prime }\widetilde{\Lambda}%
\widetilde{V}^{-1}+\frac{1}{N}e^{\prime }\widetilde{\Lambda}\widetilde{V}%
^{-1}.$ That is, $\widetilde{F}_{t}=H_{4}^{\prime }F_{t}^{0}+\widetilde{V}%
^{-1}\frac{1}{N}\widetilde{\Lambda}^{\prime }e_{t}$. So%
\begin{equation*}
\widetilde{F}_{t}-H_{4}^{\prime }F_{t}^{0}=\widetilde{V}^{-1}\frac{1}{N}%
\left( \widetilde{\Lambda}-\Lambda ^{0}Q\right) ^{\prime }e_{t}+\widetilde{V}%
^{-1}\frac{1}{N}Q\Lambda ^{0\prime }e_{t}.
\end{equation*}

We first consider the first term on the RHS of the above equation.%
\begin{equation*}
\widetilde{V}^{-1}\frac{1}{N}\left( \widetilde{\Lambda}-\Lambda ^{0}Q\right)
^{\prime }e_{t}=\widetilde{V}^{-1}\frac{1}{N}\left[ \frac{1}{T}e\left( 
\widetilde{F}-F^{0}H\right) \right] ^{\prime }e_{t}+\widetilde{V}^{-1}\frac{1%
}{N}\left( \frac{1}{T}eF^{0}H\right) ^{\prime }e_{t}\equiv W_{t}+Z_{t},\text{
say.}
\end{equation*}

For $W_{t},$ we have a further decomposition as%
\begin{eqnarray*}
W_{t} &=&\widetilde{V}^{-1}\frac{1}{N}\left( \frac{eF^{0}\Lambda ^{0\prime }e%
\widetilde{F}\widetilde{V}^{-1}}{NT^{2}}\right) ^{\prime }e_{t}+\widetilde{V}%
^{-1}\frac{1}{N}\left( \frac{ee^{\prime }\Lambda ^{0}F^{0\prime }\widetilde{F%
}\widetilde{V}^{-1}}{NT^{2}}\right) ^{\prime }e_{t}+\widetilde{V}^{-1}\frac{1%
}{N}\left( \frac{ee^{\prime }e\widetilde{F}\widetilde{V}^{-1}}{NT^{2}}%
\right) ^{\prime }e_{t} \\
&\equiv &W_{t}^{(a)}+W_{t}^{(b)}+W_{t}^{(c)},\text{ say.}
\end{eqnarray*}%
For $W_{t}^{(a)}=\frac{1}{N^{2}T^{2}}\widetilde{V}^{-2}\widetilde{F}^{\prime
}e^{\prime }\Lambda ^{0}F^{0\prime }e^{\prime }e_{t},$ we first have $\frac{1%
}{NT}F^{0\prime }e^{\prime }e_{t}=O_{p}\left( \left( NT\right)
^{-1/2}+T^{-1}\right) $ by Assumption 3 (vi). Also recall that $\left\Vert
e^{\prime }\Lambda ^{0}\right\Vert =O_{p}\left( N^{\alpha
_{1}/2}T^{1/2}\right) .$ Hence,%
\begin{eqnarray*}
W_{t}^{(a)} &=&\frac{1}{NT}\widetilde{V}^{-2}O_{p}\left( T^{\frac{1}{2}%
}\right) O_{p}\left( N^{\frac{\alpha _{1}}{2}}T^{\frac{1}{2}}\right)
O_{p}\left( \frac{1}{\sqrt{NT}}+\frac{1}{T}\right) \\
&=&\widetilde{V}^{-2}O_{p}\left( N^{\frac{\alpha _{1}-3}{2}}T^{-\frac{1}{2}%
}+N^{\frac{\alpha _{1}}{2}-1}T^{-1}\right) .
\end{eqnarray*}%
For $W_{t}^{(b)}$, by Lemma \ref{lm:eelffv},%
\begin{equation*}
W_{t}^{(b)}=\widetilde{V}^{-1}\frac{1}{NT}A^{-\frac{1}{2}}O_{p}(N+T)O_{p}(%
\sqrt{N})=A^{-\frac{1}{2}}\widetilde{V}^{-1}O_{p}\left( N^{-\frac{1}{2}}+N^{%
\frac{1}{2}}T^{-1}\right) .
\end{equation*}%
For $W_{t}^{(c)}$,%
\begin{eqnarray*}
W_{t}^{(c)} &=&\widetilde{V}^{-2}\frac{1}{N^{2}T^{2}}\left\Vert \widetilde{F}%
\right\Vert \left\Vert e\right\Vert _{sp}^{3}\left\Vert e_{t}\right\Vert =%
\widetilde{V}^{-2}\frac{1}{N^{2}T^{2}}O_{p}\left( T^{\frac{1}{2}}\right)
O_{p}\left( N^{\frac{3}{2}}+T^{\frac{3}{2}}\right) O_{p}\left( N^{\frac{1}{2}%
}\right) \\
&=&\widetilde{V}^{-2}O_{p}\left( N^{-\frac{3}{2}}+T^{-\frac{3}{2}}\right) .
\end{eqnarray*}

Turning to $Z_{t},$ given that $\left( NT\right) ^{-1}F^{0\prime }e^{\prime
}e_{t}=O_{p}\left( \left( NT\right) ^{-1/2}+T^{-1}\right) $ and $%
H=O_{p}\left( 1\right) ,$ it follows that $Z_{t}=\widetilde{V}%
^{-1}O_{p}\left( \left( NT\right) ^{-1/2}+T^{-1}\right) .$ Therefore, 
\begin{eqnarray*}
A^{\frac{1}{2}}Z_{t} &=&O_{p}\left( N^{1-\frac{\alpha _{r}}{2}}\right)
O_{p}\left( \frac{1}{\sqrt{NT}}+\frac{1}{T}\right) \\
&=&O_{p}\left( N^{\frac{1-\alpha _{r}}{2}}T^{-\frac{1}{2}}\right)
+O_{p}\left( N^{1-\frac{\alpha _{r}}{2}}T^{-1}\right) =o_{p}\left( 1\right) ,
\end{eqnarray*}%
as $N^{1-\frac{\alpha _{r}}{2}}T^{-1}\rightarrow 0$ by the condition of $%
N^{1-\alpha _{r}}T^{-1}\rightarrow 0$ implied by Assumption 5, and $N^{\frac{%
3}{2}-\alpha _{r}}T^{-1}\rightarrow 0$ implied by Assumption 7.

We next show that $A^{\frac{1}{2}}W_{t}=o_{p}\left( 1\right) $ by showing $%
A^{\frac{1}{2}}W_{t}^{(k)}=o_{p}\left( 1\right) ,$ for $k=a,b,c.$ First,%
\begin{align*}
A^{\frac{1}{2}}W_{t}^{(a)}& =O_{P}\left( A\widetilde{V}^{-2}\right)
O_{P}\left( N^{\frac{\alpha _{1}-3}{2}}T^{-\frac{1}{2}}+N^{\frac{\alpha _{1}%
}{2}-1}T^{-1}\right) \\
& =O_{p}\left( N^{2-\frac{3}{2}\alpha _{r}}\right) O_{p}\left( N^{\frac{%
\alpha _{1}-3}{2}}T^{-\frac{1}{2}}+N^{\frac{\alpha _{1}}{2}-1}T^{-1}\right)
\\
& =O_{p}\left( N^{\frac{1}{2}+\frac{1}{2}\alpha _{1}-\frac{3}{2}\alpha
_{r}}T^{-\frac{1}{2}}\right) +O_{p}\left( N^{1+\frac{1}{2}\alpha _{1}-\frac{3%
}{2}\alpha _{r}}T^{-1}\right) \\
& =O_{p}\left( N^{\frac{1-\alpha _{r}}{2}}T^{-\frac{1}{2}}N^{\frac{1}{2}%
\alpha _{1}-\alpha _{r}}\right) +O_{p}\left( N^{\frac{1}{2}\alpha
_{1}-\alpha _{r}}N^{1-\frac{1}{2}\alpha _{r}}T^{-1}\right) \\
& =o_{p}(1),
\end{align*}%
where we have used $N^{1-\alpha _{r}}T^{-1}\rightarrow 0$ and $N^{1-\frac{%
\alpha _{r}}{2}}T^{-1}\rightarrow 0$. Second, 
\begin{equation*}
A^{\frac{1}{2}}W_{t}^{(b)}=O_{p}\left( \widetilde{V}^{-1}\right) O_{p}\left(
N^{-\frac{1}{2}}+N^{\frac{1}{2}}T^{-1}\right) =O_{p}\left( N^{\frac{1}{2}%
-\alpha _{r}}\right) +O_{p}\left( N^{\frac{3}{2}-\alpha _{r}}T^{-1}\right)
=o_{p}\left( 1\right) \text{,}
\end{equation*}%
by Assumption 7. Third, 
\begin{align*}
A^{\frac{1}{2}}W_{t}^{(c)}& =O_{p}\left( A^{\frac{1}{2}}\widetilde{V}%
^{-2}\right) O_{p}\left( T^{-\frac{3}{2}}+N^{-\frac{3}{2}}\right) \\
& =O_{p}\left( N^{2-\frac{3}{2}\alpha _{r}}\right) O_{p}\left( T^{-\frac{3}{2%
}}+N^{-\frac{3}{2}}\right) \\
& =O_{p}\left( N^{2-\frac{3}{2}\alpha _{r}}T^{-\frac{3}{2}}+N^{\frac{1}{2}-%
\frac{3}{2}\alpha _{r}}\right) \\
& =O_{p}\left[ \left( N^{\frac{3}{2}-\alpha _{r}}T^{-1}\right) \left( N^{%
\frac{1}{2}-\frac{1}{2}\alpha _{r}}T^{-\frac{1}{2}}\right) \right]
+O_{p}\left( N^{\frac{1}{2}(1-3\alpha _{r})}\right) =o_{p}(1).
\end{align*}%
So we have come to that%
\begin{align}
A^{\frac{1}{2}}\left( \widetilde{F}_{t}-H_{4}^{\prime }F_{t}^{0}\right) &
=A^{\frac{1}{2}}\left( W_{t}+Z_{t}\right) +A^{\frac{1}{2}}\widetilde{V}^{-1}%
\frac{1}{N}Q\Lambda ^{0\prime }e_{t}  \notag \\
& =A^{\frac{1}{2}}\widetilde{V}^{-1}\frac{1}{N}Q\Lambda ^{0\prime
}e_{t}+o_{p}(1).  \label{eq:onetwo}
\end{align}

As for the term $\psi _{t}\equiv A^{\frac{1}{2}}\widetilde{V}^{-1}Q\frac{1}{N%
}\Lambda ^{0\prime }e_{t}=\underbrace{A^{\frac{1}{2}}\widetilde{V}^{-1}Q%
\frac{1}{N}A^{\frac{1}{2}}}_{:=\Psi }\left( A^{-\frac{1}{2}}\Lambda
^{0\prime }e_{t}\right) ,$\ given Assumption 8, we come to that $\psi _{t}%
\overset{d}{\rightarrow }\Psi \times N\left( 0,\Gamma _{t}\right) $.

Here it may be interesting to investigate the $r\times r$ matrix $\Psi $.
Note that $A^{\frac{1}{2}}\widetilde{V}^{-1}/N\asymp _{p}\func{diag}\left(
N^{-\frac{\alpha _{1}}{2}},\cdots ,N^{-\frac{\alpha _{r}}{2}}\right) .$ Also
recall that $Q$ is a (block) upper triangular matrix (in probability) such
that%
\begin{equation*}
Q(l,k)\asymp _{p}N^{\alpha _{l}-\alpha _{k}}\text{ for }1\leq k\leq l\leq r,
\end{equation*}%
and $Q(l,k)=O_{p}(1),$ for $r\geq k>l\geq 1.$ It then follows that $\Psi
\left( l,k\right) \asymp _{p}N^{-\frac{\alpha _{l}}{2}}Q(l,k)N^{\frac{\alpha
_{k}}{2}}=N^{\frac{\alpha _{k}-\alpha _{l}}{2}}Q(l,k).$ Hence,%
\begin{equation*}
\Psi \left( l,k\right) 
\begin{cases}
\asymp _{p}N^{\frac{\alpha _{k}-\alpha _{l}}{2}}N^{\alpha _{l}-\alpha
_{k}}=N^{\frac{\alpha _{l}-\alpha _{k}}{2}},\text{ for }1\leq k\leq l\leq r,
\\ 
\asymp _{p}1,\text{ for }1\leq k=l\leq r, \\ 
=O_{p}\left( N^{\frac{\alpha _{k}-\alpha _{l}}{2}}\right) ,\text{ for}\
r\geq k>l\geq 1.%
\end{cases}%
\ 
\end{equation*}%
So obviously, $\Psi $ is an asymptotically (block) diagonal matrix with full
rank.

Finally, we conclude that $A^{\frac{1}{2}}\left( \widetilde{F}%
_{t}-H_{4}^{\prime }F_{t}^{0}\right) \overset{d}{\rightarrow }N\left( 0,\Psi
^{0}\Gamma _{t}\Psi ^{0\prime }\right) .$\qed\bigskip

\noindent \textbf{Proof of Theorem \ref{thm: lamdi_dist}. }Recall that $%
\widetilde{\lambda }_{i}-Q\lambda _{i}^{0}=T^{-1}H^{\prime }F^{0\prime
}e_{i}+T^{-1}\left( \widetilde{F}-F^{0}H\right) ^{\prime }e_{i}.$ As for the
second term, 
\begin{equation*}
\frac{1}{T}\left( \widetilde{F}-F^{0}H\right) ^{\prime }e_{i}=\widetilde{V}%
^{-1}\left( \frac{1}{T}\widetilde{F}^{\prime }F^{0}\right) \frac{\Lambda
^{0\prime }ee_{i}}{NT}+\widetilde{V}^{-1}\widetilde{F}^{\prime }e^{\prime }%
\frac{\Lambda ^{0}F^{0\prime }e_{i}}{NT^{2}}+\widetilde{V}^{-1}\widetilde{F}%
^{\prime }e^{\prime }\frac{ee_{i}}{NT^{2}}.
\end{equation*}%
The first term on the RHS is 
\begin{equation*}
\left( \widetilde{V}^{-1}QAN^{-1}\right) \left( A^{-1}\frac{\Lambda
^{0\prime }ee_{i}}{T}\right) =O_{p}\left( 1\right) \left( A^{-1}\frac{%
\Lambda ^{0\prime }ee_{i}}{T}\right) =O_{p}\left( N^{-\alpha
_{r}}+N^{-\alpha _{r}/2}T^{-1/2}\right) ,
\end{equation*}%
where is first equality is by Lemma \ref{lm:Xi} and the second equality is
by Assumption 2 and 3(iii). So we have%
\begin{eqnarray*}
\left\Vert \frac{1}{T}\left( \widetilde{F}-F^{0}H\right) ^{\prime
}e_{i}\right\Vert &=&O_{p}\left( N^{-\alpha _{r}}+N^{-\alpha
_{r}/2}T^{-1/2}\right) +O_{p}\left( N^{{1-\alpha _{r}}}\right) O_{p}\left( 
\sqrt{T}\right) O_{p}\left( N^{\frac{^{a_{1}}}{2}-1}\right) O_{p}\left( 
\frac{1}{T^{\frac{3}{2}}}\right) \\
&&+O_{p}\left( N^{1-\alpha _{r}}\right) O_{p}\left( \sqrt{T}\right)
O_{p}\left( \frac{N+T}{NT}\right) \frac{1}{T}O_{p}\left( \sqrt{T}\right) \\
&=&O_{p}\left( N^{-\alpha _{r}}+N^{-\alpha _{r}/2}T^{-1/2}\right)
+O_{p}\left( N^{\frac{\alpha _{1}}{2}-\alpha _{r}}T^{-1}\right) +O_{p}\left(
N^{-\alpha _{r}}+N^{1-\alpha _{r}}T^{-1}\right) .
\end{eqnarray*}%
Together with Assumption 8, we come to that $\sqrt{T}\left( \widetilde{%
\lambda }_{i}-Q\lambda _{i}^{0}\right) =H^{\prime
}T^{-1/2}\sum_{t=1}^{T}F_{t}^{0}e_{it}+o_{p}\left( 1\right) .$ Given
Assumption 6.(ii) and the proved result that $H^{\prime }=\left( Q^{\prime
}\right) ^{-1}+o_{p}\left( 1\right) ,$ we conclude that $\sqrt{T}\left( 
\widetilde{\lambda }_{i}-Q\lambda _{i}^{0}\right) \overset{d}{\rightarrow }%
N\left( 0,Q^{\prime -1}\Phi _{i}Q^{-1}\right) .$\qed\bigskip

\noindent \textbf{Proof of Theorem \ref{thm: Cit_dist}. }By definition of $%
C_{it}$, we have the following decomposition:%
\begin{eqnarray*}
\widetilde{C}_{it}-C_{it}^{0} &=&\left( \widetilde{F}_{t}-H^{\prime
}F_{t}^{0}\right) ^{\prime }H^{-1}\lambda _{i}^{0}+F_{t}^{0\prime }H\left( 
\widetilde{\lambda}_{i}-H^{-1}\lambda _{i}^{0}\right) +\left( \widetilde{F}%
_{t}-H^{\prime }F_{t}^{0}\right) ^{\prime }\left( \widetilde{\lambda}%
_{i}-H^{-1}\lambda _{i}^{0}\right) \\
&\equiv &a_{it}+b_{it}+\varphi _{it},\text{ say.}
\end{eqnarray*}

For $a_{it},$ define $d_{t}\equiv N^{-1}\left( \widetilde{\Lambda }-\Lambda
^{0}Q^{\prime }\right) ^{\prime }e_{t},$ and then 
\begin{equation*}
a_{it}=\left( \widetilde{V}^{-1}Q\frac{\Lambda ^{0\prime }e_{t}}{N}\right)
^{\prime }H^{-1}\lambda _{i}^{0}+\left( \widetilde{V}^{-1}d_{t}\right)
^{\prime }H^{-1}\lambda _{i}^{0}=\frac{e_{t}^{\prime }\Lambda ^{0}}{N}%
Q^{\prime }\widetilde{V}^{-1}H^{-1}\lambda _{i}^{0}+\left( \widetilde{V}%
^{-1}d_{t}\right) ^{\prime }H^{-1}\lambda _{i}^{0}.
\end{equation*}%
Notice that $Q^{\prime }\widetilde{V}^{-1}H^{-1}=Q^{\prime }\widetilde{V}%
^{-1}\widetilde{V}\left( Q^{\prime }\right) ^{-1}\left( \Lambda ^{0\prime
}\Lambda ^{0}/N\right) ^{-1}=\left( \frac{\Lambda ^{0^{\prime }}\Lambda ^{0}%
}{N}\right) ^{-1}.$ So, 
\begin{eqnarray*}
a_{it} &=&\frac{e_{t}^{\prime }\Lambda ^{0}}{N}\left( \Lambda ^{0\prime
}\Lambda ^{0}/N\right) ^{-1}\lambda _{i}^{0}+\left( \widetilde{V}%
^{-1}d_{t}\right) ^{\prime }H^{-1}\lambda _{i}^{0} \\
&=&\lambda _{i}^{0\prime }A^{-1/2}\left( A^{-1/2}\Lambda ^{0\prime }\Lambda
^{0}A^{-1/2}\right) ^{-1}\left( A^{-\frac{1}{2}}\Lambda ^{0\prime
}e_{t}\right) +\lambda _{i}^{0\prime }H^{\prime -1}\widetilde{V}^{-1}d_{t}.
\end{eqnarray*}%
Consider 
\begin{equation*}
N^{\frac{\alpha _{r}}{2}}a_{it}=\lambda _{i}^{0\prime }\left( N^{\frac{%
\alpha _{r}}{2}}A^{-1/2}\right) \left( A^{-1/2}\Lambda ^{0\prime }\Lambda
^{0}A^{-1/2}\right) ^{-1}\left( A^{-\frac{1}{2}}\Lambda ^{0\prime
}e_{t}\right) +\lambda _{i}^{0\prime }H^{\prime -1}N^{\frac{\alpha _{r}}{2}}%
\widetilde{V}^{-1}d_{t.}
\end{equation*}%
Note that $N^{\frac{\alpha _{r}}{2}}A^{-1/2}\rightarrow S^{\dagger }\equiv %
\diag(\underbrace{0,\ldots ,0}_{r-r_{G}},\underbrace{1\ldots ,1}_{r_{G}}).$
Also note that 
\begin{equation*}
\lambda _{i}^{0\prime }H^{\prime -1}N^{\frac{\alpha _{r}}{2}}\widetilde{V}%
^{-1}d_{t}\leq \left\Vert \lambda _{i}^{0\prime }H^{\prime -1}\right\Vert
N^{1-\frac{\alpha _{r}}{2}}\left\Vert d_{t}\right\Vert =o_{p}\left( 1\right)
,
\end{equation*}%
where the last equality is due to $\widetilde{V}^{-1}d_{t}=W_{t}+Z_{t},$ and 
$A^{1/2}\left( W_{t}+Z_{t}\right) =o_{p}\left( 1\right) \ $which has been
proved as in \eqref{eq:onetwo} when proving \ref{thm: Ft_dist}. So it
follows that $N^{\frac{\alpha _{r}}{2}}a_{it}\overset{d}{\rightarrow }%
N\left( 0,\lambda _{i}^{0\prime }S^{\dagger }\Sigma _{\Lambda }^{\ast
-1}\Gamma _{t}\Sigma _{\Lambda }^{\ast -1}S^{\dagger }\lambda
_{i}^{0}\right) .$

For $b_{it},$ 
\begin{eqnarray*}
b_{it} &=&F_{t}^{0\prime }H\left( \widetilde{\lambda }_{i}-H^{-1}\lambda
_{i}^{0}\right) \\
&=&F_{t}^{0\prime }H\left( \widetilde{\lambda }_{i}-Q\lambda _{i}^{0}\right)
+F_{t}^{0\prime }H\left( Q-H^{-1}\right) \lambda _{i}^{0} \\
&=&F_{t}^{0\prime }H\frac{1}{T}H^{\prime }F^{0\prime }e_{i}+F_{t}^{0\prime }H%
\frac{1}{T}\left( \widetilde{F}-F^{0}H\right) ^{\prime }e_{i}+F_{t}^{0\prime
}H\left( Q-H^{-1}\right) \lambda _{i}^{0} \\
&\equiv &F_{t}^{0\prime }HH^{\prime }\frac{1}{T}F^{0\prime }e_{i}+\nu
_{it}+\zeta _{it},\text{ say.}
\end{eqnarray*}%
Given that we have shown (a) $\sqrt{T}\left\Vert \frac{1}{T}\left( 
\widetilde{F}-F^{0}H\right) ^{\prime }e_{i}\right\Vert =o_{p}\left( 1\right)
,$ (b) $\left\Vert Q-H^{-1}\right\Vert =O_{p}\left( \gamma _{NT}\right)
=o_{p}\left( T^{-\frac{1}{2}}\right) ,$ and (c)%
\begin{eqnarray*}
HH^{^{\prime }} &=&\left( H-H_{3}+H_{3}\right) \left( H-H_{2}+H_{2}\right)
^{\prime } \\
&=&H_{3}H_{2}^{\prime }+H_{3}\left( H-H_{2}\right) ^{\prime }+\left(
H-H_{3}\right) H_{2}^{\prime }+\left( H-H_{3}\right) \left( H-H_{2}\right)
^{\prime } \\
&=&\left( \frac{\widetilde{F}^{\prime }F^{0}}{T}\right) ^{-1}\frac{%
\widetilde{F}^{\prime }F^{0}}{T}\left( \frac{F^{0\prime }F^{0}}{T}\right)
^{-1}+O_{p}\left( \gamma _{NT}\right) \\
&=&\Sigma _{F}^{-1}+O_{p}\left( \gamma _{NT}\right) =\Sigma
_{F}^{-1}+o_{p}\left( T^{-\frac{1}{2}}\right) ,
\end{eqnarray*}%
it then follows that $\sqrt{T}b_{it}=F_{t}^{0\prime }\Sigma _{F}^{-1}\left(
T^{-1/2}F^{0\prime }e_{i}\right) +o_{p}\left( 1\right) \overset{d}{%
\rightarrow }N\left( 0,F_{t}^{0\prime }\Sigma _{F}^{-1}\Phi _{i}\Sigma
_{F}^{-1}F_{t}^{0}\right) .$

The third term of $\varphi _{it}$ is easy to deal with, as it is immediate that $\varphi _{it} = o_{p}(a_{it})$ and $\varphi _{it} = o_{p}(b_{it})$ given Lemma \ref{lm: equivalent_H}, Theorem \ref{thm: Ft_dist} and Theorem \ref{thm: lamdi_dist}. Therefore, $\varphi _{it}$ is relatively negligible for the asymptotic distribution of $\widetilde{C}_{it}$.

Lastly, under the weak dependence assumption over both $i$ and $t,$ $N^{%
\frac{\alpha _{r}}{2}}a_{it}$ and $\sqrt{T}b_{it}$ are asymptotically
independent. Therefore, by a similar argument in proving Theorem 3 of Bai
(2003), we come to that 
\begin{equation*}
\frac{\widetilde{C}_{it}-C_{it}^{0}}{\sqrt{N^{-\alpha
_{r}}V_{it}+T^{-1}U_{it}}}\overset{d}{\rightarrow }N\left( 0,1\right) ,
\end{equation*}%
where $V_{it}=\lambda _{i}^{0\prime }S^{\dagger }\Sigma _{\Lambda }^{\ast
-1}\Gamma _{t}\Sigma _{\Lambda }^{\ast -1}S^{\dagger }\lambda _{i}^{0},$ and 
$U_{it}=F_{t}^{0\prime }\Sigma _{F}^{-1}\Phi _{i}\Sigma _{F}^{-1}F_{t}^{0}.$%
\ \qed\bigskip

\noindent \textbf{Proof of Theorem \ref{thm: Uniform}. (1)} We first prove
the uniform convergence rate of $\widetilde{\lambda }_{i}$. Recall 
\begin{equation*}
\widetilde{\lambda }_{i}-Q\lambda _{i}^{0}=H^{\prime }\frac{1}{T}F^{0\prime
}e_{i}+\frac{1}{T}\left( \widetilde{F}-F^{0}H\right) ^{\prime }e_{i}
\end{equation*}%
As for the second term, 
\begin{eqnarray*}
\frac{1}{T}\left( \widetilde{F}-F^{0}H\right) ^{\prime }e_{i} &=&\widetilde{V%
}^{-1}\left( \frac{1}{T}\widetilde{F}^{\prime }F^{0}\right) \frac{\Lambda
^{0\prime }ee_{i}}{NT}+\widetilde{V}^{-1}\widetilde{F}^{\prime }\frac{%
e^{\prime }\Lambda ^{0}F^{0\prime }e_{i}}{NT^{2}}+\widetilde{V}^{-1}\frac{%
\widetilde{F}^{\prime }e^{\prime }ee_{i}}{NT^{2}} \\
&\equiv &\varkappa _{1i}+\varkappa _{2i}+\varkappa _{3i}.
\end{eqnarray*}%
For $\varkappa _{1i},$%
\begin{eqnarray*}
\varkappa _{1i} &=&\widetilde{V}^{-1}\left( \frac{1}{T}\widetilde{F}^{\prime
}F^{0}\right) \frac{1}{NT}\sum_{j=1}^{N}\sum_{t=1}^{T}\lambda
_{j}^{0}e_{jt}e_{it} \\
&=&\widetilde{V}^{-1}\frac{\widetilde{F}^{\prime }F^{0}}{T}\frac{1}{NT}%
\sum_{j=1}^{N}\sum_{t=1}^{T}\lambda _{j}^{0}\left[ e_{jt}e_{it}-E\left(
e_{jt}e_{it}\right) \right] +\widetilde{V}^{-1}\frac{\widetilde{F}^{\prime
}F^{0}}{T}\frac{1}{NT}\sum_{j=1}^{N}\sum_{t=1}^{T}\lambda _{j}^{0}E\left(
e_{jt}e_{it}\right) \\
&\equiv &\varkappa _{1i,a}+\varkappa _{1i,b},\text{ say.}
\end{eqnarray*}%
Assumption 5 implies that $\left( \ln N\right) ^{\frac{2}{\gamma _{2}}%
-1}=o\left( N^{\alpha _{k}}T\right) $ for $k=1,...,r,$ and thus by Lemma A.3
of Fan et al. (2011), there exists a $C>0$ such that

\begin{equation*}
P\left( \max_{i}\left\vert \frac{1}{N^{\alpha _{k}}T}\sum_{i=1}^{N}%
\sum_{t=1}^{T}\lambda _{jk}^{0}\left[ e_{jt}e_{it}-E\left(
e_{jt}e_{it}\right) \right] \right\vert >C\sqrt{\frac{\ln N}{N^{\alpha _{k}}T%
}}\right) =O\left( \frac{1}{N^{2}}\right) .
\end{equation*}%
Now let us define $\widetilde{A}=\diag\left(
N^{1-a_{1}},...,N^{1-a_{r}}\right) ,$ and then 
\begin{equation*}
\varkappa _{1i,a}=\widetilde{V}^{-1}Q\widetilde{A}^{-1}A^{-1}\frac{1}{T}%
\sum_{j=1}^{N}\sum_{t=1}^{T}\lambda _{j}^{0}\left[ e_{jt}e_{it}-E\left(
e_{jt}e_{it}\right) \right] .
\end{equation*}%
For $\Xi =\widetilde{V}^{-1}Q\widetilde{A}^{-1},$ $\Xi $ is a full rank
(block) lower triangular matrix by Lemma \ref{lm:Xi}. Then it follows that
for the $r\times 1$ vector $\varkappa _{1i,a},$ $\max_{i}\left\vert
\varkappa _{1i,a}\left( k\right) \right\vert =O_{p}\left( \sqrt{\frac{\ln N}{%
N^{\alpha _{k}}T}}\right) $ for $k=1,...,r.$ Similarly, we have 
\begin{equation*}
\varkappa _{1i,b}=\widetilde{V}^{-1}Q\widetilde{A}^{-1}A^{-1}\frac{1}{T}%
\sum_{j=1}^{N}\sum_{t=1}^{T}\lambda _{j}^{0}\tau _{ij,t},
\end{equation*}%
and $\max_{i}\left\vert \varkappa _{1i,b}\left( k\right) \right\vert
=O_{p}\left( N^{-\alpha _{k}}\right) $ for $k=1,...,r,$ by Assumption 3
(iii). Hence, 
\begin{equation*}
\max_{i}\left\vert \varkappa _{1i}\left( k\right) \right\vert =O_{p}\left( 
\sqrt{\frac{\ln N}{N^{\alpha _{k}}T}}+N^{-\alpha _{k}}\right) .
\end{equation*}%
Next for $\varkappa _{2i},$%
\begin{equation*}
\left\Vert \varkappa _{2i}\right\Vert =O_{p}\left( N^{1-\alpha _{r}}\right)
O_{p}(\sqrt{T})O_{p}\left( N^{\frac{\alpha _{1}}{2}-1}\right) \frac{1}{T}%
\left\Vert \frac{1}{T}\sum_{t=1}^{T}F_{t}^{0}e_{it}\right\Vert .
\end{equation*}%
Given Assumption 9, there exists a $C_{1}>0$ such that\ 
\begin{equation*}
P\left( \max_{i}\left\Vert \frac{1}{T}\sum_{t=1}^{T}F_{t}^{0}e_{it}\right%
\Vert >C_{1}\sqrt{\frac{\ln N}{T}}\right) =O\left( \frac{1}{N^{2}}\right) ,
\end{equation*}%
by Lemma B.1 of Fan et al. (2011). Hence, 
\begin{equation*}
\max_{i}||\varkappa _{2i}\Vert =O_{p}\left( N^{\frac{\alpha _{1}}{2}-\alpha
_{r}}T^{-1}\sqrt{\frac{\ln N}{T}}\right) .
\end{equation*}%
Third, for $\varkappa _{3i},$%
\begin{align*}
\left\Vert \varkappa _{3i}\right\Vert & =O_{p}\left( N^{1-\alpha
_{r}}\right) O_{p}(\sqrt{T})O_{P}\left( \frac{N+T}{NT}\right) \times \frac{1%
}{T^{1/2}}\left( \frac{1}{T}\sum_{t=1}^{T}e_{it}^{2}\right) ^{\frac{1}{2}} \\
& =O_{P}\left( N^{1-\alpha _{r}}T^{-1}+N^{-\alpha _{r}}\right) \times \left[ 
\frac{1}{T}\sum_{t=1}^{T}\left[ e_{it}^{2}-E\left( e_{it}^{2}\right) \right]
+E\left( e_{it}^{2}\right) \right] ^{\frac{1}{2}} \\
& =O_{P}\left( N^{1-\alpha _{r}}T^{-1}+N^{-\alpha _{r}}\right) \left[ \frac{1%
}{T}\sum_{t=1}^{T}E\left( e_{it}^{2}\right) \right] ^{\frac{1}{2}} \\
& +O_{P}\left( N^{1-\alpha _{r}}T^{-1}+N^{-\alpha _{r}}\right) \times \left[ 
\frac{1}{T}\sum_{t=1}^{T}\left[ e_{it}^{2}-E\left( e_{it}^{2}\right) \right] %
\right] ^{\frac{1}{2}}.
\end{align*}%
Now given that $\max_{i}E\left( e_{it}^{2}\right) <\infty $ and 
\begin{equation}
P\left( \max_{i}\left\vert \frac{1}{T}\sum_{t=1}^{T}\left[
e_{it}^{2}-E\left( e_{it}^{2}\right) \right] \right\vert >\sqrt{\frac{\ln N}{%
T}}\right) =O\left( \frac{1}{N^{2}}\right) ,  \label{eq:exp_tails}
\end{equation}%
by Lemma A.3 of Fan et al. (2011), we come to%
\begin{eqnarray*}
\max_{i}||\varkappa _{3i}\Vert &=&O_{p}\left( N^{1-\alpha
_{r}}T^{-1}+N^{-\alpha _{r}}\right) +O_{p}\left( N^{1-\alpha
_{r}}T^{-1}+N^{-\alpha _{r}}\right) O_{p}\left( T^{-\frac{1}{4}}\left( \ln
N\right) ^{\frac{1}{4}}\right) \\
&=&O_{p}\left( N^{1-\alpha _{r}}T^{-1}+N^{-\alpha _{r}}\right) .
\end{eqnarray*}

Putting things together, and by the previously proved result $%
\max_{i}\left\Vert F^{0\prime }e_{i}/T\right\Vert =O_{p}\left( \sqrt{%
T^{-1}\ln N}\right) ,$ it follows that, for $\varkappa _{i}:=\widetilde{%
\lambda}_{i}-Q\lambda _{i}^{0},$%
\begin{equation*}
\max_{i}||\varkappa _{i}\Vert =O_{p}\left( \sqrt{\frac{\ln N}{T}}\right) ,
\end{equation*}%
given $\sqrt{T}N^{-\alpha _{k}}\rightarrow 0$ implied by Assumption 8.

\textbf{(2)} We next prove the uniform convergence rate of $\widetilde{F}%
_{t} $. From the linear expansion of $\widetilde{F}_{t},$

\begin{equation*}
\widetilde{F}_{t}-H_{4}^{\prime }F_{t}^{0}=\widetilde{V}^{-1}\frac{1}{N}%
\left( \widetilde{\Lambda}-\Lambda ^{0}Q^{\prime }\right) e_{t}+\widetilde{V}%
^{-1}\frac{1}{N}Q\Lambda ^{0}e_{t}.
\end{equation*}%
Recall that in proving Theorem \ref{thm: Ft_dist}, we have defined $W_{t}$
and $Z_{t}$ such that, 
\begin{eqnarray*}
\widetilde{V}^{-1}\frac{1}{N}\left( \widetilde{\Lambda }-\Lambda
^{0}Q\right) ^{\prime }e_{t} &=&\widetilde{V}^{-1}\frac{1}{N}\left[ \frac{1}{%
T}e\left( \widetilde{F}-F^{0}H\right) \right] ^{\prime }e_{t}+\widetilde{V}%
^{-1}\frac{1}{N}\left( \frac{1}{T}eF^{0}H\right) ^{\prime }e_{t} \\
&\equiv &W_{t}+Z_{t}\text{.}
\end{eqnarray*}%
Also recall that 
\begin{eqnarray*}
W_{t} &=&\widetilde{V}^{-1}\frac{1}{N}\left[ \frac{eF^{0}\Lambda ^{0\prime }e%
\widetilde{F}\widetilde{V}^{-1}}{NT^{2}}\right] ^{\prime }e_{t}+\widetilde{V}%
^{-1}\frac{1}{N}\left[ \frac{ee^{\prime }\Lambda ^{0}F^{0\prime }\widetilde{F%
}\widetilde{V}^{-1}}{NT^{2}}\right] e_{t}+\widetilde{V}^{-1}\frac{1}{N}\left[
\frac{ee^{\prime }e\widetilde{F}\widetilde{V}^{-1}}{NT^{2}}\right] ^{\prime
}e_{t} \\
&\equiv &W_{t}^{(a)}+W_{t}^{(b)}+W_{t}^{(c)}.
\end{eqnarray*}

Our purpose is to show that $\max_{t}A^{1/2}W_{t}=o_{p}\left( \sqrt{\ln T}%
\right) $ and $\max_{t}A^{1/2}Z_{t}=o_{p}\left( \sqrt{\ln T}\right) $ in the
following. To this end, we first study 
\begin{equation*}
\frac{1}{NT}F^{0\prime }e^{\prime }e_{t}=\frac{1}{NT}\sum_{s=1}^{T}%
\sum_{i=1}^{N}\left[ F_{s}^{0}e_{is}e_{it}-E\left(
F_{s}^{0}e_{is}e_{it}\right) \right] +\frac{1}{NT}\sum_{s=1}^{T}%
\sum_{i=1}^{N}E\left( F_{s}^{0}e_{is}e_{it}\right) \equiv d_{t,2a}+d_{t,2b}.
\end{equation*}%
As for the term $F_{s}^{0}e_{is}e_{it},$ by Lemma A.2 of Fan et al. (2011),
it satisfies the exponential tail condition $\frac{2s_{1}s_{3}}{3s_{1}+9s_{3}%
}$ given our Assumption 9, as well as the strong mixing condition with
parameter $s_{2}$. Hence we can apply Theorem 1 of Merlevede et al. (2011)
to show 
\begin{equation*}
\max_{t}\left\Vert d_{t,2a}\right\Vert =O_{p}\left( \sqrt{\frac{\ln T}{NT}}%
\right) .
\end{equation*}%
Meanwhile, $\max_{t}\left\Vert d_{t,2b}\right\Vert =O_{p}\left(
T^{-1}\right) $ by Assumption 3 (vi). \ Therefore $\max_{t}\left\Vert
F^{0\prime }e^{\prime }e_{t}/(NT)\right\Vert =O_{p}\left( \sqrt{\ln T/\left(
NT\right) }+T^{-1}\right) .$

Meanwhile, given that $\max_{i}E\left( e_{it}^{2}\right) <\infty $ and %
\eqref{eq:exp_tails}, we come to that $N^{-1/2}\max_{t}\left\Vert
e_{t}\right\Vert =O_{p}\left( \sqrt{\ln T}\right) .$

Then following the same argument in proving Theorem \ref{thm: Ft_dist},\
just by replacing $\left\Vert F^{0\prime }e^{\prime }e_{t}/(NT)\right\Vert
=O_{p}\left( \left( NT\right) ^{-1/2}+T^{-1}\right) $ and $%
N^{-1/2}\left\Vert e_{t}\right\Vert =O_{p}\left( 1\right) $ with\ $%
\max_{t}\left\Vert F^{0\prime }e^{\prime }e_{t}/(NT)\right\Vert =O_{p}\left( 
\sqrt{\ln T/\left( NT\right) }+T^{-1}\right) $ and $N^{-1/2}\max_{t}\left%
\Vert e_{t}\right\Vert =O_{p}\left( \sqrt{\ln T}\right) $, respectively,\
we\ can verify\ that $\max_{t}A^{1/2}W_{t}=o_{p}\left( \sqrt{\ln T}\right) $
and $\max_{t}A^{1/2}Z_{t}=o_{p}\left( \sqrt{\ln T}\right) .$ This implies
that for the $r\times 1$ vector 
\begin{equation*}
\vartheta _{t}\equiv \widetilde{V}^{-1}\frac{1}{N}\left( \widetilde{\Lambda}%
-\Lambda ^{0}Q^{\prime }\right) e_{t},
\end{equation*}%
we have $\max_{t}\left\vert \vartheta _{t}\left( k\right) \right\vert
=O_{p}\left( N^{-\frac{\alpha _{k}}{2}}\sqrt{\ln T}\right) .$

Lastly for the $r\times 1$ vector $\pi _{t}\equiv \widetilde{V}^{-1}\frac{1}{%
N}Q\Lambda ^{0}e_{t},$ we have 
\begin{equation*}
A^{\frac{1}{2}}\pi _{t}=A^{\frac{1}{2}}\widetilde{V}^{-1}Q\frac{1}{N}A^{%
\frac{1}{2}}\left( A^{-\frac{1}{2}}\Lambda ^{0}e_{t}\right) =\Psi \left( A^{-%
\frac{1}{2}}\Lambda ^{0\prime }e_{t}\right) .
\end{equation*}%
We have shown that the $r\times r$ matrix $\Psi $ is asymptotically a
(block) diagonal matrix in proving Theorem \ref{thm: Ft_dist}; meanwhile, we
can prove that%
\begin{equation*}
\max_{t}\left\Vert A^{-\frac{1}{2}}\Lambda ^{0\prime }e_{t}\right\Vert
=O_{p}\left( \sqrt{\ln T}\right)
\end{equation*}%
by using the exponential inequality. So it follows for the $r\times 1$
vector $\pi _{t},$%
\begin{equation*}
\max_{t}\left\vert \pi _{t}\left( k\right) \right\vert =O_{p}\left( A_{kk}^{-%
\frac{1}{2}}\sqrt{\ln T}\right) =O_{p}\left( N^{-\frac{\alpha _{k}}{2}}\sqrt{%
\ln T}\right) .
\end{equation*}%
So it follows that, for $F_{t}^{\ast }=H_{4}^{\prime }F_{t}^{0}$, $%
\max_{t}\left\vert \widetilde{F}_{t,k}-F_{t,k}^{\ast }\right\vert
=O_{P}\left( N^{-\frac{\alpha _{k}}{2}}\sqrt{\ln T}\right) $.

\textbf{(3)} Finally comes the uniform convergence result of $\widetilde{C}%
_{it}$. This is easy because 
\begin{align*}
\widetilde{C}_{it}-C_{it}^{0}& =\widetilde{\lambda}_{i}^{\prime }\widetilde{F%
}_{t}-\lambda _{i}^{0\prime }F_{t}^{0}=\widetilde{\lambda}_{i}^{\prime }%
\widetilde{F}_{t}-\left( H^{-1}\lambda _{i}^{0}\right) ^{\prime }H^{\prime
}F_{t}^{0} \\
& =\left( \widetilde{\lambda}_{i}-H^{-1}\lambda _{i}^{0}\right) ^{\prime
}H^{\prime }F_{t}^{0}+\left( H^{-1}\lambda _{i}^{0}\right) ^{\prime }\left( 
\widetilde{F}_{t}-H^{\prime }F_{t}^{0}\right) +\left( \widetilde{\lambda}%
_{i}-H^{\prime }\lambda _{i}^{0}\right) ^{\prime }\left( \widetilde{F}%
_{t}-H^{\prime }F_{t}^{0}\right) \\
& \equiv \Delta _{1,it}+\Delta _{2,it}+\Delta _{3,it},\text{ say.}
\end{align*}

Given that (a) 
\begin{eqnarray*}
\max_{i}\left\Vert \widetilde{\lambda }_{i}-H^{-1}\lambda
_{i}^{0}\right\Vert &\leqq &\max_{i}\left\Vert \widetilde{\lambda }%
_{i}-Q\lambda _{i}^{0}\right\Vert +\left\Vert Q-H^{-1}\right\Vert
\max_{i}\left\Vert \lambda _{i}^{0}\right\Vert \\
&=&O_{p}\left( \sqrt{\frac{\ln N}{T}}\right) +O_{p}\left( \gamma
_{NT}\right) =O_{p}\left( \sqrt{\frac{\ln N}{T}}\right) ,
\end{eqnarray*}%
(b) $\max_{t}\left\Vert H^{\prime }F_{t}^{0}\right\Vert =O_{p}\left( \left(
\ln T\right) ^{s_{3}^{-1}}\right) $ by the exponential inequality for $%
F_{t}^{0},$ and (c) $\max_{t}\left\Vert \widetilde{F}_{t}-H^{\prime
}F_{t}^{0}\right\Vert =O_{p}\left( N^{-\frac{\alpha _{r}}{2}}\sqrt{\ln T}%
\right) ,$ it follows that 
\begin{equation*}
\max_{i,t}\left\vert \widetilde{C}_{it}-C_{it}^{0}\right\vert =O_{p}\left( 
\sqrt{\frac{\ln N}{T}}(\log T)^{\frac{1}{s_{3}}}+N^{-\frac{\alpha _{r}}{2}}%
\sqrt{\ln T}\right) .\qed
\end{equation*}

\section{Proofs of main results in Section \protect\ref{sec:sparsity_reveal}}

\noindent \textbf{Proof of Proposition \ref{prop:symm_diff}. }We\textbf{\ }%
will combine the proofs for (i) and (ii) together. Recall that $\lambda
_{i}^{{\ast }}=Q\lambda _{i}^{0}.$ To begin with, let us define%
\begin{equation*}
\mathcal{L}_{k}^{\ast }\left( c\right) =\{i:\left\vert \lambda _{ik}^{\ast
}\right\vert >c\}.
\end{equation*}%
We first study $|\mathcal{L}_{k}^{\ast }\left( c_{\lambda ,NT}\right)
\triangle \mathcal{L}_{k}^{{0}}|$ in order to formally present the sparsity
for factor $k.$

Let $\zeta =\min_{1\leq k\leq G-1}\left( \alpha _{\lbrack k]}-\alpha
_{\lbrack k+1]}\right) ,$ i.e., the minimum discrepancy between distinct
factor strengths. We consider the following 3 cases. The results stated
during the proof hold implicitly with probability approaching 1.

\textbf{Case (a)}: $1\notin \omega (k)$ and $r\notin \omega (k).$ Then%
\begin{align}
\lambda _{ik}^{{\ast }}& =\sum_{l\leq \min \omega (k)-1}Q(k,l)\lambda
_{il}^{0}+\sum_{l\in \omega (k)}Q(k,l)\lambda _{il}^{0}+\sum_{l\geq \max
\omega (k)+1}Q(k,l)\lambda _{il}^{0}  \label{eq:lambda_star} \\
& =S_{ik}^{\dagger }(a)+S_{ik}^{\dagger }(b)+S_{ik}^{\dagger }(c).
\label{eq:Sikabc}
\end{align}%
Obviously, $S_{ik}^{\dagger }(a)$ represents the effects to $\lambda _{ik}^{{%
\ast }}$\ passed on from relatively strong factors, whereas $S_{ik}^{\dagger
}(c)$ represents the effects from relatively weak factors. Also, $%
S_{ik}^{\dagger }(b)$ represents the effects from factor $k$ itself and
those with the same strength, if any. A useful fact to make use of is that,
by Proposition \ref{prop:Q_upper_trian}\textbf{,}%
\begin{equation}
S_{ik}^{\dagger }(a)=O_{p}\left( \zeta \right) =o_{p}\left( c_{\lambda
,NT}\right) ,\text{ uniformly over }i.  \label{eq:Sika}
\end{equation}

We next consider the size of the following defined two sets 
\begin{eqnarray*}
\mathcal{T}_{k}^{a} &=&\left\{ i:i\in \mathcal{L}_{k}^{{0}}\text{, }i\notin 
\mathcal{L}_{k}^{\ast }\left( c_{\lambda ,NT}\right) \right\} , \\
\mathcal{T}_{k}^{b} &=&\left\{ i:i\notin \mathcal{L}_{k}^{{0}},\text{ }i\in 
\mathcal{L}_{k}^{\ast }\left( c_{\lambda ,NT}\right) \right\} ,
\end{eqnarray*}%
by two subcases below.

\begin{itemize}
\item Case (a1): $\alpha _{k}$ is unique among $\{\alpha _{1},\ldots ,\alpha
_{r}\}$. Then we can simplify $\lambda _{ik}^{{\ast }}$ a bit by 
\begin{equation*}
\lambda _{ik}^{{\ast }}=S_{ik}^{\dagger }(a)+Q(k,k)\lambda
_{ik}^{0}+S_{ik}^{\dagger }(c).
\end{equation*}%
First, for $i\in \mathcal{L}_{k}^{0},$ there exists a constant $\underline{%
\lambda }_{k}>0$ such that $\left\vert \lambda _{ik}^{0}\right\vert >%
\underline{\lambda }_{k}$ by Assumption 2. Also, $Q(k,k)\neq 0$ implied by
the full rank and upper triangular matrix of $Q.$ Hence, 
\begin{equation}
\left\vert \{i:i\in \mathcal{L}_{k}^{0},\text{ and }Q(k,k)\lambda
_{ik}^{0}>c_{\lambda ,NT}\}\right\vert =\left\vert \mathcal{L}%
_{k}^{0}\right\vert .  \label{eq:Sikb1}
\end{equation}%
As for $S_{ik}^{\dagger }(c)=\sum_{l\geq \max \omega (k)+1}Q(k,l)\lambda
_{il}^{0},$ note that for $l\geq \max \omega (k)+1,$ we have $\left\Vert
\Lambda _{\cdot l}^{0}\right\Vert _{0}\ll N^{\alpha _{k}}.$ Hence, with
probability approaching 1, 
\begin{equation}
\sum_{i\in \mathcal{L}_{k}^{0}}\mathbf{1}\left[ \left\vert S_{ik}^{\dag
}\left( c\right) \right\vert >c_{\lambda ,NT}\right] \leq \sum_{i\in 
\mathcal{L}_{k}^{0}}\mathbf{1}\left[ \left\vert S_{ik}^{\dag }\left(
c\right) \right\vert >0\right] =o_{p}\left( N^{\alpha _{k}}\right) .
\label{eq:Sikc}
\end{equation}%
Together with \eqref{eq:Sika} and \eqref{eq:Sikc}, we come to that 
\begin{equation*}
\left\vert \mathcal{T}_{k}^{a}\right\vert =o_{p}\left( N^{\alpha
_{k}}\right) .
\end{equation*}

Second, for $i\notin \mathcal{L}_{k}^{0},$ then it follows that $%
S_{ik}^{\dagger }(b)=0. $ This, again with \eqref{eq:Sika} and %
\eqref{eq:Sikb1}, implies that 
\begin{equation*}
\left\vert \mathcal{T}_{k}^{b}\right\vert =o_{p}\left( N^{\alpha
_{k}}\right) .
\end{equation*}

We therefore come to that 
\begin{equation}  \label{eq:sym_diff_a1}
\left\vert\mathcal{L}_{k}^{\ast }\left( c_{\lambda ,NT}\right)\triangle 
\mathcal{L}_{k}^{{0}}\right\vert= o_{p}\left( N^{\alpha _{k}}\right) \text{,
if }\alpha _{k}\text{ is unique};
\end{equation}

\item Case (a2): $\alpha _{k}$ is \textit{not} unique among $\{\alpha
_{1},\ldots ,\alpha _{r}\}$. Then the difference from Case (a1) is only on $%
S_{ik}^{\dagger }(b)$ defined in \eqref{eq:Sikabc}. Specifically, for $i\in 
\mathcal{L}^{0}_{k}$, we must have that, with probability approaching 1, 
\begin{equation}  \label{eq:Sikb2}
\left| \{ i: i\in \mathcal{L}^{0}_{k}, \text{ and } \left| S_{ik}^{\dagger
}(b) \right|> c_{\lambda,NT} \} \right| = \left| \{ i: i\in \mathcal{L}%
^{0}_{k}, \text{ and } \left| S_{ik}^{\dagger }(b) \right|> 0 \} \right|
\asymp \left| \mathcal{L}^{0}_{k} \right|.
\end{equation}
Otherwise, it would violate our definition for $\Lambda^{0}$ being the
sparsest representation with $R^{\ast}=I_{r}$ specified in Definition \ref%
{def:sparse-represenation}. Hence, 
\begin{equation*}
\left\vert \mathcal{T}_{k}^{a}\right\vert =o_{p}\left( N^{\alpha
_{k}}\right) .
\end{equation*}

For $i\notin \mathcal{L}_{k}^{0}$, if it holds that $\frac{\left\vert \cup
_{k^{\prime }\in \omega \left( k\right) }\mathcal{L}_{k^{\prime
}}^{0}\backslash \mathcal{L}_{k}^{0}\right\vert }{N^{\alpha _{k}}}=o(1)$, we
then have $\left(\underset{k^{\prime} \in \omega \left( k\right)}{\cup} 
\mathcal{L}^{0}_{k^{\prime}} \right) \backslash \mathcal{L}^{0}_{k} \ll
N^{\alpha_{k}} = \left| \mathcal{L}^{0}_{k} \right|$, which leads to, with
probability approaching 1, 
\begin{equation}  \label{eq:Sikb3}
\left| \{ i: i\notin \mathcal{L}^{0}_{k}, \text{ and } \left|
S_{ik}^{\dagger }(b) \right|> c_{\lambda,NT} \} \right| = \left| \{ i:
i\notin \mathcal{L}^{0}_{k}, \text{ and } \left| S_{ik}^{\dagger }(b)
\right|> 0 \} \right| \ll \left| \mathcal{L}^{0}_{k} \right|.
\end{equation}
Hence, 
\begin{equation*}
\left\vert \mathcal{T}_{k}^{b}\right\vert =o_{p}\left( N^{\alpha
_{k}}\right) .
\end{equation*}

On the other hand, if it does not holds that $\frac{\left\vert \cup
_{k^{\prime }\in \omega \left( k\right) }\mathcal{L}_{k^{\prime
}}^{0}\backslash \mathcal{L}_{k}^{0}\right\vert }{N^{\alpha _{k}}}=o(1)$,
the result in \eqref{eq:Sikb3} would simply become 
\begin{equation}  \label{eq:Sikb4}
\left| \{ i: i\notin \mathcal{L}^{0}_{k}, \text{ and } \left|
S_{ik}^{\dagger }(b) \right|> c_{\lambda,NT} \} \right| \asymp \left| 
\mathcal{L}^{0}_{k} \right|.
\end{equation}
Hence, 
\begin{equation*}
\left\vert \mathcal{T}_{k}^{b}\right\vert =O_{p}\left( N^{\alpha
_{k}}\right) .
\end{equation*}

Putting things together, by the definition that $\mathcal{L}_{k}^{\ast
}\left( c_{\lambda ,NT}\right)\triangle \mathcal{L}_{k}^{{0}} = \mathcal{T}%
_{k}^{a} \cup \mathcal{T}_{k}^{b} $, we have 
\begin{equation}  \label{eq:sym_diff_a2}
\left\vert\mathcal{L}_{k}^{\ast }\left( c_{\lambda ,NT}\right)\triangle 
\mathcal{L}_{k}^{{0}}\right\vert= 
\begin{cases}
o_{p}\left( N^{\alpha _{k}}\right), \text{ if }\alpha _{k}\text{ is not
unique and } \frac{\left\vert \cup _{k^{\prime }\in \omega \left( k\right) }%
\mathcal{L}_{k^{\prime }}^{0}\backslash \mathcal{L}_{k}^{0}\right\vert }{%
N^{\alpha _{k}}}=o(1); \\ 
O_{p}\left( N^{\alpha _{k}}\right) ,\text{ if }\alpha _{k}\text{ is not
unique and } \frac{\left\vert \cup _{k^{\prime }\in \omega \left( k\right) }%
\mathcal{L}_{k^{\prime }}^{0}\backslash \mathcal{L}_{k}^{0}\right\vert }{%
N^{\alpha _{k}}}\neq o(1).%
\end{cases}%
\end{equation}
\end{itemize}

\textbf{Case (b)}: $1\in \omega (k),$ i.e., factor $k$ is the strongest one.
Then it follows that $S_{ik}^{\dag }\left( a\right) =0,$ and $\lambda _{ik}^{%
{\ast }}=S_{ik}^{\dag }\left( b\right) +S_{ik}^{\dag }\left( c\right) .$ The
same proof from Case (a) shows that \eqref{eq:sym_diff_a2} remains true.

\textbf{Case (c)}: $r\in \omega (k),$ i.e., factor $k$ is the weakest one.
Then it follows that $S_{ik}^{\dag }\left( c\right) =0,$ and $\lambda _{ik}^{%
{\ast }}=S_{ik}^{\dag }\left( a\right) +S_{ik}^{\dag }\left( b\right) .$ The
same proof from Case (a) shows that \eqref{eq:sym_diff_a2} remains true.

It only remains to bound $\left\vert \mathcal{L}_{k}^{\ast }\left(
c_{\lambda ,NT}\right) \triangle \widehat{\mathcal{L}}_{k}\right\vert $. To
this end, let us further define two sets%
\begin{eqnarray*}
\mathcal{W}_{k}^{a} &=&\left\{ i:i\in \mathcal{L}_{k}^{\ast }\left(
c_{\lambda ,NT}\right) \text{, }i\notin \widehat{\mathcal{L}}_{k}\right\} 
\text{ and} \\
\mathcal{W}_{k}^{b} &=&\left\{ i:i\notin \mathcal{L}_{k}^{\ast }\left(
c_{\lambda ,NT}\right) ,\text{ }i\in \widehat{\mathcal{L}}_{k}\right\} .
\end{eqnarray*}%
Also notice that by \eqref{eq:lambda_star}, we have uniformly over $i$ that,
either (i) $\lambda _{ik}^{{\ast }}$ is bounded away from below, or (ii) $%
\left\vert \lambda _{ik}^{{\ast }}\right\vert =O_{p}\left( \zeta \right)
=o_{p}\left( c_{\lambda ,NT}\right) .$

Define $\widetilde{\mathcal{W}}_{k}^{a}=\left\{ i:i\in \mathcal{L}_{k}^{\ast
}\left( 2c_{\lambda ,NT}\right) \text{, }i\notin \widehat{\mathcal{L}}%
_{k}\right\} ,$ and then it follows that $\left\vert \mathcal{W}%
_{k}^{a}\triangle \widetilde{\mathcal{W}}_{k}^{a}\right\vert =\left\vert 
\mathcal{W}_{k}^{a}\backslash \widetilde{\mathcal{W}}_{k}^{a}\right\vert
=o_{p}\left( 1\right) .$ Meanwhile, it also holds that $\ \left\vert 
\widehat{\mathcal{L}}_{k}\triangle \widetilde{\mathcal{W}}%
_{k}^{a}\right\vert =o_{p}\left( 1\right) ,$ by $\sup_{i}\left\vert 
\widetilde{\lambda }_{ik}-\lambda _{ik}^{\ast }\right\vert =O_{p}\left( 
\sqrt{(\ln N)/T}\right) =o_{p}\left( c_{\lambda ,NT}\right) $ in Theorem \ref%
{thm: Uniform}\textbf{\ }(i). So $\left\vert \widehat{\mathcal{L}}%
_{k}\triangle \mathcal{W}_{k}^{a}\right\vert \leq \left\vert \mathcal{W}%
_{k}^{a}\triangle \widetilde{\mathcal{W}}_{k}^{a}\right\vert +\left\vert 
\widehat{\mathcal{L}}_{k}\triangle \widetilde{\mathcal{W}}%
_{k}^{a}\right\vert =o_{p}\left( 1\right) .$

Similarly, define $\widetilde{\mathcal{W}}_{k}^{b}=\left\{ i:i\notin 
\mathcal{L}_{k}^{\ast }\left( 0.5c_{\lambda ,NT}\right) ,\text{ }i\in 
\widehat{\mathcal{L}}_{k}\right\} ,$ and then it follows that $\left\vert 
\mathcal{W}_{k}^{b}\triangle \widetilde{\mathcal{W}}_{k}^{b}\right\vert
=\left\vert \mathcal{W}_{k}^{b}\backslash \widetilde{\mathcal{W}}%
_{k}^{b}\right\vert =o_{p}\left( 1\right) .$ Meanwhile, it also holds that$\
\left\vert \widehat{\mathcal{L}}_{k}\triangle \widetilde{\mathcal{W}}%
_{k}^{b}\right\vert =o_{p}\left( 1\right) ,$ by $\sup_{i}\left\vert 
\widetilde{\lambda }_{ik}-\lambda _{ik}^{\ast }\right\vert =O_{p}\left( 
\sqrt{(\ln N)/T}\right) =o_{p}\left( c_{\lambda ,NT}\right) $ in Theorem \ref%
{thm: Uniform}\textbf{\ }(i). So $\left\vert \widehat{\mathcal{L}}%
_{k}\triangle \mathcal{W}_{k}^{b}\right\vert \leq \left\vert \mathcal{W}%
_{k}^{b}\triangle \widetilde{\mathcal{W}}_{k}^{b}\right\vert +\left\vert 
\widehat{\mathcal{L}}_{k}\triangle \widetilde{\mathcal{W}}%
_{k}^{b}\right\vert =o_{p}\left( 1\right) .$

By the definition that $\mathcal{L}_{k}^{\ast }\left( c_{\lambda ,NT}\right)
\triangle \widehat{\mathcal{L}}_{k}=\mathcal{W}_{k}^{a}\cup \mathcal{W}%
_{k}^{b}$, we have%
\begin{equation}
\left\vert \mathcal{L}_{k}^{\ast }\left( c_{\lambda ,NT}\right) \triangle 
\widehat{\mathcal{L}}_{k}\right\vert =o_{p}\left( 1\right) .
\label{eq:sym_diff2}
\end{equation}%
Finally, given \eqref{eq:sym_diff_a1}, \eqref{eq:sym_diff_a2}, %
\eqref{eq:sym_diff2}, and the fact that $\left\vert \mathcal{L}_{k}^{{0}%
}\triangle \widehat{\mathcal{L}}_{k}\right\vert \leq \left\vert \mathcal{L}%
_{k}^{\ast }\left( c_{\lambda ,NT}\right) \triangle \mathcal{L}_{k}^{{0}%
}\right\vert +\left\vert \mathcal{L}_{k}^{\ast }\left( c_{\lambda
,NT}\right) \triangle \widehat{\mathcal{L}}_{k}\right\vert ,$ the proofs of
(i) and (ii) are complete. \qed%
\bigskip

\noindent \textbf{Proof of Theorem \ref{thm: factor_strength}. }Recall that $%
\widetilde{\lambda }_{i}-Q\lambda _{i}^{0}=H^{\prime }F^{0\prime
}e_{i}/T+T^{-1}\left( \widetilde{F}-F^{0}H\right) ^{\prime }e_{i}.$ Also
recall that we have previously shown that $\max_{i}\left\vert \widetilde{%
\lambda }_{ik}-\lambda _{ik}^{{\ast }}\right\vert =O_{p}\left( \sqrt{(\ln
N)/T}\right) ,$ for $k=1,...,r,$ where $\lambda _{ik}^{{\ast }%
}=\sum_{l=1}^{r}Q\left( k,l\right) \lambda _{il}^{0}.$ To accommodate the
possible case where there are multiple factors with equal strength, recall
what we have defined in proving Proposition \textbf{\ref{prop:symm_diff}}: $%
\zeta =\min_{1\leq k\leq G-1}\left( \alpha _{\lbrack k]}-\alpha _{\lbrack
k+1]}\right) .$ For a given index $k\in \left\{ 1,...,r\right\} ,$ we
consider the following 3 cases.

\textbf{Case (a)}: $1\notin \omega (k)$ and $r\notin \omega (k).$ Then 
\begin{align*}
\lambda _{ik}^{{\ast }}& =\sum_{l\leq \min \omega (k)-1}Q(k,l)\lambda
_{il}^{0}+\sum_{l\in \omega (k)}Q(k,l)\lambda _{il}^{0}+\sum_{l\geq \max
\omega (k)+1}Q(k,l)\lambda _{il}^{0} \\
& \equiv S_{ik}^{\dagger }(a)+S_{ik}^{\dagger }(b)+S_{ik}^{\dagger }(c).
\end{align*}

First, for $S_{ik}^{\dagger }(a),$ by the (block) upper triangular matrix of 
$Q,$%
\begin{equation}
Q\left( k,l\right) =O_{p}\left( N^{\alpha _{k}-\alpha _{l}}\right)
=O_{p}\left( N^{-\zeta }\right) ,\text{ for }l\leq \min \omega (k)-1.
\label{eq:Sika2}
\end{equation}%
So $S_{ik}^{\dagger }(a)=O_{p}\left( N^{-\zeta }\right) $ uniformly over $i.$

Second for $S_{ik}^{\dagger}(c),$ note that relatively to the $k$th factor,
the $l$th factor is weaker, implying that 
\begin{equation}
\sum_{i=1}^{N}\mathbf{1}\left( S_{ik}^{\dag }\left( c\right) \neq 0\right)
\leq \sum_{i=1}^{N}\mathbf{1}\left( \lambda _{il}^{0}\neq 0\right) \asymp
N^{\alpha _{l}}\ll N^{\alpha _{k}},\text{ for }l\geq \max \omega (k)+1.
\label{eq:Sikc2}
\end{equation}

Third, for $S_{ik}^{\dagger }(b),$ we will show a statement used later in
this proof: for any diminishing sequence $\widetilde{c}_{NT}=o(1),$%
\begin{equation}  \label{eq:Sikb22}
\sum_{i=1}^{N}\mathbf{1}\left( S_{ik}^{\dag }\left( b\right) >\widetilde{c}%
_{NT}\right) \asymp _{p}N^{\alpha _{k}}.
\end{equation}%
To prove it, first note that it is obvious that%
\begin{equation*}
\sum_{i=1}^{N}\mathbf{1}\left( S_{ik}^{\dag }\left( b\right) >\widetilde{c}%
_{NT}\right) \leq \sum_{i=1}^{N}\mathbf{1}\left( S_{ik}^{\dag }\left(
b\right) \neq 0\right) =O_{p}\left( N^{\alpha _{k}}\right) ,\text{ for }l\in
\omega (k).
\end{equation*}%
Also, we can exclude that $\sum_{i=1}^{N}\mathbf{1}\left( S_{ik}^{\dag
}\left( b\right) >\widetilde{c}_{NT}\right) =o_{p}\left( N^{\alpha
_{k}}\right) .$ To see why, suppose that $\sum_{i=1}^{N}\mathbf{1}\left(
S_{ik}^{\dag }\left( b\right) >\widetilde{c}_{NT}\right) =o_{p}\left(
N^{\alpha _{k}}\right) ,$ and then by \eqref{eq:Sika2} and \eqref{eq:Sikc2},
it can only be the case that $\left\Vert \Lambda _{k}^{{\ast }}\right\Vert
_{0}=o_{p}\left( N^{\alpha _{k}}\right) ,$ which violates the very
Definition \ref{def:sparse-represenation} of the sparsest $\Lambda^{0}$ with 
$R^{\ast}=I_{r}$. Thus, \eqref{eq:Sikb22} must hold.

Now with $c_{\lambda ,NT}=\frac{1}{\sqrt{\ln \left( NT\right) }},$ we can
write 
\begin{align*}
\sum_{i=1}^{N}\mathbf{1}\left( \left\vert \lambda _{ik}^{{\ast }}\right\vert
>c_{\lambda ,NT}\right) & \leq \sum_{i=1}^{N}\mathbf{1}\left( \left\vert
S_{ik}^{\dag }(a)\right\vert >\frac{1}{3}c_{\lambda ,NT}\right)
+\sum_{i=1}^{N}\mathbf{1}\left( \left\vert S_{ik}^{\dag }(b)\right\vert >%
\frac{1}{3}c_{\lambda ,NT}\right) \\
& +\sum_{i=1}^{N}\mathbf{1}\left( \left\vert S_{ik}^{\dag }(c)\right\vert >%
\frac{1}{3}c_{\lambda ,NT}\right) \\
& =o_{p}(1)+O_{p}\left( N^{\alpha _{k}}\right) +o_{p}\left( N^{\alpha
_{k}}\right) =O_{p}\left( N^{\alpha _{k}}\right) .
\end{align*}%
Meanwhile, we also have 
\begin{eqnarray*}
&&\sum_{i=1}^{N}\mathbf{1}\left[ \left\vert \lambda _{ik}^{{\ast }%
}\right\vert >c_{\lambda ,NT}\right] \\
&\geq &\sum_{i=1}^{N}\mathbf{1}\left[ \left\vert S_{ik}^{\dag }\left(
b\right) \right\vert -\left\vert S_{ik}^{\dag }\left( a\right) +S_{ik}^{\dag
}\left( c\right) \right\vert >c_{\lambda ,NT}\right] \\
&\geq &\sum_{i=1}^{N}\mathbf{1}\left[ \left\vert S_{ik}^{\dag }\left(
b\right) \right\vert >2c_{\lambda ,NT}\text{ and }\left\vert S_{ik}^{\dag
}\left( a\right) +S_{ik}^{\dag }\left( c\right) \right\vert <c_{\lambda ,NT}%
\right] \\
&=&\sum_{i=1}^{N}\left\{ \mathbf{1}\left[ \left\vert S_{ik}^{\dag }\left(
b\right) \right\vert >2c_{\lambda ,NT}\right] -\mathbf{1}\left[ \left\vert
S_{ik}^{\dag }\left( b\right) \right\vert >2c_{\lambda ,NT}\text{ and }%
\left\vert S_{ik}^{\dag }\left( a\right) +S_{ik}^{\dag }\left( c\right)
\right\vert \geq c_{\lambda ,NT}\right] \right\} \\
&\geq &\sum_{i=1}^{N}\mathbf{1}\left[ \left\vert S_{ik}^{\dag }\left(
b\right) \right\vert >2c_{\lambda ,NT}\right] -\sum_{i=1}^{N}\mathbf{1}\left[
\left\vert S_{ik}^{\dag }\left( a\right) +S_{ik}^{\dag }\left( c\right)
\right\vert \geq c_{\lambda ,NT}\right] \\
&\geq &\sum_{i=1}^{N}\mathbf{1}\left[ \left\vert S_{ik}^{\dag }\left(
b\right) \right\vert >2c_{\lambda ,NT}\right] -\sum_{i=1}^{N}\mathbf{1}\left[
\left\vert S_{ik}^{\dag }\left( a\right) \right\vert >\frac{1}{2}c_{\lambda
,NT}\right] -\sum_{i=1}^{N}\mathbf{1}\left[ \left\vert S_{ik}^{\dag }\left(
c\right) \right\vert >\frac{1}{2}c_{\lambda ,NT}\right] \\
& = &\sum_{i=1}^{N}\mathbf{1}\left[ \left\vert S_{ik}^{\dag }\left( b\right)
\right\vert >2c_{\lambda ,NT}\right] -o_{p}(1)-o_{p}\left( N^{\alpha
_{k}}\right) .
\end{eqnarray*}%
Now given \eqref{eq:Sikb22}, it then follows that, 
\begin{equation}
\sum_{i=1}^{N}1\left[ \left\vert \lambda _{ik}^{{\ast }}\right\vert
>c_{\lambda ,NT}\right] \asymp _{p}N^{\alpha _{k}}.
\label{eq:strength_bound}
\end{equation}

\textbf{Case (b)}: $1\in \omega (k).$ Then it follows that $S_{ik}^{\dag
}\left( a\right) =0,$ and $\lambda _{ik}^{{\ast }}=S_{ik}^{\dag }\left(
b\right) +S_{ik}^{\dag }\left( c\right) .$ The same proof from Case (a)
shows that $\sum_{i=1}^{N}1\left[ \left\vert \lambda _{ik}^{{\ast }%
}\right\vert >c_{\lambda ,NT}\right] \asymp _{p}N^{\alpha _{k}}.$

\textbf{Case (c)}: $r\in \omega (k).$ Then it follows that $S_{ik}^{\dag
}\left( c\right) =0,$ and $\lambda _{ik}^{{\ast }}=S_{ik}^{\dag }\left(
a\right) +S_{ik}^{\dag }\left( b\right) .$ The same proof from Case (a)
shows that $\sum_{i=1}^{N}1\left[ \left\vert \lambda _{ik}^{{\ast }%
}\right\vert >c_{\lambda ,NT}\right] \asymp _{p}N^{\alpha _{k}}.$

Finally, recall that 
\begin{equation*}
\widehat{D}_{k}=\sum_{i=1}^{N}1\left( \left\vert \widetilde{\lambda }%
_{ik}\right\vert >c_{\lambda ,NT}\right) =\sum_{i=1}^{N}1\left[ \left\vert 
\widetilde{\lambda }_{ik}-\lambda _{ik}^{{\ast }}+\lambda _{ik}^{{\ast }%
}\right\vert >c_{\lambda ,NT}\right] .
\end{equation*}%
Given that $\max_{i}\left\vert \widetilde{\lambda }_{ik}-\lambda _{ik}^{{%
\ast }}\right\vert =O_{p}\left( \sqrt{(\ln N)/T}\right) =o_{p}\left(
c_{\lambda ,NT}\right) ,$ it follows that 
\begin{equation*}
\widehat{D}_{k}\asymp _{p}\sum_{i=1}^{N}1\left[ \left\vert \lambda _{ik}^{{%
\ast }}\right\vert >c_{\lambda ,NT}\right] \asymp _{p}N^{\alpha _{k}}.
\end{equation*}%
But this just implies that there exists constants $c^{\dag }>0$ and $C^{\dag
}>0,$ both independent of $N$ and $T$, such that with probability
approaching 1,%
\begin{equation*}
c^{\dag }N^{\alpha _{k}}\leq \widehat{D}_{k}\leq C^{\dag }N^{\alpha _{k}},
\end{equation*}%
that is,%
\begin{equation*}
\alpha _{k}+\frac{\ln c^{\dag }}{\ln N}\leq \frac{\ln \widehat{D}_{k}}{\ln N}%
\leq \alpha _{k}+\frac{\ln C^{\dag }}{\ln N}.
\end{equation*}%
So we come to that%
\begin{equation*}
\widehat{\alpha }_{k}\equiv \frac{\ln \widehat{D}_{k}}{\ln N}\overset{p}{%
\rightarrow }\alpha _{k},
\end{equation*}%
for $k=1,...,r.$ \qed

\section{Proofs of main results in Section \protect\ref{sec:no_f}}

\noindent \textbf{Proof of Theorem \ref{thm: number of factors}.} The proof
will be done by showing that both (a) $\widehat{r}\geq r$ and (b) $\widehat{r%
}\leq r$ hold in probability.

First, recall that we have shown that $\widetilde{V}_{k}\asymp N^{\alpha
_{k}-1}$ in probability for $k=1,\ldots ,r$, and also recall that 
\begin{equation*}
\widehat{r}=\max \left\{ k:\widetilde{V}_{k}^{r_{\max }}\geq \widehat{\sigma 
}^{2}N^{-1/2}\left( \ln \ln N\right) ^{1/2}\right\} .
\end{equation*}%
It is easy to show that $\widehat{\sigma }^{2}$ is a consistent estimator of 
$(NT)^{-1}\sum_{i=1}^{N}\sum_{t=1}^{T}E\left( e_{it}^{2}\right) $ which is
bounded away from zero and finite. It follows that $\widetilde{V}_{r}\asymp
_{p}N^{\alpha _{r}-1}\gg N^{-1/2}\left( \ln \ln N\right) ^{1/2}\asymp _{p}$ $%
\widehat{\sigma }^{2}N^{-1/2}\left( \ln \ln N\right) ^{1/2}$, implying that $%
r\leq \widehat{r}$. So (a) holds.

Second, suppose (b) does not hold, i.e., there exists some $\dot{r}$ such
that $r_{\max }\geq \dot{r}\geq r+1$ and that 
\begin{equation*}
\dot{r}=\max \left\{ k:\widetilde{V}_{k}^{r_{\max }}\geq \widehat{\sigma }
^{2}N^{-1/2}\left( \ln \ln N\right) ^{1/2}\right\}
\end{equation*}%
in probability not approaching 0. We will show that it leads to a
contradiction.

To this end, let us first investigate the rate of $T^{-1}\Vert \widetilde{F}%
-F^{0}H\Vert ^{2}$ without the true number of factors $r$. This implies that
some of the previously derived results assuming that $r$ is known, e.g.,
Lemma \ref{lm:G_full_rank}, cannot be applied here. So the rate derived
below is expected to be slower than stated in Proposition \ref{prop:
factor_consistent}.

Recall that 
\begin{equation*}
\widetilde{F}-F^{0}H=\left[ \frac{1}{N}\left( e^{\prime }\Lambda ^{0}\right)
\left( \frac{1}{T}F^{0\prime }\widetilde{F}\right) +\frac{1}{NT}F^{0}\Lambda
^{0\prime }e\widetilde{F}+\frac{1}{NT}e^{\prime }e\widetilde{F}\right] 
\widetilde{V}^{-1}\equiv \left( a_{1}+a_{2}+a_{3}\right) \widetilde{V}^{-1}.
\end{equation*}%
Or alternatively, $\left( \widetilde{F}-F^{0}H\right) \widetilde{V}%
=a_{1}+a_{2}+a_{3}.$ It then follows that 
\begin{eqnarray*}
\left\Vert a_{1}+a_{2}\right\Vert &\leqslant &\left\Vert a_{1}\right\Vert
+\left\Vert a_{2}\right\Vert \leqslant \frac{2}{T}\Vert F\Vert \Vert 
\widetilde{F}\Vert \frac{1}{N}\left\Vert \Lambda ^{0\prime }e\right\Vert \\
&=&O_{p}(1)\left[ \sum_{p=1}^{r}\sum_{t=1}^{T}\left( \frac{1}{N}
\sum_{i=1}^{N}\lambda _{ip}^{0\prime }e_{it}\right) ^{2}\right]
^{1/2}=O_{p}\left( T^{\frac{1}{2}}N^{\frac{\alpha _{1}}{2}-1}\right) .
\end{eqnarray*}%
Meanwhile, 
\begin{equation*}
\left\Vert a_{3}\right\Vert \leq \frac{1}{NT}\left\Vert e^{\prime
}e\right\Vert _{sp}\left\Vert \widetilde{F}\right\Vert =O_{p}(\frac{1}{NT}
)O_{p}(N+T)O_{p}\left( T^{\frac{1}{2}}\right) =O_{P}\left( T^{-\frac{1}{2}
}+N^{-1}T^{\frac{1}{2}}\right) .
\end{equation*}%
So 
\begin{equation}
\frac{1}{T}\left\Vert \widetilde{F}-F^{0}H\right\Vert ^{2}\leqslant \Vert 
\widetilde{V}\Vert ^{-2}O_{p}\left( N^{\alpha _{1}-2}+T^{-2}+N^{-2}\right)
=\Vert \widetilde{V}\Vert ^{-2}O_{p}\left( N^{\alpha _{1}-2}+T^{-2}\right) .
\label{F_L2bd}
\end{equation}

Going back to (b) with $\dot{r}\geq r+1$, now define $\dot{H}=\left( \Lambda
^{\prime }\Lambda /N\right) \left( F^{\prime }\widetilde{F}^{\dot{r}%
}/T\right) \widetilde{V}^{\dot{r}-1}$, where $\widetilde{F}^{\dot{r}}$ and $%
\widetilde{V}^{\dot{r}}$, analogue to $\widetilde{F}$ and $\widetilde{V}$,
respectively, are estimators under $\dot{r}$ factors. Given (\ref{F_L2bd}),
we immediately have, 
\begin{align*}
\frac{1}{T}\left\Vert \widetilde{F}^{\dot{r}}-F^{0}\dot{H}\right\Vert ^{2}&
\leq \left( \min_{1\leq k\leq r}\widetilde{V}_{k}\right) ^{-2}O_{p}\left(
N^{\alpha _{1}-2}+T^{-2}\right) \\
& \leq O_{P}(N)O_{P}\left( N^{\alpha _{1}-2}+T^{-2}\right) =O_{p}\left(
N^{\alpha _{1}-1}+NT^{-2}\right) =o_{p}(1)\text{, }
\end{align*}%
given that $N/T^{2}\rightarrow 0$ implied by Assumption 7. This implies that
we can follow in proving Lemma \ref{lm:G_full_rank} to show that%
\begin{equation*}
\frac{\dot{H}^{\prime }F^{0\prime }\widetilde{F}^{\dot{r}}}{T}=\frac{1}{T} 
\widetilde{F}^{\dot{r}\prime }\widetilde{F}^{\dot{r}}+o_{p}\left( 1\right)
=I_{\dot{r}}+o_{p}\left( 1\right) .
\end{equation*}%
But this is a contradiction, since $\rank\left( \dot{H}^{\prime }F^{0\prime }%
\widetilde{F}^{\dot{r}}/T\right) \leq r$ given $\dot{H}$ is $r\times \dot{r}$
whereas $\rank\left( I_{\dot{r}}\right) =\dot{r}\geq r+1.$ \qed 

\section{Proofs of auxiliary lemmas}

\label{sec:auxiliary}

In this section, we provide proofs of the auxiliary lemmas used in this
paper.

\noindent \textbf{Proof of Lemma \ref{lm:LeF}}. By Proposition \ref{prop:
factor_consistent}, we have 
\begin{align*}
\Lambda^{0 \prime} e \widetilde{F} & = \Lambda^{0 \prime} e FH + \Lambda^{0
\prime} e \left( \widetilde{F} - FH \right) \\
& = O_{p} \left(N^{\alpha_{1}/2} T^{1/2} \right) + O_{p}
\left(N^{\alpha_{1}/2} T^{1/2} \right) \sqrt{T} O_{p} \left( N^{\frac{%
\alpha_{1}}{2}-\alpha_{r}} + N^{1-\alpha_{r}} T^{-1} \right) \\
& = O_{p} \left( N^{\alpha_{1}/2} T^{1/2} + N^{\alpha_{1}-\alpha_{r}} T +
N^{1+\frac{\alpha_{1}}{2}-\alpha_{r}} \right). \qed
\end{align*}

\noindent \textbf{Proof of Lemma \ref{lm:Xi}}. Recall $Q(l,k)\asymp
_{p}N^{\alpha _{l}-\alpha _{k}}$, for $1\leqslant k\leqslant l\leqslant r$,
by Proposition \ref{prop:Q_upper_trian}. Then by the result of Proposition %
\ref{prop: V_rate}, we have%
\begin{equation*}
\Xi (l,k) 
\begin{cases}
\asymp _{p}N^{1-\alpha _{l}}N^{\alpha _{l}-\alpha _{k}}N^{\alpha
_{k}-1}\asymp _{p}1\text{, for }1\leqslant k<l\leqslant r, \\ 
=O_{p}\left( N^{1-\alpha _{l}}\right) O_{p}\left( N^{\alpha _{k}-1}\right)
=O_{p}\left( N^{\alpha _{k}-\alpha _{l}}\right) =o_{p}(1),\text{ for }
r\geqslant k>l>1.\qed%
\end{cases}%
\end{equation*}%
\bigskip

\noindent \textbf{Proof of Lemma \ref{lm:elffv}}. Write 
\begin{align*}
\frac{1}{NT}e^{\prime }\Lambda ^{0}F^{0\prime }\widetilde{F}\widetilde{V}
^{-1}& =\left( \frac{1}{N}e^{\prime }\Lambda ^{0}\right) \left( \frac{1}{T}
F^{0\prime }\widetilde{F}\right) \widetilde{V}^{-1}=e^{\prime }\Lambda
^{0}A^{-\frac{1}{2}}A^{\frac{1}{2}}N^{-1}Q^{\prime }\widetilde{V}^{-1} \\
& =\left( e^{\prime }\Lambda ^{0}A^{-\frac{1}{2}}\right) A^{-\frac{1}{2}
}\left( AN^{-1}Q^{\prime }\widetilde{V}^{-1}\right) .
\end{align*}%
Given Assumptions 2-3, we have $E(N^{-\frac{\alpha _{k}}{2}%
}\sum_{i=1}^{N}\lambda _{ik}^{0}e_{it})^{2}<M$, for each $t.$ Therefore, $%
\left\Vert e^{\prime }\Lambda ^{0}A^{-\frac{1}{2}}\right\Vert =O_{p}\left(
T^{\frac{1}{2}}\right) .$ Also, by Lemma \ref{lm:Xi}, we come to 
\begin{equation*}
\frac{1}{NT}\left\Vert e^{\prime }\Lambda ^{0}F^{0\prime }\widetilde{F} 
\widetilde{V}^{-1}\right\Vert =O_{p}\left( T^{\frac{1}{2}}\right) A^{-\frac{%
1 }{2}}O_{p}(1)=O_{p}\left( T^{\frac{1}{2}}\right) O_{p}\left( N^{-\frac{
\alpha _{r}}{2}}\right) =O_{p}\left( N^{-\frac{\alpha _{r}}{2}}T^{\frac{1}{2}
}\right) .\qed
\end{equation*}%
\bigskip

\noindent \textbf{Proof of Lemma \ref{lm:eelffv}}. We have 
\begin{eqnarray*}
\frac{1}{NT}ee^{\prime }\Lambda ^{0}F^{0\prime }\widetilde{F}\widetilde{V}
^{-1} &=&\left( ee^{\prime }\right) \Lambda ^{0}A^{-\frac{1}{2}}A^{-\frac{1}{
2}}\left( AN^{-1}Q^{\prime }\widetilde{V}^{-1}\right) =O_{p}(N+T)O(1)A^{- 
\frac{1}{2}}O_{p}(1) \\
&=&A^{-\frac{1}{2}}O_{p}(N+T).\qed
\end{eqnarray*}%
\bigskip

\noindent \textbf{Proof of Lemma \ref{lm:G_full_rank}}. By the so defined
matrix $H=\left( \Lambda ^{0\prime }\Lambda ^{0}/N\right) \left( F^{0\prime }%
\widetilde{F}/T\right) \widetilde{V}^{-1}$, we have%
\begin{equation*}
I_{r}=\frac{\widetilde{F}^{\prime }\widetilde{F}}{T}=\frac{\left( \widetilde{
F}-F^{0}H+F^{0}H\right) ^{\prime }\widetilde{F}}{T}=\frac{\left( \widetilde{%
F }-F^{0}H\right) ^{\prime }\widetilde{F}}{T}+\frac{H^{\prime }F^{0\prime } 
\widetilde{F}}{T}.
\end{equation*}%
Here, 
\begin{equation*}
\frac{1}{T}\left\Vert \left( \widetilde{F}-F^{0}H\right) ^{\prime } 
\widetilde{F}\right\Vert \leq \left( \frac{1}{T}\left\Vert \widetilde{F}
-F^{0}H\right\Vert ^{2}\right) ^{1/2}\left( \frac{1}{T}\left\Vert \widetilde{
F}\right\Vert ^{2}\right) ^{1/2}=O_{p}\left( \left[ \frac{1}{T}\left\Vert 
\widetilde{F}-F^{0}H\right\Vert ^{2}\right] ^{1/2}\right) .
\end{equation*}%
As for $\widetilde{F}-F^{0}H$, we here just provide a rough bound so as to
decompose it as 
\begin{align*}
\widetilde{F}_{t}-H^{\prime }F_{t}& =\widetilde{V}^{-1}\left( \frac{1}{T}
\sum_{s=1}^{T}\widetilde{F}_{s}\gamma _{N}(s,t)+\frac{1}{T}\sum_{s=1}^{T} 
\widetilde{F}_{s}\zeta _{s,t}+\frac{1}{T}\sum_{s=1}^{T}\widetilde{F}_{s}\eta
_{st}+\frac{1}{T}\sum_{s=1}^{T}\widetilde{F}_{s}\xi _{st}\right) \\
& \equiv \widetilde{V}^{-1}\left( I_{t}+II_{t}+III_{t}+IV_{t}\right) ,
\end{align*}%
where $\zeta _{s,t}\equiv e_{s}^{\prime }e_{t}/N-\gamma _{N}(s,t)$, $\eta
_{st}\equiv F_{s}^{0\prime }\Lambda ^{0\prime }e_{t}/N$, and $\xi
_{st}\equiv F_{t}^{0\prime }\Lambda ^{0\prime }e_{s}/N$. Bai and Ng (2002)
have proved in their Theorem 1 that, $T^{-1}\sum_{t=1}^{T}(\left\Vert
I_{t}\right\Vert ^{2}+\left\Vert II_{t}\right\Vert ^{2}+\left\Vert
III_{t}\right\Vert ^{2}+\left\Vert IV_{t}\right\Vert ^{2})=O_{p}\left(
N^{-1}+T^{-1}\right)$, and note that the result holds under either strong or
weak factors. So it follows that 
\begin{align}
\frac{1}{T}\left\Vert \widetilde{F}-F^{0}H\right\Vert ^{2}& =\frac{1}{T}
\sum_{t=1}^{T}\left\Vert \widetilde{V}^{-1}\left(
I_{t}+II_{t}+III_{t}+IV_{t}\right) \right\Vert ^{2}  \notag \\
& \leq \left\Vert \widetilde{V}^{-1}\right\Vert ^{2}\frac{1}{T}
\sum_{t=1}^{T}\left\Vert I_{t}+II_{t}+III_{t}+IV_{t}\right\Vert ^{2}  \notag
\\
& = O_{p}\left( N^{2\left( 1-\alpha _{r}\right) }\right)\frac{1}{T}
\sum_{t=1}^{T}\left\Vert I_{t}+II_{t}+III_{t}+IV_{t}\right\Vert ^{2} , 
\notag \\
& =O_{p}\left( N^{2\left( 1-\alpha _{r}\right) }\right) O_{p}\left(
T^{-1}+N^{-1}\right) =o_{p}(1),  \label{eq:Ftilde_FH}
\end{align}%
where the second equality is due to $\widetilde{V}$ being a diagonal matrix
and by Proposition \ref{prop: V_rate}, and the last equality is due to that $%
\alpha _{k}\in (1/2,1]\ $and $N^{1-\alpha _{r}}/T^{1/2}\rightarrow 0$ by
Assumption 5.

We will show next that the result $T^{-1}\left\Vert \widetilde{F}%
-F^{0}H\right\Vert ^{2}=o_{p}(1)$ implies two useful conditions: (a) $%
H=O_{p}(1)$, and (b) $H$ is of full rank $r$.

For (a), recall that $H=\frac{\Lambda^{0 \prime}\Lambda^{0}}{N} \frac{F^{0
\prime} \widetilde{F}}{T} \widetilde{V}^{-1} = \Sigma_{N,\Lambda} \frac{F^{0
\prime} \widetilde{F}}{T} \widetilde{V}^{-1}$, and note that $\frac{F^{0
\prime} \widetilde{F}}{T} = O_{p}(1)$, and that $\Sigma_{N,\Lambda}(j,k) =
O(N^{\alpha_{j} \wedge \alpha_{k}-1})$ which can be readily verified by the
definition of $\Lambda^{0}$ and $\alpha_{k}$. So, 
\begin{align*}
H(j,k) = & \sum_{l=1}^{r} \Sigma_{N,\Lambda}(j,l) Q(k,l) \widetilde{V}%
^{-1}(k,k) \\
= & \sum_{l=1}^{r} O(N^{\alpha_{j} \wedge \alpha_{l}-1}) O_{p}(1)
O_{p}(N^{1-\alpha_{k}}).
\end{align*}
We see that for $H(j,k) \neq O_{p}(1)$, it can only be the case that $H(j,k)
\neq O_{p}(1)$ with $j<k$. Yet we will prove that this can never be true via
contradiction as below.

Let us consider $F_{\cdot ,k}^{\ast }$ with $F^{\ast }=F^{0}H$. Note that $%
T^{-1} \left\Vert \widetilde{F}_{\cdot,k}-F_{\cdot,k}^{\ast }\right\Vert
^{2}\leq T^{-1} \left\Vert \widetilde{F}-F^{0}H\right\Vert ^{2}=o_{p}(1).$
By the Minkowski inequality, this implies that $\left\vert
T^{-1}\sum_{t=1}^{T}\widetilde{F}_{tk}^{2}-T^{-1}\sum_{t=1}^{T}F_{tk}^{\ast
2}\right\vert =o_{p}(1),$ which leads to 
\begin{equation}
T^{-1}\sum_{t=1}^{T}F_{tk}^{\ast 2}\overset{p}{\rightarrow }1.
\label{eq:average_F*}
\end{equation}

Suppose $H(j,k)\neq O_{p}(1)$ for some $j$ and $k$ such that $j<k$. Then
this means that $\forall M>0$, $\exists$ a constant $\nu>0$ and a
subsequence $T_{m}$ ($m=1,2,\ldots$), such that on an event $\mathcal{A}_{m}$
with positive probability $P(\mathcal{A}_{m})>\nu$, we have $\left| H(j,k)
\right| > M$. Meanwhile, recall that $T^{-1}F^{0\prime }F^{0} \rightarrow
_{p}\Sigma _{F}$ as $T\rightarrow \infty $ for some p.d. matrix $\Sigma _{F}$
by Assumption 1, implying that $T^{-1}\sum_{t=1}^{T}(F_{tj}^{0})^{2}
\rightarrow_{p} \Sigma_{F}(j,j)$ for $\Sigma_{F}(j,j)$ being \textit{bounded
away} from 0, $j=1,\ldots, r$. This in turn implies that 
\begin{equation}  \label{eq:average_F0}
T_{m}^{-1} \sum_{t=1}^{T_{m}}(F_{tj}^{0})^{2} \rightarrow_{p}
\Sigma_{F}(j,j)>0.
\end{equation}
Given that $F_{t}^{0} = O_{p}(1)$ implied by Assumption 1, and the fact that 
$F_{tk}^{\ast }=\sum_{j=1}^{r}F_{tj}^{0}H(j,k)$, it follows from %
\eqref{eq:average_F0} that, $\exists a,b>0$ with $a$ bounded away from 0,
such that on the event $\mathcal{A}_{m}$ we must have $T_{m}^{-1}%
\sum_{t=1}^{T_{m}}F_{tk}^{\ast 2} >aM^{2}+b$. But this contradicts with %
\eqref{eq:average_F*} since $M$ is arbitrarily big. So (a) is proved.

For (b), it is obvious to hold when $r=1$. We will show in detail the proof
for when $r=2$, and extends it later to $r>2$.

When $r=2$, \ 
\begin{eqnarray}
o_{p}\left( 1\right) &=&\frac{1}{T}\left\Vert \widetilde{F}%
-F^{0}H\right\Vert ^{2}  \notag \\
&=&\frac{1}{T}\left\Vert \widetilde{F}_{\cdot ,1}-F_{\cdot ,1}^{\ast
}\right\Vert ^{2}+\frac{1}{T}\left\Vert \widetilde{F}_{\cdot ,2}-F_{\cdot
,2}^{\ast }\right\Vert ^{2}.  \label{eq:F4}
\end{eqnarray}%
Suppose $H$ is not full rank. Then we have $F_{\cdot ,2}^{\ast }=aF_{\cdot
,1}^{\ast }$ for some $a\neq 0.$ So we must have $\frac{1}{T}\left\Vert a%
\widetilde{F}_{\cdot ,1}-aF_{\cdot ,1}^{\ast }\right\Vert ^{2}=o_{p}\left(
1\right) $ and $\frac{1}{T}\left\Vert \widetilde{F}_{\cdot ,2}-F_{\cdot
,2}^{\ast }\right\Vert ^{2}=o_{p}(1)$ from \eqref{eq:F4}. Then by the
Minkowski inequality,%
\begin{eqnarray*}
\frac{1}{T}\left\Vert \widetilde{F}_{\cdot ,2}-a\widetilde{F}_{\cdot
,1}\right\Vert ^{2} &=&\frac{1}{T}\left\Vert \left( F_{\cdot ,2}^{\ast }-%
\widetilde{F}_{\cdot ,2}\right) +\left( a\widetilde{F}_{\cdot ,1}-aF_{\cdot
,1}^{\ast }\right) \right\Vert ^{2} \\
&\leq &\frac{1}{T}\left\Vert F_{\cdot ,2}^{\ast }-\widetilde{F}_{\cdot
,2}\right\Vert ^{2}+\frac{1}{T}\left\Vert a\widetilde{F}_{\cdot
,1}-aF_{\cdot ,1}^{\ast }\right\Vert ^{2} \\
&=&o_{p}(1).
\end{eqnarray*}%
But given the orthogonality between $\widetilde{F}_{\cdot ,1}$ and $%
\widetilde{F}_{\cdot ,2}$,%
\begin{eqnarray*}
\frac{1}{T}\left\Vert \widetilde{F}_{\cdot ,2}-a\widetilde{F}_{\cdot
,1}\right\Vert ^{2} &=&\frac{1}{T}\sum_{t=1}^{T}\left( \widetilde{F}_{t2}-a%
\widetilde{F}_{t1}\right) ^{2} \\
&=&\frac{1}{T}\sum_{t=1}^{T}\widetilde{F}_{t2}^{2}+\frac{a^{2}}{T}%
\sum_{t=1}^{T}\widetilde{F}_{t1}^{2} \\
&=&1+a^{2},
\end{eqnarray*}%
which is clearly a contradiction. So (b) holds when $r=2$.

For the case when $r>2$, the argument is almost identical. We just need to
replace $F_{\cdot ,1}^{\ast }$ with all $\left( F_{\cdot ,1}^{\ast
},F_{\cdot ,3}^{\ast },...,F_{\cdot ,r}^{\ast }\right) ,$ and replace $%
aF_{\cdot ,1}^{\ast }$ with $\left( F_{\cdot ,1}^{\ast },F_{\cdot ,3}^{\ast
},...,F_{\cdot ,r}^{\ast }\right) \times a$ for an $(r-1)\times 1$ nonzero
vector $a$ (similarly with $a\widetilde{F}_{t1}$)$.$ So proof of (b) is
complete.

Finally, note that \eqref{eq:Ftilde_FH} also leads to that $T^{-1}\left\Vert
\left( \widetilde{F}-F^{0}H\right) ^{\prime }\widetilde{F}\right\Vert
=o_{p}(1)$, implying that $H^{\prime }\left( F^{0\prime }\widetilde{F}%
/T\right) =I_{r}+o_{p}(1).$ Now we have that with probability approaching 1, 
\begin{equation*}
\rank\left( H^{\prime }\frac{F^{0\prime }\widetilde{F}}{T}\right)
=r\Longrightarrow \rank\left( \frac{F^{0\prime }\widetilde{F}}{T}\right)
\geq r\Longrightarrow \rank\left( \frac{F^{0\prime }\widetilde{F}}{T}\right)
=r,
\end{equation*}%
by noticing that $F^{0\prime }\widetilde{F}/T$ is an $r\times r$ matrix. So
the matrix $Q=\widetilde{F}^{\prime } F^{0}/T$ is of full rank $r$ with
probability approaching 1 as desired.\ \qed

\bigskip

\textbf{References}\setlength{\baselineskip}{11pt}

\begin{description}

\item Bai, J. (2003) Inferential theory for factor models of large
dimensions. \textit{Econometrica} 71(1), 135--171.

\item Bai, J. \&  S. Ng (2002) Determining the number of factors in
approximate factor models. \textit{Econometrica} 70(1), 191--221.

\item Fan, J., Y. Liao, \&  M.  Mincheva (2011) High-dimensional covariance
matrix estimation in approximate factor models. \textit{The} \textit{Annals
of Statistics} 39(6), 3320--3356.

\item Merlevede, F., M. Peligrad, \&  E. Rio (2011) A Bernstein type
inequality and moderate deviations for weakly dependent sequences. \textit{\
Probability Theory and Related Fields} 151, 435--474.
\end{description}

\newpage
\section{Additional simulation results}

\subsection{Simulation results under $r=5$}

\begin{table}[tbph]
	\caption{Estimating the number of factors when $r=5$}
	\label{tab:no_factor_r5}\centering
	\resizebox{\linewidth}{!}{
		\begin{tabular}{ccccccccccccccc}
			\hline
			&       & \multicolumn{6}{c}{RMSE}                      &       & \multicolumn{6}{c}{Bias} \\
			\cline{3-8}\cline{10-15}    $N$     & $T$     & WZ    & BN    & GCT   & FR    & ED    & AH    &       & WZ    & BN    & GCT   & FR    & ED    & AH \\
			\hline
			100   & 100   & 0.240  & 0.720  & 2.721  & 3.972  & 0.508  & 4.000  &     & 0.029  & -0.482  & -2.668  & -3.956  & -0.054  & -4.000  \\
			& 200   & 0.249  & 0.432  & 2.431  & 3.978  & 0.318  & 4.000  &    & -0.052  & -0.176  & -2.341  & -3.963  & 0.048  & -4.000  \\
			& 400   & 0.285  & 0.274  & 2.101  & 3.982  & 0.830  & 4.000  &     & -0.080  & -0.074  & -1.993  & -3.971  & 0.249  & -4.000  \\
			200   & 100   & 0.177  & 0.590  & 2.494  & 3.947  & 0.246  & 4.000  &     & 0.023  & -0.334  & -2.421  & -3.898  & 0.046  & -4.000  \\
			& 200   & 0.134  & 0.221  & 1.840  & 3.907  & 0.274  & 4.000  &     & -0.014  & -0.046  & -1.748  & -3.819  & 0.057  & -4.000  \\
			& 400   & 0.167  & 0.100  & 1.386  & 3.927  & 0.320  & 4.000  &     & -0.025  & -0.007  & -1.303  & -3.856  & 0.062  & -4.000  \\
			400   & 100   & 0.157  & 0.638  & 2.292  & 3.811  & 0.231  & 4.000  &     & 0.024  & -0.399  & -2.217  & -3.629  & 0.049  & -4.000  \\
			& 200   & 0.055  & 0.166  & 1.544  & 3.313  & 0.225  & 4.000  &    & -0.003  & -0.028  & -1.457  & -2.744  & 0.044  & -4.000  \\
			& 400   & 0.097  & 0.045  & 1.048  & 2.898  & 0.225  & 3.997  &    & -0.010  & -0.002  & -1.029  & -2.100  & 0.046  & -3.994  \\
			\hline
	\end{tabular}	}
\end{table}

\begin{table}[tbph]
	\caption{Estimation of factor models when $r=5$}
	\label{tab:FM_estimate_r5}\centering
	\resizebox{\linewidth}{!}{
		\begin{tabular}{cccccccccccccccc}
			\hline
			&       & \multicolumn{4}{c}{$TR^{F}$}     &       & \multicolumn{4}{c}{$TR^{\Lambda}$}     &       & \multicolumn{4}{c}{$RMSE^C$} \\
			\cline{3-6}\cline{8-11}\cline{13-16}   $N$     & $T$     & PC    & Ada   & Deb   & Res   &       & PC    & Ada   & Deb   & Res   &       & PC    & Ada   & Deb   & Res \\
			\hline
			100   & 100   & \textbf{0.946 } & 0.919  & 0.922  & 0.922  &       & \textbf{0.774 } & 0.685  & 0.769  & 0.766  &       & \textbf{1.612 } & 1.653  & 1.620  & 1.622  \\
			& 200   & \textbf{0.953 } & 0.938  & 0.941  & 0.941  &       & \textbf{0.820 } & 0.761  & 0.818  & 0.817  &       & \textbf{1.599 } & 1.633  & 1.610  & 1.611  \\
			& 400   & \textbf{0.956 } & 0.948  & 0.950  & 0.950  &       & 0.844  & 0.814  & \textbf{0.845 } & \textbf{0.845 } &       & \textbf{1.594 } & 1.615  & 1.603  & 1.603  \\
			200   & 100   & \textbf{0.969 } & 0.947  & 0.949  & 0.949  &       & 0.794  & 0.699  & 0.796  & \textbf{0.800 } &       & \textbf{1.500 } & 1.574  & 1.527  & 1.529  \\
			& 200   & \textbf{0.973 } & 0.960  & 0.963  & 0.963  &       & 0.837  & 0.777  & 0.839  & \textbf{0.843 } &       & \textbf{1.496 } & 1.541  & 1.513  & 1.514  \\
			& 400   & \textbf{0.976 } & 0.969  & 0.971  & 0.971  &       & 0.860  & 0.829  & 0.862  & \textbf{0.864 } &       & \textbf{1.497 } & 1.516  & 1.500  & 1.501  \\
			400   & 100   & \textbf{0.980 } & 0.961  & 0.963  & 0.963  &       & 0.798  & 0.704  & 0.804  & \textbf{0.816 } &       & \textbf{1.438 } & 1.497  & 1.444  & 1.446  \\
			& 200   & \textbf{0.983 } & 0.972  & 0.974  & 0.974  &       & 0.841  & 0.780  & 0.844  & \textbf{0.852 } &       & \textbf{1.430 } & 1.467  & 1.434  & 1.435  \\
			& 400   & \textbf{0.985 } & 0.979  & 0.980  & 0.980  &       & 0.866  & 0.833  & 0.867  & \textbf{0.872 } &       & \textbf{1.425 } & 1.447  & 1.428  & 1.429  \\
			\hline
	\end{tabular}}
\end{table}

\begin{table}[htbp]
	\caption{FDR and Power under $r=5$}
	\label{tab:FDR_r5}\centering
	\resizebox{\linewidth}{!}{
		\begin{tabular}{ccccccccccccccc}
			\hline
			$N$ 	&   $T$    & \multicolumn{1}{c}{$\text{FDR}_{1}$} & \multicolumn{1}{c}{$\text{FDR}_{2}$} & \multicolumn{1}{c}{$\text{FDR}_{3}$} & \multicolumn{1}{c}{$\text{FDR}_{4}$} & \multicolumn{1}{c}{$\text{FDR}_{5}$} & \multicolumn{1}{c}{$\overline{\text{FDR}}$} &       & \multicolumn{1}{c}{$\text{Power}_{1}$} & \multicolumn{1}{c}{$\text{Power}_{2}$} & \multicolumn{1}{c}{$\text{Power}_{3}$} & \multicolumn{1}{c}{$\text{Power}_{4}$} & \multicolumn{1}{c}{$\text{Power}_{5}$} & \multicolumn{1}{c}{$\overline{\text{Power}}$} \\
			\hline
			&      & \multicolumn{13}{c}{Panel A: PC+Screening} \\
			\cline{3-15}    100   & 100   & 0     & 0.239  & 0.371  & 0.558  & 0.520  & 0.222  &       & 0.802  & 0.522  & 0.530  & 0.363  & 0.427  & 0.617  \\
			& 200   & 0     & 0.241  & 0.376  & 0.538  & 0.496  & 0.221  &       & 0.817  & 0.540  & 0.544  & 0.386  & 0.452  & 0.633  \\
			& 400   & 0     & 0.242  & 0.370  & 0.529  & 0.467  & 0.218  &       & 0.834  & 0.549  & 0.569  & 0.396  & 0.486  & 0.650  \\
			200   & 100   & 0     & 0.265  & 0.383  & 0.532  & 0.456  & 0.209  &       & 0.813  & 0.531  & 0.538  & 0.412  & 0.497  & 0.645  \\
			& 200   & 0     & 0.255  & 0.393  & 0.488  & 0.440  & 0.205  &       & 0.832  & 0.547  & 0.555  & 0.444  & 0.509  & 0.664  \\
			& 400   & 0     & 0.256  & 0.388  & 0.471  & 0.401  & 0.202  &       & 0.837  & 0.566  & 0.578  & 0.473  & 0.563  & 0.680  \\
			400   & 100   & 0     & 0.278  & 0.376  & 0.471  & 0.327  & 0.189  &       & 0.820  & 0.551  & 0.565  & 0.489  & 0.618  & 0.679  \\
			& 200   & 0     & 0.273  & 0.375  & 0.432  & 0.229  & 0.178  &       & 0.836  & 0.558  & 0.595  & 0.531  & 0.667  & 0.698  \\
			& 400   & 0     & 0.265  & 0.384  & 0.412  & 0.192  & 0.175  &       & 0.849  & 0.576  & 0.604  & 0.557  & 0.721  & 0.715  \\
			&       & \multicolumn{13}{c}{Panel B: SOFAR\_Adaptive} \\
			\cline{3-15}    100   & 100   & 0     & 0.076  & 0.153  & 0.279  & 0.401  & 0.117  &       & 0.741  & 0.729  & 0.712  & 0.674  & 0.664  & 0.722  \\
			& 200   & 0     & 0.087  & 0.197  & 0.313  & 0.418  & 0.138  &       & 0.812  & 0.799  & 0.760  & 0.724  & 0.738  & 0.787  \\
			& 400   & 0     & 0.305  & 0.507  & 0.597  & 0.691  & 0.349  &       & 0.939  & 0.766  & 0.771  & 0.703  & 0.682  & 0.827  \\
			200   & 100   & 0     & 0.040  & 0.091  & 0.179  & 0.301  & 0.065  &       & 0.744  & 0.733  & 0.720  & 0.705  & 0.694  & 0.731  \\
			& 200   & 0     & 0.036  & 0.088  & 0.189  & 0.285  & 0.064  &       & 0.808  & 0.794  & 0.786  & 0.759  & 0.753  & 0.793  \\
			& 400   & 0     & 0.343  & 0.544  & 0.581  & 0.653  & 0.342  &       & 0.937  & 0.772  & 0.769  & 0.731  & 0.712  & 0.838  \\
			400   & 100   & 0     & 0.295  & 0.367  & 0.445  & 0.472  & 0.189  &       & 0.859  & 0.536  & 0.549  & 0.473  & 0.531  & 0.687  \\
			& 200   & 0     & 0.335  & 0.476  & 0.502  & 0.487  & 0.251  &       & 0.903  & 0.661  & 0.675  & 0.616  & 0.665  & 0.775  \\
			& 400   & 0     & 0.383  & 0.543  & 0.586  & 0.553  & 0.320  &       & 0.931  & 0.767  & 0.782  & 0.751  & 0.778  & 0.846  \\
			&       & \multicolumn{13}{c}{Panel C: SOFAR\_Resparsified} \\
			\cline{3-15}    100   & 100   & 0     & 0.075  & 0.172  & 0.302  & 0.456  & 0.130  &       & 0.800  & 0.806  & 0.783  & 0.746  & 0.733  & 0.789  \\
			& 200   & 0     & 0.071  & 0.169  & 0.298  & 0.435  & 0.125  &       & 0.862  & 0.858  & 0.822  & 0.778  & 0.802  & 0.842  \\
			& 400   & 0     & 0.306  & 0.518  & 0.647  & 0.705  & 0.377  &       & 0.950  & 0.831  & 0.849  & 0.737  & 0.767  & 0.869  \\
			200   & 100   & 0     & 0.060  & 0.148  & 0.291  & 0.423  & 0.106  &       & 0.804  & 0.808  & 0.804  & 0.766  & 0.764  & 0.800  \\
			& 200   & 0     & 0.052  & 0.128  & 0.251  & 0.380  & 0.091  &       & 0.859  & 0.860  & 0.854  & 0.832  & 0.829  & 0.854  \\
			& 400   & 0     & 0.343  & 0.534  & 0.651  & 0.659  & 0.365  &       & 0.946  & 0.847  & 0.860  & 0.761  & 0.795  & 0.883  \\
			400   & 100   & 0     & 0.373  & 0.546  & 0.687  & 0.699  & 0.342  &       & 0.884  & 0.540  & 0.541  & 0.545  & 0.597  & 0.707  \\
			& 200   & 0     & 0.362  & 0.480  & 0.589  & 0.591  & 0.303  &       & 0.919  & 0.761  & 0.754  & 0.711  & 0.764  & 0.831  \\
			& 400   & 0     & 0.388  & 0.497  & 0.620  & 0.579  & 0.329  &       & 0.941  & 0.843  & 0.876  & 0.842  & 0.867  & 0.895  \\
			\hline
	\end{tabular}}
\end{table}

\begin{table}[htbp]
	\caption{Estimation of factor strength when $r=5$ with $\protect\alpha %
		=\left(1, 0.9, 0.8, 0.7,0.6\right)$}
	\label{tab:fct_strength_r5}\centering
	\resizebox{\linewidth}{!}{
		\begin{tabular}{ccccccccccccc}
			\hline
			&  & \multicolumn{5}{c}{RMSE} &  & \multicolumn{5}{c}{Bias} \\ 
			\cline{3-7}\cline{9-13}
			$N$ & $T$ & $\widehat{\alpha}_{1}$ & $\widehat{\alpha}_{2}$ & $\widehat{\alpha}_{3}$ & $\widehat{\alpha}_{4}$ & $\widehat{\alpha}_{5}$ &  & $\widehat{\alpha}_{1}$ & $\widehat{\alpha}_{2}$ & $\widehat{\alpha}_{3}$ & $\widehat{\alpha}_{4}$ & $\widehat{\alpha}_{5}$ \\ \hline\cline{3-13}
			\hline
			&       & \multicolumn{11}{c}{Panel A: PC+Screening} \\
			\cline{3-13}    100   & 100   & 0.036  & 0.045  & \textbf{0.041 } & 0.076  & 0.156  &       & -0.034  & -0.035  & 0.020  & 0.050  & 0.111  \\
			& 200   & 0.034  & 0.043  & \textbf{0.043 } & \textbf{0.076 } & 0.187  &       & -0.033  & -0.034  & \textbf{0.021 } & \textbf{0.047 } & \textbf{0.081 } \\
			& 400   & 0.033  & 0.040  & \textbf{0.043 } & \textbf{0.078 } & 0.223  &       & -0.032  & \textbf{-0.030 } & 0.023  & \textbf{0.045 } & \textbf{0.058 } \\
			200   & 100   & 0.031  & 0.032  & 0.047  & 0.083  & 0.139  &       & -0.030  & -0.023  & 0.035  & 0.065  & 0.115  \\
			& 200   & 0.030  & 0.030  & \textbf{0.047 } & 0.076  & 0.159  &       & -0.029  & -0.021  & \textbf{0.035 } & \textbf{0.052 } & 0.095  \\
			& 400   & 0.029  & \textbf{0.027 } & \textbf{0.049 } & \textbf{0.076 } & 0.172  &       & -0.028  & \textbf{-0.018 } & \textbf{0.037 } & \textbf{0.053 } & \textbf{0.079 } \\
			400   & 100   & 0.027  & 0.020  & 0.053  & 0.089  & 0.117  &       & -0.027  & \textbf{-0.009 } & 0.045  & 0.077  & 0.101  \\
			& 200   & 0.026  & 0.018  & 0.052  & 0.084  & 0.107  &       & -0.026  & -0.008  & 0.042  & 0.070  & 0.065  \\
			& 400   & 0.025  & \textbf{0.015 } & \textbf{0.053 } & \textbf{0.082 } & 0.124  &       & -0.025  & \textbf{-0.006 } & \textbf{0.044 } & \textbf{0.067 } & \textbf{0.051 } \\
			&       & \multicolumn{11}{c}{Panel B: SOFAR\_Debiased} \\
			\cline{3-13}    100   & 100   & 0.029  & 0.056  & 0.045  & \textbf{0.055 } & \textbf{0.062 } &       & -0.027  & -0.045  & \textbf{-0.009 } & \textbf{0.003 } & \textbf{0.021 } \\
			& 200   & 0.020  & \textbf{0.024 } & 0.061  & 0.089  & \textbf{0.106 } &       & -0.019  & \textbf{0.000 } & 0.050  & 0.077  & 0.088  \\
			& 400   & 0.015  & \textbf{0.037 } & 0.097  & 0.142  & \textbf{0.157 } &       & -0.013  & 0.033  & 0.093  & 0.136  & 0.144  \\
			200   & 100   & 0.027  & 0.055  & \textbf{0.041 } & \textbf{0.046 } & \textbf{0.046 } &       & -0.026  & -0.047  & \textbf{-0.013 } & \textbf{-0.016 } & \textbf{0.018 } \\
			& 200   & 0.019  & \textbf{0.020 } & 0.059  & \textbf{0.070 } & \textbf{0.089 } &       & -0.018  & \textbf{0.002 } & 0.048  & 0.058  & \textbf{0.076 } \\
			& 400   & 0.013  & 0.037  & 0.097  & 0.127  & \textbf{0.136 } &       & -0.013  & 0.035  & 0.094  & 0.123  & 0.127  \\
			400   & 100   & 0.027  & 0.053  & \textbf{0.039 } & \textbf{0.045 } & \textbf{0.031 } &       & -0.026  & -0.048  & \textbf{-0.023 } & \textbf{-0.031 } & \textbf{-0.002 } \\
			& 200   & 0.018  & \textbf{0.017 } & \textbf{0.051 } & \textbf{0.051 } & \textbf{0.061 } &       & -0.017  & \textbf{0.002 } & \textbf{0.041 } & \textbf{0.040 } & \textbf{0.046 } \\
			& 400   & 0.012  & 0.039  & 0.095  & 0.107  & \textbf{0.103 } &       & -0.012  & 0.037  & 0.092  & 0.103  & 0.095  \\
			&       & \multicolumn{11}{c}{Panel C: SOFAR\_Resparsified} \\
			\cline{3-13}    100   & 100   & \textbf{0.024 } & \textbf{0.021 } & 0.053  & 0.092  & 0.100  &       & \textbf{-0.023 } & \textbf{0.003 } & 0.043  & 0.080  & 0.084  \\
			& 200   & \textbf{0.017 } & 0.036  & 0.087  & 0.136  & 0.135  &       & \textbf{-0.016 } & 0.033  & 0.082  & 0.130  & 0.122  \\
			& 400   & \textbf{0.012 } & 0.056  & 0.121  & 0.181  & 0.179  &       & \textbf{-0.011 } & 0.055  & 0.119  & 0.178  & 0.168  \\
			200   & 100   & \textbf{0.022 } & \textbf{0.018 } & 0.048  & 0.081  & 0.098  &       & \textbf{-0.022 } & \textbf{0.010 } & 0.040  & 0.074  & 0.092  \\
			& 200   & \textbf{0.016 } & 0.039  & 0.076  & 0.116  & 0.115  &       & \textbf{-0.015 } & 0.038  & 0.071  & 0.111  & 0.110  \\
			& 400   & \textbf{0.011 } & 0.058  & 0.111  & 0.163  & 0.153  &       & \textbf{-0.011 } & 0.058  & 0.108  & 0.160  & 0.148  \\
			400   & 100   & \textbf{0.021 } & \textbf{0.019 } & 0.041  & 0.068  & 0.087  &       & \textbf{-0.021 } & 0.016  & 0.034  & 0.063  & 0.085  \\
			& 200   & \textbf{0.015 } & 0.042  & 0.066  & 0.095  & 0.098  &       & \textbf{-0.014 } & 0.041  & 0.062  & 0.092  & 0.095  \\
			& 400   & \textbf{0.010 } & 0.060  & 0.097  & 0.139  & 0.123  &       & \textbf{-0.010 } & 0.059  & 0.095  & 0.137  & 0.121  \\
			\hline
	\end{tabular}}
\end{table}


\subsection{Robust check for factor strength estimation} \label{sec:additional_MC}

The next two tables below report the estimation results of factor strength
when $r=3$ or $5$ respectively, where $\widetilde{c}_{\lambda ,NT}=c[\ln
(NT)]^{-1/2}$ takes values $c=0.8$ and $1.2$.

\begin{table}[tbph]
\caption{Estimation of factor strength when $r=3$ with $\protect\widetilde{c}%
_{\protect\lambda ,NT}=c[\ln (NT)]^{-1/2}$}
\label{tab-robust_c_r3}\centering
\begin{tabular}{ccccccccc}
\hline
&  & \multicolumn{3}{c}{RMSE} &  & \multicolumn{3}{c}{Bias} \\ 
\cline{3-5}\cline{7-9}
$N$ & $T$ & $\widehat{\alpha}_{1}$ & $\widehat{\alpha}_{2}$ & $\widehat{
\alpha} _{3}$ &  & $\widehat{\alpha}_{1}$ & $\widehat{\alpha}_{2}$ & $%
\widehat{\alpha }_{3}$ \\ \hline\cline{3-9}\hline
&  & \multicolumn{7}{c}{Panel A: $c=0.8$} \\ \cline{3-9}
100 & 100 & 0.018 & 0.067 & 0.183 &  & 0.012 & 0.050 & 0.161 \\ 
& 200 & 0.017 & 0.063 & 0.200 &  & 0.011 & 0.043 & 0.125 \\ 
& 400 & 0.018 & 0.064 & 0.233 &  & 0.013 & 0.042 & 0.102 \\ 
200 & 100 & 0.016 & 0.074 & 0.183 &  & 0.013 & 0.066 & 0.172 \\ 
& 200 & 0.013 & 0.070 & 0.182 &  & 0.011 & 0.061 & 0.136 \\ 
& 400 & 0.014 & 0.069 & 0.197 &  & 0.011 & 0.059 & 0.128 \\ 
400 & 100 & 0.014 & 0.085 & 0.171 &  & 0.013 & 0.081 & 0.167 \\ 
& 200 & 0.011 & 0.078 & 0.150 &  & 0.010 & 0.074 & 0.134 \\ 
& 400 & 0.011 & 0.076 & 0.159 &  & 0.010 & 0.072 & 0.120 \\ 
&  & \multicolumn{7}{c}{Panel B: $c=1.2$} \\ \cline{3-9}
100 & 100 & 0.019 & 0.065 & 0.097 &  & -0.010 & -0.034 & 0.041 \\ 
& 200 & 0.016 & 0.062 & 0.150 &  & -0.008 & -0.033 & 0.016 \\ 
& 400 & 0.016 & 0.057 & 0.175 &  & -0.007 & -0.029 & 0.018 \\ 
200 & 100 & 0.013 & 0.041 & 0.086 &  & -0.008 & -0.010 & 0.053 \\ 
& 200 & 0.012 & 0.040 & 0.105 &  & -0.007 & -0.010 & 0.042 \\ 
& 400 & 0.010 & 0.036 & 0.140 &  & -0.005 & -0.005 & 0.029 \\ 
400 & 100 & 0.010 & 0.035 & 0.073 &  & -0.007 & 0.007 & 0.046 \\ 
& 200 & 0.009 & 0.029 & 0.080 &  & -0.006 & 0.008 & 0.026 \\ 
& 400 & 0.007 & 0.028 & 0.095 &  & -0.004 & 0.010 & 0.023 \\ \hline
\end{tabular}%
\end{table}

\begin{table}[tbph]
\caption{Estimation of factor strength when $r=5$ with $\protect\widetilde{c}%
_{\protect\lambda ,NT}=c[\ln (NT)]^{-1/2}$}
\label{tab-robust_c_r5}\centering
\begin{tabular}{ccccccccccccc}
\hline
&  & \multicolumn{5}{c}{RMSE} &  & \multicolumn{5}{c}{Bias} \\ 
\cline{3-7}\cline{9-13}
$N$ & $T$ & $\widehat{\alpha}_{1}$ & $\widehat{\alpha}_{2}$ & $\widehat{%
\alpha}_{3}$ & $\widehat{\alpha}_{4}$ & $\widehat{\alpha}_{5}$ &  & $%
\widehat{\alpha}_{1}$ & $\widehat{\alpha}_{2}$ & $\widehat{\alpha}_{3}$ & $%
\widehat{\alpha}_{4}$ & $\widehat{\alpha}_{5}$ \\ \hline\cline{3-13}\hline
&  & \multicolumn{11}{c}{Panel A: $c=0.8$} \\ \cline{3-13}
100 & 100 & 0.028 & 0.027 & 0.064 & 0.118 & 0.205 &  & -0.027 & -0.009 & 
0.055 & 0.107 & 0.171 \\ 
& 200 & 0.027 & 0.027 & 0.065 & 0.108 & 0.238 &  & -0.026 & -0.009 & 0.055 & 
0.092 & 0.136 \\ 
& 400 & 0.026 & 0.027 & 0.066 & 0.110 & 0.269 &  & -0.025 & -0.008 & 0.055 & 
0.091 & 0.099 \\ 
200 & 100 & 0.025 & 0.020 & 0.074 & 0.121 & 0.193 &  & -0.024 & 0.002 & 0.068
& 0.111 & 0.184 \\ 
& 200 & 0.024 & 0.018 & 0.073 & 0.112 & 0.198 &  & -0.023 & 0.002 & 0.066 & 
0.098 & 0.159 \\ 
& 400 & 0.023 & 0.019 & 0.072 & 0.111 & 0.210 &  & -0.022 & 0.004 & 0.064 & 
0.095 & 0.136 \\ 
400 & 100 & 0.022 & 0.021 & 0.081 & 0.133 & 0.173 &  & -0.021 & 0.012 & 0.076
& 0.127 & 0.165 \\ 
& 200 & 0.021 & 0.020 & 0.078 & 0.123 & 0.158 &  & -0.020 & 0.014 & 0.072 & 
0.113 & 0.135 \\ 
& 400 & 0.020 & 0.019 & 0.079 & 0.125 & 0.151 &  & -0.020 & 0.014 & 0.074 & 
0.115 & 0.107 \\ 
&  & \multicolumn{11}{c}{Panel B: $c=1.2$} \\ \cline{3-13}
100 & 100 & 0.043 & 0.069 & 0.041 & 0.064 & 0.123 &  & -0.042 & -0.061 & 
-0.014 & -0.001 & 0.036 \\ 
& 200 & 0.042 & 0.067 & 0.040 & 0.061 & 0.183 &  & -0.041 & -0.059 & -0.010
& -0.003 & 0.010 \\ 
& 400 & 0.039 & 0.060 & 0.039 & 0.058 & 0.192 &  & -0.038 & -0.053 & -0.008
& 0.001 & 0.011 \\ 
200 & 100 & 0.037 & 0.053 & 0.031 & 0.053 & 0.095 &  & -0.037 & -0.047 & 
0.004 & 0.016 & 0.056 \\ 
& 200 & 0.035 & 0.047 & 0.031 & 0.054 & 0.116 &  & -0.035 & -0.042 & 0.005 & 
0.011 & 0.036 \\ 
& 400 & 0.035 & 0.043 & 0.031 & 0.050 & 0.142 &  & -0.034 & -0.038 & 0.007 & 
0.012 & 0.029 \\ 
400 & 100 & 0.033 & 0.037 & 0.034 & 0.056 & 0.074 &  & -0.033 & -0.031 & 
0.014 & 0.035 & 0.039 \\ 
& 200 & 0.032 & 0.033 & 0.032 & 0.050 & 0.073 &  & -0.032 & -0.029 & 0.014 & 
0.024 & 0.017 \\ 
& 400 & 0.031 & 0.030 & 0.032 & 0.052 & 0.082 &  & -0.031 & -0.026 & 0.018 & 
0.026 & 0.005 \\ \hline
\end{tabular}%
\end{table}

\newpage

\subsection{Factor number estimation with different $r_{\max}$}

\begin{table}[htbp]
	\centering
	\caption{Estimation of factor numbers with different $r_{max}$ under $r=3$}\label{rmax_r3}
	\begin{tabular}{ccccccc}
		\hline
		&       & \multicolumn{2}{c}{$r_{\max}$=7} &       & \multicolumn{2}{c}{$r_{\max}$=9} \\
		\cline{3-4}\cline{6-7}    $N$     & $T$     & \multicolumn{1}{c}{RMSE} & \multicolumn{1}{c}{Bias} &       & \multicolumn{1}{c}{RMSE} & \multicolumn{1}{c}{Bias} \\
		\hline
		100   & 100   & 0.248  & 0.046  &     & 0.391  & 0.140  \\
		& 200   & 0.238  & -0.043  &     & 0.200  & -0.020  \\
		& 400   & 0.309  & -0.092  &     & 0.258  & -0.061  \\
		200   & 100   & 0.177  & 0.026  &     & 0.184  & 0.031  \\
		& 200   & 0.158  & -0.019  &     & 0.140  & -0.010  \\
		& 400   & 0.200  & -0.040  &     & 0.179  & -0.028  \\
		400   & 100   & 0.128  & 0.017  &     & 0.155  & 0.023  \\
		& 200   & 0.071  & -0.003  &     & 0.071  & 0.000  \\
		& 400   & 0.105  & -0.011  &     & 0.087  & -0.007  \\
		\hline
	\end{tabular}%
	\label{tab:addlabel}%
\end{table}%

\begin{table}[htbp]
	\centering
	\caption{Estimation of factor numbers with different $r_{max}$ under $r=5$}\label{rmax_r5}
	\begin{tabular}{ccccccc}
		\hline
		&       & \multicolumn{2}{c}{$r_{\max}$=7} &       & \multicolumn{2}{c}{$r_{\max}$=9} \\
		\cline{3-4}\cline{6-7}    $N$     & $T$     & \multicolumn{1}{c}{RMSE} & \multicolumn{1}{c}{Bias} &       & \multicolumn{1}{c}{RMSE} & \multicolumn{1}{c}{Bias} \\
		\hline
		100   & 100   & 0.252  & 0.014  &       & 0.272  & 0.056  \\
		& 200   & 0.280  & -0.071  &       & 0.219  & -0.037  \\
		& 400   & 0.346  & -0.115  &       & 0.294  & -0.084  \\
		200   & 100   & 0.163  & 0.011  &       & 0.177  & 0.024  \\
		& 200   & 0.170  & -0.022  &       & 0.107  & -0.010  \\
		& 400   & 0.212  & -0.043  &       & 0.166  & -0.025  \\
		400   & 100   & 0.124  & 0.014  &       & 0.150  & 0.023  \\
		& 200   & 0.084  & -0.004  &       & 0.059  & -0.002  \\
		& 400   & 0.120  & -0.015  &       & 0.100  & -0.009  \\
		\hline
	\end{tabular}%
	\label{tab:addlabel}%
\end{table}%

\newpage

\subsection{FDR and Power with smaller gaps of factor strengths}

\begin{table}[h]
	\centering
	\caption{FDR and Power with $\alpha=(0.8,0.75,0.7)$ under $r=3$} \label{smaller_gap}
	\begin{tabular}{ccccccccccc}
		\hline
		$N$ & $T$ & \multicolumn{1}{c}{$\text{FDR}_{1}$} & \multicolumn{1}{c}{$\text{
				FDR}_{2}$} & \multicolumn{1}{c}{$\text{FDR}_{3}$} & \multicolumn{1}{c}{$%
			\overline{\text{FDR}}$} &  & \multicolumn{1}{c}{$\text{Power}_{1}$} & 
		\multicolumn{1}{c}{$\text{Power}_{2}$} & \multicolumn{1}{c}{$\text{Power}
			_{3} $} & \multicolumn{1}{c}{$\overline{\text{Power}}$} \\ \hline
		100   & 100   & 0.331  & 0.454  & 0.606  & 0.407  &       & 0.802  & 0.411  & 0.332  & 0.592  \\
		& 200   & 0.322  & 0.443  & 0.587  & 0.397  &       & 0.836  & 0.421  & 0.370  & 0.619  \\
		& 400   & 0.318  & 0.440  & 0.591  & 0.393  &       & 0.873  & 0.440  & 0.374  & 0.644  \\
		200   & 100   & 0.337  & 0.458  & 0.606  & 0.409  &       & 0.829  & 0.422  & 0.365  & 0.622  \\
		& 200   & 0.331  & 0.462  & 0.570  & 0.400  &       & 0.856  & 0.418  & 0.394  & 0.640  \\
		& 400   & 0.328  & 0.443  & 0.593  & 0.397  &       & 0.891  & 0.442  & 0.398  & 0.666  \\
		400   & 100   & 0.338  & 0.470  & 0.578  & 0.406  &       & 0.846  & 0.428  & 0.422  & 0.653  \\
		& 200   & 0.331  & 0.457  & 0.590  & 0.400  &       & 0.878  & 0.433  & 0.427  & 0.672  \\
		& 400   & 0.328  & 0.463  & 0.580  & 0.397  &       & 0.916  & 0.438  & 0.446  & 0.698  \\
		\hline
	\end{tabular}%
	\label{tab:addlabel}%
\end{table}%

\subsection{The estimation results for PC\ estimators when $r$ is unknown}

While the results reported in Tables \ref{tab:FM_estimate_r3} and \ref%
{tab:FM_estimate_r5} are under known numbers of factors, we also experiment
with estimated numbers of factors by each proposed approach, which is more
realistic and reflects more precisely how the estimation of factor numbers
may affect consequent estimators, and report the results at Tables \ref%
{tab-FM_estimate_unknownr3} and \ref{tab-FM_estimate_unknownr5}. The results
suggest that the main conclusion basically still holds, except for the
comparison of RMSE$^{C}$ when both $N$ and $T$ are relatively small.

\begin{table}[tbph]
\caption{Estimation of factor models when $r=3$ is unknown}
\label{tab-FM_estimate_unknownr3}\centering
\resizebox{\linewidth}{!}{
		\begin{tabular}{cccccccccccccccc}
			\hline
			&       & \multicolumn{4}{c}{$TR^{F}$}     &       & \multicolumn{4}{c}{$TR^{\Lambda}$}     &       & \multicolumn{4}{c}{$RMSE^C$} \\
			\cline{3-6}\cline{8-11}\cline{13-16}    $N$     & $T$     & PC    & Ada   & Deb   & Res   &       & PC    & Ada   & Deb   & Res   &       & PC    & Ada   & Deb   & Res \\
			\hline
  100   & 100   & \textbf{0.922 } & 0.896  & 0.903  & 0.903  &       & 0.715  & 0.897  & 0.873  & \textbf{0.899 } &       & 0.956  & 0.647  & 0.625  & \textbf{0.620 } \\
& 200   & \textbf{0.935 } & 0.904  & 0.906  & 0.906  &       & 0.789  & 0.920  & 0.907  & \textbf{0.921 } &       & 0.941  & 0.617  & 0.599  & \textbf{0.596 } \\
& 400   & \textbf{0.941 } & 0.934  & 0.926  & 0.926  &       & \textbf{0.825 } & 0.796  & 0.805  & 0.809  &       & \textbf{0.935 } & 0.969  & 0.953  & 0.953  \\
200   & 100   & \textbf{0.955 } & 0.936  & 0.935  & 0.935  &       & 0.743  & \textbf{0.908 } & 0.867  & 0.902  &       & 0.883  & 0.588  & 0.565  & \textbf{0.558 } \\
& 200   & \textbf{0.965 } & 0.944  & 0.946  & 0.946  &       & 0.810  & \textbf{0.926 } & 0.904  & 0.924  &       & 0.874  & 0.562  & 0.549  & \textbf{0.544 } \\
& 400   & \textbf{0.969 } & 0.962  & 0.965  & 0.965  &       & 0.852  & 0.836  & 0.863  & \textbf{0.872 } &       & 0.871  & 0.883  & \textbf{0.866 } & \textbf{0.866 } \\
400   & 100   & \textbf{0.969 } & 0.929  & 0.857  & 0.857  &       & \textbf{0.749 } & 0.616  & 0.597  & 0.617  &       & \textbf{0.812 } & 0.892  & 0.871  & 0.870  \\
& 200   & \textbf{0.976 } & 0.960  & 0.966  & 0.966  &       & 0.815  & 0.751  & 0.823  & \textbf{0.848 } &       & \textbf{0.804 } & 0.843  & 0.809  & 0.808  \\
& 400   & \textbf{0.980 } & 0.975  & 0.977  & 0.977  &       & 0.858  & 0.842  & 0.865  & \textbf{0.880 } &       & \textbf{0.797 } & 0.815  & 0.798  & 0.798  \\
			\hline
	\end{tabular}}
\end{table}

\begin{table}[tbph]
\caption{Estimation of factor models when $r=5$ is unknown}
\label{tab-FM_estimate_unknownr5}\centering
\resizebox{\linewidth}{!}{
		\begin{tabular}{cccccccccccccccc}
			\hline
			&       & \multicolumn{4}{c}{$TR^{F}$}     &       & \multicolumn{4}{c}{$TR^{\Lambda}$}     &       & \multicolumn{4}{c}{$RMSE^C$} \\
			\cline{3-6}\cline{8-11}\cline{13-16}   $N$     & $T$     & PC    & Ada   & Deb   & Res   &       & PC    & Ada   & Deb   & Res   &       & PC    & Ada   & Deb   & Res \\
			\hline
 100   & 100   & \textbf{0.946 } & 0.912  & 0.914  & 0.914  &       & 0.774  & 0.890  & 0.895  & \textbf{0.905 } &       & 1.585  & 0.995  & \textbf{0.950 } & \textbf{0.950 } \\
& 200   & \textbf{0.953 } & 0.922  & 0.924  & 0.924  &       & 0.819  & 0.913  & 0.917  & \textbf{0.923 } &       & 1.586  & 0.963  & \textbf{0.937 } & \textbf{0.937 } \\
& 400   & \textbf{0.956 } & 0.945  & 0.952  & 0.952  &       & 0.844  & 0.808  & \textbf{0.849 } & \textbf{0.849 } &       & \textbf{1.588 } & 1.608  & 1.590  & 1.590  \\
200   & 100   & \textbf{0.968 } & 0.945  & 0.947  & 0.947  &       & 0.792  & 0.898  & 0.893  & \textbf{0.908 } &       & 1.495  & 0.940  & 0.897  & \textbf{0.895 } \\
& 200   & \textbf{0.973 } & 0.953  & 0.954  & 0.954  &       & 0.835  & 0.921  & 0.918  & \textbf{0.927 } &       & 1.498  & 0.907  & 0.883  & \textbf{0.882 } \\
& 400   & \textbf{0.976 } & 0.969  & 0.970  & 0.970  &       & 0.860  & 0.830  & 0.862  & \textbf{0.864 } &       & 1.497  & 1.514  & \textbf{1.496 } & 1.497  \\
400   & 100   & \textbf{0.980 } & 0.953  & 0.961  & 0.961  &       & 0.800  & 0.689  & 0.803  & \textbf{0.815 } &       & \textbf{1.420 } & 1.493  & 1.434  & 1.436  \\
& 200   & \textbf{0.983 } & 0.971  & 0.972  & 0.972  &       & 0.842  & 0.773  & 0.844  & \textbf{0.851 } &       & \textbf{1.418 } & 1.482  & 1.442  & 1.443  \\
& 400   & \textbf{0.985 } & 0.979  & 0.980  & 0.980  &       & 0.865  & 0.833  & 0.867  & \textbf{0.872 } &       & 1.422  & 1.439  & \textbf{1.420 } & \textbf{1.420 } \\
			\hline
	\end{tabular}}
\end{table}


\end{document}